\documentstyle[11pt]{article}

\newcommand\eq[1]{Eq.~(\ref{#1})}
\newcommand\eqs[2]{Eqs.~(\ref{#1}) and (\ref{#2})}
\newcommand\eqss[3]{Eqs.~(\ref{#1}), (\ref{#2}) and (\ref{#3})}
\newcommand\eqsss[4]{Eqs.~(\ref{#1}), (\ref{#2}), (\ref{#3})
and (\ref{#4})}
\newcommand\eqssss[5]{Eqs.~(\ref{#1}), (\ref{#2}), (\ref{#3}),
(\ref{#4}) and (\ref{#5})}

\newcommand\rfrac[2]{\left(\frac{#1}{#2}\right)}

\newcommand{\sub}[1]{_{\mbox{\scriptsize#1}}}

\newcommand{\subsub}[1]{_{\mbox{\tiny #1}}}

\newcommand\ee{\end{equation}}
\newcommand\be{\begin{equation}}
\newcommand\eea{\end{eqnarray}}
\newcommand\bea{\begin{eqnarray}}
\newcommand\eqa{\!\!\!&=&\!\!\!}
\newcommand\equiva{\!\!\!&\equiv&\!\!\!}

\newcommand\simeqa{\!\!\!&\simeq &\!\!\!}

\newcommand\sunit{\,\mbox{s}}
\newcommand\TeV{\,\mbox{TeV}}
\newcommand\GeV{\,\mbox{GeV}}
\newcommand\MeV{\,\mbox{MeV}}
\newcommand\eV{\,\mbox{eV}}
\newcommand\km{\,\mbox{km}}

\newcommand\Mpc{\,\mbox{Mpc}}

\newcommand\mone{^{-1}}
\newcommand\mtwo{^{-2}}
\newcommand\mthree{^{-3}}
\newcommand\mfour{^{-4}}
\newcommand\mhalf{^{-1/2}}
\newcommand\mthreehalf{^{-3/2}}
\newcommand\mthird{^{-1/3}}

\newcommand\half{^{1/2}}
\newcommand\threehalf{^{3/2}}
\newcommand\third{^{1/3}}
\newcommand\twothird{^{2/3}}
\newcommand\quarter{^{1/4}}

\newcommand\msun{M_\odot}
\newcommand\mpl{m_{Pl}}

\newcommand\del{{\mbox{\boldmath $\nabla$}}}
\newcommand\bfk{{\bf k}}
\newcommand\bfr{{\bf r}}
\newcommand\bfv{{\bf v}}
\newcommand\bfe{{\bf e}}
\newcommand\bfx{{\bf x}}
\newcommand\bfy{{\bf y}}
\newcommand\sk{_{\mbox{\scriptsize \bf k}}}

\newcommand\pa{\partial}
\newcommand\pdif[2]{\frac{\pa #1}{\pa #2}}

\newcommand\lsim{\mathrel{\rlap{\lower4pt\hbox{\hskip1pt$\sim$}}
    \raise1pt\hbox{$<$}}}
\newcommand\gsim{\mathrel{\rlap{\lower4pt\hbox{\hskip1pt$\sim$}}
    \raise1pt\hbox{$>$}}}

\begin{document}

\begin{titlepage}
\begin{flushright}
SUSSEX-AST 92/8-2; LANC-TH 8-2-92\\
(Original: August 1992)\\
(This final version: March 1993)
\end{flushright}
\begin{center}
{\Large \bf The Cold Dark Matter Density Perturbation\footnote{To appear, {\em
Physics Reports}. Original version entitled `The Spectral Slope in the Cold
Dark Matter Cosmogony'.}\\}
\vspace{.3in}
{\large Andrew R.~Liddle$^{\dagger}$ and David H.~Lyth$^*$\\}
\vspace{.4 cm}
{\em $^{\dagger}$Astronomy Centre, \\ Division of Physics and Astronomy, \\
University of Sussex, \\ Brighton BN1 9QH.~~~U.~K.}\\
\vspace{.4 cm}
{\em $^*$School of Physics and Materials,\\ Lancaster University,\\
Lancaster LA1 4YB.~~~U.~K.}\\
\end{center}

\vspace{.6cm}
\begin{abstract}
\noindent
The outstanding problem in cosmology today is undoubtedly the origin and
evolution of large scale structure. In this context, no model has proved as
successful as the standard Cold Dark Matter (CDM) model, based on a flat
(scale-invariant) spectrum of density fluctuations growing under gravitational
instability. Nevertheless, pressure has recently been exerted on the model
both from analyses of large scale clustering in the galaxy distribution and
from the measurement of the level of microwave background anisotropies by the
Cosmic Background Explorer (COBE) satellite.

In this report we aim to give a unified view of the Cold Dark Matter model,
beginning with the creation of perturbations during an inflationary epoch and
pursuing it right through to comparison with a host of observations. We
discuss in detail the evolution of density perturbations in Friedmann
universes, utilising the fluid flow approach pioneered by Hawking, and provide
a simple derivation of the Sachs--Wolfe effect giving large angle microwave
background anisotropies. We illustrate the means by which inflation provides
an initial spectrum of inhomogeneities, the spectrum having a shape which can
be readily calculated in a given inflationary model. We also include a
discussion of the generation of long wavelength gravitational waves, which
have recently been recognised as having a potentially important role with
regard to microwave background anisotropies.

Although the standard CDM model is based on a scale-invariant spectrum, the
generic prediction of simple inflationary models is for a power-law spectrum,
tilted away from scale-invariance to provide extra large-scale power. For many
models such as chaotic inflation, this tilting is rather modest. However, in
several models, such as power-law inflation, extended inflation and natural
inflation, the tilting of the spectrum can be more dramatic, and potentially
useful in the light of observations indicating a deficit of large-scale power
in the galaxy distribution. In the former two of these, there is the
interesting extra of a substantial production of gravitational waves.

We then discuss observational constraints on cold dark matter cosmogonies
based on power-law spectra. We examine a range of phenomena, including large
angle microwave background fluctuations, clustering in the galaxy
distribution, peculiar velocity flows, the formation of high redshift quasars
and the epoch of structure formation. One of our aims is to compute the
current constraints on both the shape of the spectrum as defined by the
spectral index $n$, and its normalisation as defined by the usual quantity
$\sigma_8\equiv 1/b_8$.

We end by discussing briefly some variants on the CDM model, such as the
incorporation of a hot dark matter component, or the introduction of a
cosmological constant term. We do not however investigate them in the depth
that we do the tilted CDM models, though the techniques illustrated throughout
the paper provide the background required for such an undertaking.

\end{abstract}

E-mail addresses: arl @ uk.ac.sussex.starlink~~;~~
lyth @ uk.ac.lancs.ph.v1
\end{titlepage}

\tableofcontents
\newpage

\section{Introduction}
\label{INTRO}
\setcounter{equation}{0}
\renewcommand\theequation{\thesection.\arabic{equation}}

For many (and probably most) cosmologists, the outstanding problem at present
is to develop an understanding of the origin and evolution of large scale
structure in the universe, with the ultimate goal of explaining such phenomena
as the epoch of galaxy formation, the clustering in the galaxy distribution
and the amplitude and form of anisotropies in the microwave background. The
present time is a particularly exciting one, as experiments through the last
decade have presented great leaps in our knowledge of the universe around us,
and experiments proposed for the years to come promise to further
revolutionise our view. Attempts to model the formation of large scale
structure already face an impressive array of constraints they must obey.

Gravity being the only universally attractive long range force, the formation
of structure must begin with a small perturbation in the otherwise uniform
distribution of matter in the early universe. Two alternative origins have
been proposed for this perturbation. The first is that it originated as a
quantum fluctuation during inflation which became `frozen-in' after horizon
exit; such a mechanism typically yields a Gaussian, adiabatic perturbation
with a more or less scale-invariant spectrum, of the kind which has long been
favoured on grounds of simplicity (Harrison 1970; Zel'dovich 1970). The second
is that it was caused by topological defects like cosmic strings (Kibble 1976;
Zel'dovich 1980; Vilenkin 1981) leading to a more complicated type of
perturbation.

Partly because of its simplicity, the first possibility has been investigated
in far more detail than the second, and it forms the subject of the present
report. A crucial question for this hypothesis is the nature and amount of the
dark matter in the universe. Soon after the need for such matter came to be
widely accepted in the early 1980's, it became clear that the hypothesis
fails completely if the dark matter consists of massive neutrinos, because
their thermal motion wipes out small scale structure (White, Frenk \& Davis
1983). Given the failure of this hot dark matter (HDM) model, attention turned
to the other extreme, of matter which has by definition negligible random
motion. In its standard form this cold dark matter (CDM) model assumes that the
universe has flat spatial geometry, critical matter density and a spectrum
which is precisely scale invariant. Then it is defined by the present Hubble
parameter $H_0$, the baryonic matter density $\Omega_B$ and the normalisation
of the spectrum. For the Hubble parameter, a value
$H_0=50\km\sunit\mone\Mpc\mone$ is usually taken, both in order to have an
acceptably old universe and to make the model itself reasonably viable. The
baryonic content is taken as accurately determined by the standard big bang
nucleosynthesis picture.

For the first few years of its existence, this standard form of the CDM model
was very successful in describing the observed features of structure in the
universe, from the scale of galaxies upwards
(Peebles 1982b; Blumenthal {\it et al } 1984; Bardeen {\it et al} 1986; White
{\it et al} 1987; Frenk {\it et al} 1988; Efstathiou 1990). More recently,
though, the model has run into difficulties and is widely felt to be in need of
modification. One possibility is to `tilt' the spectrum away from its scale
invariant form. Another is to allow the cold dark matter density to be less
than the critical value, making up the difference with either a cosmological
constant or hot dark matter. Of course there are other possibilities, but these
are the ones that have received the most attention at the present time.

\subsection{Generalities}

The standard hot big bang model (see {\it eg} Kolb \& Turner 1990), which
provides the framework in which gravitational instability scenarios are set,
consists of a collection of parameters which are more or less well known. The
rate of expansion of the universe is described by the Hubble parameter $H =
\dot{a}/a$ ($a$ being the scale factor), whose present value is
commonly parametrised as
\begin{equation}
H_0 = 100 \, h \; {\rm km \, s}^{-1} \, {\rm Mpc}^{-1} \; \; \; , \; \;
	h \in [0.4, 1]
\end{equation}
With $c=1$, this can be written accurately as $H_0 = h/3000 \;
{\rm Mpc}^{-1}$. The energy density is usually written in terms of the
critical density
\begin{equation}
\rho_c = \frac{3 \mpl^2 H^2}{8 \pi}
\end{equation}
as $\Omega = \rho/\rho_c$. It is convenient to regard any cosmological constant
as a time independent contribution to $\rho$, and then the case $\Omega=1$
corresponds to flat spatial sections.

Different direct measurements of the Hubble parameter give $h$ in the range
$.4$ to $1$. The present value of $\Omega$ is even more uncertain than $h$,
with any value between roughly 0.1 and 1 permitted by various experiments. The
theorist's favourite, for fairly well founded reasons we shall examine later,
is the $\Omega = 1$ case which describes a universe with flat spatial
sections. Perhaps surprisingly, the density in ordinary matter is far better
known, provided one takes the reasonable step of believing the phenomenally
successful nucleosynthesis calculations (see {\it eg} Walker {\it et al}
1991), as lying in the range\footnote{In the context of cosmology ordinary
matter is usually referred to as baryonic matter since the baryons (protons
and neutrons) far outweigh the electrons. Thus $\Omega_B$ is referred to as
the baryon density.}
\be 0.010 \leq \Omega_B h^2 \leq 0.016 \ee
It is immediately clear that for the theorist's $\Omega = 1$ option to be
correct, the bulk of the universe must be in some as-yet-undetected form of
{\em dark matter}. The nature of this dark matter turns out to be a crucial
ingredient in models of structure formation.

The age of the universe has also been estimated, most reliably from the ages
of globular clusters as upwards of $13$ Gyr. In a flat ($\Omega = 1$)
universe, the age is simply $t = 2H_0^{-1}/3 = 6.5/h$ Gyr, and so one requires
$h<.5 $ for consistency. Combined with the result $h>.4$ from direct
observation, a value $h\simeq.5$ is therefore preferred over higher
values. Should measurements
giving a high Hubble parameter prove correct, one would be forced either to
let $\Omega$ be significantly less than $1$ or to introduce unusual dynamics
such as a cosmological constant in order to reconcile them.

In itself, the hot big bang model has nothing to say on structure formation,
remaining homogeneous for all time. However, prompted by certain conceptual
problems with the hot big bang --- why should the initial conditions be so
homogeneous, why should the universe be so close to the critical density, what
prevented an overabundance of magnetic monopoles, etc --- cosmologists devised
the paradigm of {\em inflation}, whereby the universe in its earliest stages
undergoes a period of accelerated expansion $\ddot{a} > 0$
(Guth 1981; Kolb \& Turner 1990; Linde 1990). It was later
realised that inflation, by a
mechanism we shall shortly discuss, could generate small density
irregularities on large (superhorizon) scales, and that these would
naturally lead to the gaussian, adiabatic, scale invariant density
perturbation which had already been proposed by Harrison and Zel'dovich, and
adopted on grounds of simplicity by most large scale structure workers. With
inflation to provide an initial spectrum of density inhomogeneities, the
spotlight is transferred to developing an understanding of whether or not such
a spectrum can evolve so as to reproduce large scale structure observations.
The evolution of the spectrum depends sensitively on the assumptions (or
observations) one makes as to the matter content of the universe.

In attempting to explain the increasing range of observations, no model has
provided as dramatic an advance in understanding as that which has come to be
known as the standard Cold Dark Matter (CDM) model (White {\it et al} 1987;
Frenk {\it et al} 1988; Efstathiou 1990; Ostriker 1993).
 As implied by its name, the dark
matter is assumed to be cold, which for most purposes can be taken to mean
non-relativistic. By definition,
dark matter does not interact significantly with more
conventional forms of matter by any means other than gravitationally, and in
particular is beneficial for structure formation in that it is not subject to
pressure forces from interaction with radiation which prevent baryonic density
inhomogeneities on scales smaller than superclusters from collapsing before
radiation decouples from matter. Structure can thus start to form earlier in
the dark matter providing initial gravitational wells to kick-start structure
formation in baryonic matter after decoupling.

Over the last decade, the cold dark matter model has been extensively
explored, most notably by Davis, Efstathiou, Frenk and White in a series of
papers. The assumptions above provide a scenario in which calculations and
predictions are readily framed. As long as they remain small, the
inhomogeneities can be split up into each comoving scale evolving separately,
enabling analytic progress to be made. On scales much larger than $8 h^{-1}$
Mpc, such calculations are still applicable today. On smaller scales the
perturbations reach the nonlinear regime and become mode-coupled, and more
complex methods, primarily large $N$-body computer simulations, must be
employed to produce predictions. In the late eighties, the CDM model could
boast impressive successes in reproduction of galaxy clustering statistics,
structure formation epochs and peculiar velocity flows, while keeping
fluctuations in the microwave background at a level safely below the
observational upper limits. The only addition to the model which observation
required is that which remains the most controversial to the present day ---
the question of {\em biassing}.

The concept of biassing arose from the realisation that in the theory one
calculates the distribution of matter in the universe, but in observations one
measures the distribution of galaxies. In general these need not be the same;
the question is whether or not `light traces mass'. The unbiased CDM models
appeared unable to produce enough clustering in the galaxies, and the standard
assumption became that light does not trace mass. The clustering of galaxies,
as measured by the two-point correlation function, was assumed to be greater
than that of the mass by a multiplicative constant, the bias parameter $b$,
originally thought to lie anywhere in the range $1$ to perhaps $2.5$. With the
incorporation of the extra free parameter, CDM was able to reproduce galaxy
distributions, at least out to $10 h^{-1}$ Mpc.

More recently, standard CDM model has come under fire from two directions, and
though neither is as yet completely conclusive these are certainly troubled
times for CDM. The first problem is that on scales $10$ to $100\Mpc$ the
standard CDM model does not give the correct scale-dependence for the galaxy
correlation function; it seems to have too little power on the large scales in
this range, relative to the power on smaller scales. Perhaps the most
prominent data leading to this conclusion is the APM survey (Maddox {\it et
al} 1990, 1991) giving the galaxy angular correlation function, but also
important are the QDOT survey (Saunders {\it et al} 1991), observations of
$X$-ray galaxy
clusters (Lahav {\it et al} 1989) and of radio galaxies (Peacock 1991;
Peacock \& Nicholson 1991), and optical galaxy redshift surveys such as CfA
(Vogeley {\it et al} 1992) and the Southern Sky Redshift Survey (Park, Gott \&
da Costa 1992).

The second problem for standard CDM is that there seems to be a conflict
between the normalisation of the spectrum of the perturbation, as specified
for instance by the usual quantity $\sigma_8\equiv 1/b_8$, which is required
by different types of observation. Perhaps the most reliable normalisation
comes from the first positive measurement of microwave background
anisotropies, reported by the COBE DMR collaboration in April 1992 (Smoot {\it
et al} 1992). It probes the density perturbation on scales of order
$10^3\Mpc$, and requires $\sigma_8=.95\pm.2$. Second, there are surveys of the
galaxy distribution and bulk flow which probe the normalisation on scales of
order $10$ to $50\Mpc$, notably from the QDOT (Kaiser {\it et al} 1991) and
POTENT (Dekel {\it et al} 1992) groups; these seem to require $\sigma_8$ in
the range $.7$ to $1.1$, which is compatible with the COBE normalisation. The
problem comes with observations of the pairwise velocity dispersion of
galaxies on scales $\lsim 3\Mpc$, which according to most calculations seem to
require $\sigma_8<.5$. Gelb, Gradwohl and Frieman (1993) have claimed recently
that through non-linear effects these velocity dispersion observations probe
the normalisation of the density perturbation on a wide range of scales,
extending into the regime $10$ to $50\Mpc$ probed by the bulk flow data. If
that is so, the problem is of a different type from that represented by the
galaxy correlation function data, and cannot cannot really be paraphrased by
stating that there is a relative lack of power on large scales. Recent
calculations of the predicted galaxy cluster abundance (White, Efstathiou \&
Frenk 1993) also indicate less small-scale power than indicated by the bulk
flow data.

Faced with troubling observations but a generally successful model, the
impulse of the community is to re-examine, and if necessary relax, the
underlying assumptions of the standard CDM model. Several means are available
to achieve this, which we shall discuss later. One of the purposes of this
paper is to examine the option which, while not necessarily the most useful,
is to the particle cosmologist the most compelling. That is to examine the
form of the primeval spectrum.

Standard CDM is based on a scale-invariant spectrum. Whilst initially this was
based entirely on aesthetics (it is for instance the only option which does
not require a choice of normalisation scale), it was later justified on the
basis of being a prediction of inflationary cosmology. Nevertheless, it has
long been known, at least within the inflation community, that the predicted
spectrum is only approximately flat, and that within any given model there are
readily calculable deviations from scale invariance. An example is provided by
the standard `chaotic' inflation scenario, (Linde 1983, 1987, 1990), wherein a
logarithmic correction (see {\it eg} Salopek, Bond \& Bardeen 1989; Mukhanov,
Feldman \& Brandenberger 1992; Schaefer \& Shafi 1992) adds some power at
large scales, giving a density contrast at horizon crossing which is
5--10\% greater
at $1000h^{-1}$ Mpc than at $10h^{-1}$ Mpc. Hitherto, when our knowledge of
the primeval spectrum was restricted to studies of local clustering up to
scales of perhaps $100 h^{-1}$ Mpc, these corrections where rightly regarded
as insignificant. The recent results (Smoot {\it et al} 1992) from the Cosmic
Background Explorer satellite (COBE) have changed this picture, by providing
for the first time good estimates of the amplitude of the spectrum on scales
of upwards of $1000 h^{-1}$ Mpc. The time has come for these correction terms
to be taken seriously, and possibly the real test of inflation lies not in how
flat the primeval spectrum is, but in the type and size of deviations from it.

Some inflationary models predict an even more dramatic change from the
Harrison--Zel'dovich case. The simplest is power-law inflation (Abbott \& Wise
1984b; Lucchin \& Matarrese 1985), which can be realised via a scalar field
evolving in a potential of exponential form and which leads to an expansion of
the universe which is not the conventional nearly exponential growth, but
rather has the scale factor growing as a rapid power-law in time. A second
option is provided by the so-called extended inflation scenarios (La \&
Steinhardt 1989; Kolb, Salopek \& Turner 1990; Kolb 1991), wherein
modifications to Einsteinian gravity allow inflation to proceed via a
first-order phase transition. Yet another option is provided by `natural
inflation' (Freese, Frieman \& Olinto 1990; Adams {\it et al} 1993), a
specific case of the more general situation of a scalar field evolving near
the top of an inverted harmonic oscillator potential which also gives a
power-law spectrum. And finally, it is worth emphasising that although
technically logarithmically corrected, models such as chaotic inflation
produce spectra which are excellently approximated across large scale
structure scales by a power-law, though typically much closer to the
scale-invariant case than the genuine power-law models allow. Only in
exceptional cases (we discuss a two-scale inflation model) does inflation seem
capable of providing spectra tilted in the opposite direction, giving extra
small-scale power.

Another potentially vital aspect of inflationary models is that they can
generate long wavelength gravitational waves (Fabbri \& Pollock 1983; Abbott
\& Wise 1985; Starobinsky 1985). Recently, a host of papers
(Krauss \& White 1992; Salopek 1992a, 1992b; Davis {\it et al} 1992b;
Liddle \& Lyth 1992; ; Lidsey \& Coles 1992; Lucchin, Matarrese \& Mollerach
1992; Souradeep \& Sahni 1992) have served to remind the community that while
this can be a very small effect in some inflationary models, it is not
necessarily the case that this be so. We discuss the generation of these modes
during inflation, and emphasise the inflationary models in which this effect
is most important. Typically, a strong gravitational wave component is
associated with tilt, though the converse --- that tilt necessarily implies
gravitational waves --- need not be true.

\subsection{About this paper}

The primary aim of this report is to provide a unified view of the cold dark
matter model based on inflationary cosmology, examining the assumptions made
and testing the observational viability. In the first half of the paper, we
concentrate on developing the theory of the origin and evolution of
perturbations in Friedmann universes. To do this, we utilise an approach,
initiated by Hawking (1966), which is somewhat different from the usual metric
perturbation approach of Lifshitz (1946). In the perfect fluid approximation
it leads to a straightforward derivation of the evolution of the density
perturbation and related quantities, using fluid flow equations which closely
resemble the non-relativistic ones. They cover a variety of physical
situations, ranging from pressure-free adiabatic perturbations to isocurvature
perturbations in two component fluids. On large scales they are applicable in
a very general setting, making no assumptions about the type of dark matter.
We include detailed discussions of peculiar velocities, which we use to
provide a simple derivation of the Sachs--Wolfe effect giving the large scale
cosmic microwave background radiation (cmb) anisotropy arising from the
density perturbation. We also discuss the anisotropy which may arise from
gravitational waves. Then we go on to discuss the origin of these
perturbations in inflationary cosmology, emphasising that their spectra
typically exhibit deviations from scale invariance. We calculate the extent of
the spectral tilt for the most popular models, and also calculate the relative
contribution of gravitational waves to the cmb anisotropy.

In the second half of the paper we move on to the observational side. We
concentrate on CDM models, generalising standard CDM by incorporating
arbitrary power-law spectra as motivated by inflation. Parametrising the
inflation--generated spectrum of density perturbations by its normalisation
$\sigma_8$ {\em and} by its degree of tilt as specified by the spectral index
$n$, we first map out forbidden regions in the $n$-$\sigma_8$ plane coming
from different types of observation. We examine a range of observations
spanning the range of scales from the present horizon size (microwave
background anisotropies) down to the scale of typical galaxies, and find that
if the observations are all taken at face value no choice of the parameter
pair $n$-$\sigma_8$ can fit them all. We end by briefly discussing the
possibility that the cold dark matter makes up only part of the critical
density, with the remainder either a cosmological constant or hot dark matter.
The emphasis in this second half is on providing an illustration of the
techniques used to obtain the constraints, and we caution the reader that the
detailed results may in places be rapidly superceded by improved observations.

\subsection{Some basic facts and assumptions}

Before commencing, let us note some important cosmological parameters and
scales we shall need.

For ease of comparison with other work we often take $h\mone\Mpc$ for the
distance unit. The units are taken to be such that $c=1$ and (in Section
\ref{INFL}) $\hbar=1$, and in these units the Planck scale is defined by
$m_{Pl}=G\mhalf =1.2\times10^{19}\GeV$.

The total energy density is taken to be critical, $\Omega=1$, and except in
Section \ref{MDM} we discount the possibility of a cosmological contribution
to it. The present Hubble parameter is taken to be given by $h=0.5$. The most
important scales are the following (Hogan, Kaiser \& Rees 1982). Expressions
are quoted first with the relevant factors of $h$ to show the scaling, and
then for our choice of $h=0.5$. The mean density at the present epoch is
$\rho_0 = 3 H^2/8\pi G$, which in convenient astrophysical units is
\begin{eqnarray}
\rho_0 & = & 2.78 h^{-1} \times 10^{11} \msun (h^{-1} {\rm Mpc})^{-3}\\
       & = & 6.94 \times 10^{10} \msun {\rm Mpc}^{-3} \quad {\rm for} \; \;
	h=0.5
\end{eqnarray}
where $\msun$ is the solar mass. We normalise the scalar factor to be $a=1$ at
the present, so that it is related to redshift simply by $a = (1+z)^{-1}$.
With this choice, comoving and physical scales coincide at the present, with
physical scales growing proportional to $a$. Thus, comoving scales are
specified by their physical size at the present. The present horizon
size\footnote{This is somewhat notional, being the distance light could travel
were the standard hot big bang extrapolated back to its origin. With
inflation, the true horizon size can be vastly greater, while in practice the
last scattering surface provides a limit to the distance we can see which is
close to this value.} is simply $2H_0^{-1} = 6000 h^{-1} {\rm Mpc} = 12000$
Mpc. The last scattering surface is centred at a redshift $z=1080$, taking on
roughly the form of a gaussian of width $\Delta z = 80$ (Jones \& Wyse 1985).
In comoving units, this width is around $7 h^{-1} {\rm Mpc} = 14$ Mpc, and for
microwave features of this size or smaller the width must be taken into
account. Matter--radiation equality occurred at a redshift $z = 24000 h^2$.

We shall be defining horizon crossing at a given epoch as $k = aH$. We can
thus give the horizon-crossing wavenumbers for these different important
epochs. They are (with $h=.5$)
\begin{eqnarray}
k_{{\rm hor}}^{-1} & = & 6000 {\rm Mpc}\\
k_{{\rm dec}}^{-1} & = & 180 {\rm Mpc}\\
k_{{\rm eq}}^{-1} & = & 79 {\rm Mpc}
\end{eqnarray}

\section{The density perturbation}
\label{DENSPER}
\setcounter{equation}{0}
\renewcommand\theequation{\thesection.\arabic{equation}}

\subsection{Cosmological perturbation theory}

Departures from homogeneity and isotropy are small in the early universe, and
even at the present epoch on sufficiently large scales. In this situation one
can use cosmological perturbation theory, which develops linear equations for
the perturbations which are valid to first order. This means that the
equations become exact if each perturbation is multiplied by a common
parameter which tends to zero.
A perturbation can be either a small change in some quantity which is non-zero
in the limit of homogeneity and isotropy, or the entire value of a quantity
which vanishes in that limit.

Two essentially different ways of handling the perturbations have been
developed in the literature. The usual one, due to Lifshitz (1946),
works with coordinates so that perturbations of the metric tensor
components are involved. The other, due to Hawking (1966), is a
coordinate-free approach. The former approach is described in many
textbooks and reviews (Weinberg 1972; Peebles 1980; Mukhanov, Feldman
\& Brandenberger 1992) and we shall use it now to explain how
cosmological perturbations may be divided into three independent modes.
Then we shall abandon it in favour of Hawking's approach, which is
simpler for the mode which involves the density perturbation.

\subsubsection*{The metric perturbation approach}

In discussing perturbations, one
can employ any coordinate system which reduces to one of the standard choices
in the limit where the perturbations vanish.
In this limit the universe is homogeneous and isotropic, and its energy
momentum tensor necessarily has the perfect fluid form
\be T^{\mu\nu}=p g^{\mu\nu}+ (\rho+p) u^\mu u^\nu \ee
where $\rho$ is the energy density, $p$ is the pressure and $u^\mu$ is the
fluid four-velocity. The usual choice of coordinates in this limit
corresponds to the line element
\be ds^2=dt^2-a^2(t)\delta_{ij} dx^i dx^j  \ee
Then $\rho$ and $p$ depend only on $t$, and $u^\mu=(1,0,0,0)$.

The perturbations in the metric tensor components may then be defined by
\be ds^2=(1+h_{00})dt^2+ 2h_{0i}dt dx^i - a^2(t)(\delta_{ij} +h_{ij})dx^i
	dx^j  \label{10} \ee
Ignoring the phenomena of particle diffusion and free-streaming
the energy-momentum tensor continues to have the perfect fluid form,
its perturbation being defined by
the perturbations $\delta\rho$, $\delta p$ and $\delta u^\mu=(0,v^i)$,
where $v^i$ is
the three-velocity of the fluid in the chosen coordinate system.

Einstein's field equation yields a set of linear partial differential
equations involving the metric perturbations, $\delta\rho$, $\delta p$ and
$v^i$. They were written down first by Lifshitz (1946). For the case of
critical density which we are considering here, each perturbation `lives' in
flat space, and can be expanded as a Fourier series in a comoving box. One
then has ordinary differential equations which do not couple separate Fourier
coefficients. The following three modes
propagate independently
\begin{itemize}
\item The perturbations $\delta\rho$ and $\delta p$, and the
irrotational part of $v^i$.
\item
The rotational part of $v^i$ (vorticity).
\item
Gravitational waves, characterised by the traceless transverse
part of $h_{ij}$.
\end{itemize}
Each of the first two modes is also associated with a definite type of
metric perturbation, which we have not bothered to specify.
These modes are usually referred to as respectively scalar, vector and
tensor modes because of the spatial transformation properties of the
metric components, but more instructive labels are `density',
`vorticity' and `gravitational wave' modes.

Vorticity decays with time, making the early universe less homogeneous
than the present one, so it is presumably absent. Gravitational waves
on cosmological scales might be generated during inflation and we shall
need to consider them later. For the moment, though, our interest is
in the mode involving the density perturbation. This mode describes
the perturbation in the density and motion of the matter and radiation
in the universe, and can be generalised to include the effects of
particle diffusion and free streaming  (in principle
it then couples to the other modes, but in practice the coupling is
completely negligible).

For a given perturbed space-time the perturbations in the metric and the
energy-momentum tensor are uniquely defined only when a definite coordinate
choice has been made. A coordinate choice, or
more loosely a family of coordinate choices,
is called a gauge. A coordinate choice can be thought of as a slicing of
spacetime into spacelike hypersurfaces (with constant time coordinate) plus a
threading of spacetime into timelike lines (with constant space coordinates).
For the density mode the choice of gauge has a big influence on the appearance
of the equations, and largely for this reason the metric perturbation
formalism of Lifshitz was worked over by many subsequent authors as described
in several text books and reviews (Weinberg 1972; Peebles 1980; Mukhanov,
Feldman \& Brandenberger 1992).

The most widely used gauge, at least until recently, was the synchronous
gauge. In full generality the synchronous gauge employs arbitrarily chosen
geodesics and the hypersurfaces orthogonal to them.
Thus in \eq{10} $h_{0i}=0$ and $h_{00}=1$. In practice, solutions of the
synchronous gauge equations are selected which correspond to geodesics which
become
comoving in the limit of very early times (Lyth \& Stewart 1990b), and from
now on the term `synchronous gauge' will denote this particular choice.

In an influential paper, Bardeen (1980) considered several other gauges, among
them the `comoving' gauge in which comoving worldlines are used instead of
geodesics, along with the hypersurfaces orthogonal to them which are
conveniently termed `comoving hypersurfaces'. The comoving gauge was shown by
Lyth (1985) to lead, on each scale, to particularly simple equations well
before horizon entry as will be explained below.
Bardeen also developed a `gauge invariant' formalism
which we shall not need.

\subsubsection*{The fluid flow approach}

A description of perturbations which makes no mention of the metric
perturbation was proposed by Hawking (1966). For the density mode of the
perturbation it is simpler than the metric perturbation formalism, at least
when particle diffusion and free streaming are negligible. For that mode, it
boils down to evolving the quantities of interest along each comoving
worldline, these quantities being the energy density $\rho$, the pressure $p$
and the locally defined Hubble parameter $H$, all measured with respect to the
worldline. The evolution equations are a pair of fluid flow equations, which
hardly differ in form from their non-relativistic counterparts.

To implement Hawking's approach for this type of perturbation, one has to
introduce spacelike hypersurfaces, on which each quantity is given as the sum
of an average plus a perturbation. As in the metric perturbation
approach, one can derive differential equations for the Fourier components of
the perturbations. The form of the equations is uniquely determined once the
hypersurfaces have been chosen, and is of course the same as that obtained
from the metric perturbation approach, using any gauge whose time coordinate
is constant on each hypersurface. The big advantage of this approach for the
density mode is that it eliminates all mention of the perturbation in the
choice of space coordinates which is defined by threading timelike lines
through the spacelike hypersurfaces, and of the associated metric
perturbations $h_{ij}$ and $h_{i0}$. One needs only the perturbation $h_{00}$,
specifying the proper time separation $d\tau$ of adjacent hypersurfaces,
which is obviously defined once the slicing into hypersurfaces has been
defined.
Because of this, we shall from now on use the term `gauge' to denote simply a
slicing of spacetime into hypersurfaces.

Hawking's original implementation of his approach was valid only for the case
of zero pressure gradient. It was carried through for nonzero pressure
gradient by Olson (1976) in what amounted to the synchronous gauge, and by
Lyth and Mukherjee (1988) in the comoving gauge. The approach has since been
discussed in detail, and extended to the case of non-critical density and
non-isotropic stress, by Ellis, Bruni and others, in a series of papers which
can be traced from that of Bruni, Dunsby and Ellis (1992).

However it is treated, the list of perturbations given above is incomplete
because it does not mention the constituents of the universe. For our purpose
these are nuclei, electrons and cold dark matter particles (matter), as well
as  photons and neutrinos (radiation). Until Section \ref{MDM} the neutrinos
are assumed to be massless, corresponding to the absence of hot dark matter.
The usual assumption is that long before horizon entry the perturbations
satisfy the {\em adiabatic} condition. This implies that there is a common
radiation density contrast and a common matter density contrast, related by
$\delta\rho_m/\rho_m= \frac34\delta\rho_r/\rho_r$. A less natural initial
condition is the {\it isocurvature} condition, that there is negligible
perturbation in the total energy density. The most general initial density
perturbation is the sum of an adiabatic and an isocurvature one. For an
adiabatic perturbation, the cosmological perturbation theory described above
gives an unambiguous description of very large scale perturbations, which
enter the horizon well after matter domination. In particular it allows one to
describe their effect on the cosmic microwave background radiation (cmb),
which is their only observable signature, the necessary formalism being worked
out first by Sachs and Wolfe (1969).

On smaller scales additional considerations become necessary. At the very
least, one has to take into account the separate evolution of the
perturbations of the radiation and the dark matter, on the assumption that
both are perfect fluids. The necessary equations in the synchronous gauge are
given in, for example, the text of Peebles (1980). Simpler equivalent
equations in the comoving gauge were derived by Kodama and Sasaki (1984, 1987)
using the metric perturbation approach, and rederived by Lyth and Stewart
(1990b) using the fluid flow approach. To obtain accurate results one usually
has to go beyond the perfect fluid approximation, to deal with free streaming
and diffusion of the radiation. The necessary formalism is given in the text
of Peebles (1980) and the review of Efstathiou (1990), in the synchronous
gauge.

In this report we give a treatment of the perturbations which is complete as
far as large scales are concerned and is also rather simple. Using fluid flow
equations and the comoving gauge, we use Hawking's approach to track an
adiabatic density perturbation from a supposed inflationary origin to the
present. A simple treatment of large scales is given, including the
Sachs--Wolfe effect.  Then the same thing is done for `isocurvature'
perturbations, using a `two-fluid' approach. Finally, a possible contribution
to the large scale cmb anisotropy from gravitational waves is described,
starting with the supposed inflationary origin of such waves and evolving them
to the present.

On smaller scales our treatment is less complete because the effects of free
streaming and diffusion are not described (save in a very simple
approximation). Nevertheless we are able to deduce the qualitative behaviour
of the transfer function relating the final, evolved density perturbation to
the primeval one. Parametrisations of the adiabatic transfer function which
various authors have obtained by taking into account free-streaming and
(sometimes) diffusion are considered at this point, with an eye to the second
half of the paper where theory and observation are compared. The CDM
prediction for the smaller scale cmb anisotropy is also displayed, again
without detailed discussion.

Although the cold dark matter scenario is the focus of the report as a whole,
the part of it dealing with cosmological perturbation theory has much wider
application. In particular, the treatment of perturbations remains valid
before horizon entry even if the dark matter is not cold. Provided that there
is no cosmological constant it remains valid to the present, for those very
large scales which dominate the large scale cmb anisotropy. The effect of a
cosmological constant is easy to evaluate using the formalism, though we have
not given the details.

\subsection{Fluid flow equations}

We are supposing that the early universe may be regarded as a fluid, with a
smoothly varying four-velocity field $u^\mu$. At each point in space-time, a
comoving observer is defined as one with this four-velocity. Relative to a
comoving observer, the momentum density is zero. More formally, at each point
in space-time the time-space components $T_{0i}$ of energy-momentum tensor
vanish, in an orthonormal  basis such that $u^\mu=(1,0,0,0)$. We shall call
such a basis a comoving one. At least at sufficiently early and late times one
can use the perfect fluid approximation, which means that in a comoving basis
the space-space components have the form $T_{ij}=-p\delta_{ij}$,
where $p$ is the
pressure. The energy density $T_{00}$ in this basis is called $\rho$.

Around a given point in space-time, it is convenient to use a locally inertial
coordinate system in which the fluid is momentarily at rest. Then $u^\mu=
(1,0,0,0)$ at the point, and in a small region around it the space components
$u^i$ define the three velocity of the fluid. The remaining component is given
by
\be u^0=(1-u^i u_i)\half \label{11} \ee
At the point, the Einstein field equation is equivalent to
\bea
\dot\rho\eqa-3H(\rho+p) \label{12} \\
\dot u_i\eqa-\frac{\pa_i p}{\rho+p}
\label{13} \eea
Here, $\pa_i$ ($i$=1,2,3) denotes differentiation with respect to the locally
inertial space coordinates, and the dot denotes differentiation with respect
to the locally inertial time coordinate, which may be identified with proper
time along the comoving worldline passing through the point. The first
equation is equivalent to the energy conservation equation $dE=-p dV$ for the
energy $E$ in a comoving volume. The second is the familiar `acceleration
equals force divided by mass', except that the mass density is replaced by
$\rho+p$ (not just by the energy density $\rho$).

The spatial derivatives of the three-velocity may be split uniquely into an
antisymmetric vorticity $\omega_{ij}$, a symmetric traceless shear
$\sigma_{ij}$, and a locally defined Hubble parameter $H$, \be
\pa_i u_j=\omega_{ij}+\sigma_{ij}+\delta_{ij} H \label{14} \ee From \eq{11},
the
derivatives of the time component vanish at the point where the fluid is at
rest, \be \dot u_0 = \pa_i u_0 =0 \label{15} \ee These equations hold at each
space-time point, in a locally inertial coordinate system in which the fluid
is instantaneously at rest.

If the vorticity vanishes throughout some region of space-time, there exist
hypersurfaces orthogonal to the flow lines, which we shall call comoving
hypersurfaces. They are the surfaces of simultaneity in the neighbourhood of
each comoving observer.

\subsubsection*{The unperturbed universe}

In a perfectly homogeneous and isotropic universe, the vorticity and shear are
zero, and $H$, $\rho$ and $P$ are constant on each comoving hypersurface.
Between any two comoving hypersurfaces, the proper time interval along a
comoving worldline is independent of position, and may be used as a time
coordinate labelling the hypersurfaces. From the equation of motion \eq{13}
one may derive the Raychaudhuri equation (Raychaudhuri 1955, 1979; Ehlers
1961)
\be \dot H= -H^2-\frac{4\pi G}{3} (\rho+3p) \label{16} \ee
Combining it with the energy conservation equation \eq{12} leads to the
Friedmann equation
\be H^2=\frac{8\pi G}{3} \rho-\frac{K}{a^2} \label{17} \ee
where $K$ is a constant. As usual $a$ denotes the scale factor of the
universe, satisfying $H=\dot a/a$ and normalised so that $a=1$ at the present
epoch. We are assuming critical density, corresponding to $K=0$, so
\be H^2=\frac{8\pi G}{3} \rho \label{17a} \ee
As explained in Appendix A, the
quantity $K$ is related to the intrinsic curvature scalar of comoving
hypersurfaces by
\be R^{(3)}=6\frac{K}{a^2} \label{18} \ee
For critical density $R^{(3)}$ vanishes and the comoving hypersurfaces are
flat (the curvature scalar determines their curvature entirely, because they
are homogeneous). As a result, the space coordinates can be taken to be
Cartesian. It is convenient to use comoving coordinates $x^i$, which are
related to Cartesian coordinates $y^i$ by $y^i=a x^i$, and are constant along
each comoving worldline.

{}From \eqs{16}{17a}, it follows that during any era when $w\equiv p/\rho$ is
constant,
\be H=\frac{2}{3+3w} t\mone \label{19} \ee
and therefore $a\propto t^{2/(3+3w)}$. We are interested in matter domination,
$w=0$, and radiation domination, $w=1/3$.

\subsection{The perturbed fluid flow equations}

Now we consider perturbations away from homogeneity and isotropy. The
vorticity is expected to be negligible, and can in any case be eliminated by
redefining the fluid four-velocity as indicated in Section \ref{DENSPER}.8.
This having being done, comoving hypersurfaces exist. On each
hypersurfaces one can define
average plus a perturbation,
\bea \rho(\bfx,t)\eqa \bar\rho(t)+\delta\rho(\bfx,t)
\label{110} \\
p(\bfx,t)\eqa \bar p(t)+\delta p(\bfx,t)
\label{110a} \\
H(\bfx,t)\eqa \overline H(t)+\delta H(\bfx,t)
\label{110b} \eea
Here $t$ is the time coordinate labelling the hypersurfaces, and ${\bf
x}=(x^1,x^2,x^3)$ are space coordinates. We would like to choose them to be
comoving coordinates, related to Cartesian coordinates by $y^i=a x^i$, with
$a$ the average scale given by $\dot a/a=\overline
H$ (we shall not have occasion
to define a perturbed scale factor). This cannot be done exactly, because the
comoving hypersurfaces are not flat in the  presence of perturbations. However
the departure from flatness is of first order, and can therefore be ignored
when consider perturbations which are themselves of first order. In other
words {\it all perturbations `live' in flat space.}

\subsubsection*{Independent scales}

Each perturbation $f$ can be written as a Fourier series,
defined in a comoving box much bigger than the observable
universe
\be \delta f({\bfx},t)=\sum\sk \delta f\sk(t)
	e^{i
{\mbox{\scriptsize{\bf k}.{\bf x}}}
} \label{116} \ee
The beauty of this expansion is that each Fourier mode propagates
independently, as long as the cosmological perturbation theory that we are
developing here is valid. The inverse wavenumber $a/k$ is said to define a
scale, which is specified by giving its present value $k\mone$. The term scale
is, of course, appropriate because a feature --- such as a density enhancement
that will later become a galaxy --- with size $r$ is dominated by wavenumbers
of order $1/r$.

\subsubsection*{The evolution of the density perturbation}

Now we derive differential equations for the perturbations. In doing so
we have to remember that the comoving worldlines are not in general
geodesics, because of the pressure gradient. As a result,
the proper time interval $d\tau$ between a pair of
comoving hypersurfaces is position dependent. Its average may be identified
with the coordinate time, and then its variation with position is given by
\be \frac{d\tau}{dt}=\left(1-\frac{\delta p}{\rho+p} \right)
\label{111} \ee
A simple proof of this important formula is given in Appendix A. When
considering perturbations the variation is a second order effect and can be
ignored, so that coordinate time is identified with proper time. On the other
hand, the variation is a first order effect when considering the unperturbed
(average) value of a quantity, and has to be included. This fact was not
recognised in Hawking's pioneering development of the fluid flow approach.

The energy conservation condition \eq{12} is not affected by the perturbations,
but the Raychaudhuri equation \eq{16} becomes to first order
(Lyth \& Stewart 1990b)
\be \dot H= -H^2-\frac{4\pi G}{3} (\rho+3p) -\frac13 \frac
	{\nabla^2 \delta p}{\rho +p} \label{112} \ee
Here $\nabla^2$ is the Laplacian on a comoving hypersurface, given in terms of
comoving coordinates by
\be \nabla^2=a\mtwo \delta^{ij} \pa_i \pa_j \label{113} \ee

The energy conservation equation \eq{12} and the Raychaudhuri equation \eq{112}
determine the evolution of the energy density and the Hubble parameter along
each worldline, including first order perturbations away homogeneity and
isotropy. Averaging them over a comoving hypersurface, one verifies that
$\bar\rho$ and $\overline H$ satisfy the unperturbed equations \eqs{12}{16}.
Subtracting the averages and using \eq{111} gives
\bea (\delta \rho)\dot{}\eqa -3(\rho+p)\delta H-3H \delta\rho
\label{114} \\
(\delta H)\dot{} \eqa 2 H \delta H-\frac{4\pi G}{3}\delta\rho
-\frac13\frac{\nabla^2\delta p}{\rho+p}
\label{115} \eea
The dots denote differentiation with respect to time.

Remarkably, the pressure perturbation appears in these equations only through
its spatial gradient, because of the relation \eq{111} between proper time and
coordinate time on comoving hypersurfaces. For any other choice, such as a
synchronous one, there would have been additional terms which involve $\delta
P$ itself, not just its gradient.

For the Fourier components \eq{115} becomes
\be
(\delta H \sk)\dot{}= 2 H \delta H\sk
-\frac{4\pi G}{3}\delta\rho\sk
+\frac13\rfrac{k}{a}^2\frac{\delta p\sk}{\rho+p}
\label{117} \ee
Eliminating $\delta H\sk$ with \eq{114} gives a linear
second order differential equation for each Fourier component of the density
perturbation. Often it is convenient to use the {\em density contrast}
\be \delta\equiv \frac{\delta\rho}{\rho} \ee
and in terms of this quantity the equation is
\be H\mtwo\ddot{\delta}\sk  +[2-3(2w-c_s^2)]H\mone \dot{\delta}\sk
	-\frac32(1-6c_s^2+8w-3w^2)\delta\sk =
	-\left(\frac{k}{aH}\right)^2 \frac{\delta p\sk }{\rho}
	\label{118} \ee
where $w=p/\rho$ and $c_s^2=\dot p/\dot \rho$. This equation was first given
(with a different choice of variables) by Bardeen (1980).

\subsubsection*{Horizon entry}

The right hand side of \eq{118} involves the ratio $aH/k$ of the scale
$a/k$ to the Hubble distance $H\mone$. As long as gravity is
attractive, so that $\ddot a<0$, it is a decreasing function of time.  Each
scale is said to {\it enter the horizon} at the epoch $aH/k=1$. Before horizon
entry, when $aH/k>1$ it is said to be outside the horizon, and afterwards to
be inside the horizon. As we shall see, the factor $aH/k$ is ubiquitous in
cosmological perturbation theory, and quite different physics operates in the
eras before and after horizon entry.

\subsection{The case of zero pressure gradient}

The right hand side of \eq{118}, which involves the pressure gradient, will be
negligible well before horizon entry unless $\delta p$ is far bigger than
$\delta \rho$. As we shall see this seemingly bizarre circumstance does occur
if the initial conditions are what are called isocurvature, but it does not
occur with the standard adiabatic initial condition which we shall take for
granted unless otherwise stated. The pressure gradient is negligible also
after matter domination, simply because there is no significant source of
pressure. In these circumstances \eq{118} can be reduced to a first order
equation, which has a very simple interpretation (Lyth 1985; Lyth \& Mukherjee
1987).

To derive the solution and its interpretation, it is best to go back to the
original fluid flow equations \eqs{12}{112}. They are valid in the presence of
first order perturbations, and we see that if the pressure gradient is
negligible they reduce to the unperturbed equations. {\it Each comoving region
of the universe evolves as if there were no perturbation, when the pressure
gradient is negligible.} As a result, the unperturbed Friedmann equation is
valid for every comoving region, with a time independent value of $K$ which is
still related to the spatial curvature scalar by \eq{18}. We are assuming that
the average of $K$ on each hypersurface vanishes (ie., the average density is
critical with respect to the average Hubble parameter), so that the Friedmann
equation becomes
\be H^2=\frac{8\pi G}{3}\rho -\frac{\delta K}{a^2}
\label{119} \ee
This equation is valid to first order in the perturbations, and to that order
we can ignore the perturbation of $a$ in the last term. The perturbations
therefore satisfy
\be 2H\delta H\sk =\frac{8\pi G}{3} \delta \rho \sk
-\frac {\delta K\sk }{a^2} \label{120}  \ee

Combining \eqs{114}{120} one finds a {\it first} order differential equation
for the density contrast,
\be \frac{2H\mone}{5+3w} \frac{d}{dt}
\left[ \rfrac{aH}{k}^2 \delta\sk \right]
+ \rfrac{aH}{k}^2 \delta\sk
=\frac{2+2w}{5+3w} {{\cal R}}\sk  \label{121} \ee
where $w=p/\rho$.

The {\it time-independent} quantity ${{\cal R}}\sk $ on the right hand side
of this equation is defined by
\be {{\cal R}} \sk =\frac32 \frac{\delta K\sk }{k^2}=
\frac14 \rfrac{a}{k}^2 R^{(3)} \sk  \label{122} \ee
The second equality uses \eq{18}, which as we show in Appendix A remains valid
in the presence of the perturbation. According to this equality ${{\cal
R}}\sk$ is a measure of the perturbation in the spatial curvature of comoving
hypersurfaces. Although ${{\cal R}}$ is extremely useful in cosmological
perturbation there is unfortunately no standard notation for it. The first
author to introduce it seems to have been Bardeen (1980), who called it
$\phi_m$. He noted that it is constant outside the horizon, but did not relate
it to the comoving density perturbation. Kodama and Sasaki (1984) called it
${{\cal R}}_m$ and noted (in somewhat obscure language) its simple relation
\eq{532} with the inflaton field perturbation. On scales far outside the
horizon it coincides with $\zeta/3$ of Bardeen, Steinhardt \& Turner (1983)
and the $\zeta$ of Salopek, Bond \& Bardeen (1989).

During any era when $w$ is constant, \eq{121} has a solution
\be \rfrac{aH}{k}^2 \delta\sk =\frac{2+2w}{5+3w} {{\cal R}} \sk
\label{123} \ee
The most general solution is this one, plus a solution of the homogeneous
equation
\be \rfrac{aH}{k}^2 \delta\sk \propto t^{-[(5+3w)/(3+3w)]}
\label{124} \ee
The latter decays, so that a few Hubble times after the onset of the epoch in
question the density contrast will be given by \eq{123}.

\eq{124} is valid only when the pressure gradient is negligible. Thus it
is valid after matter domination,
\be \rfrac{aH}{k}^2 \delta\sk = \frac25 {{\cal R}}\sk (\mbox{final})
\label{126} \ee
and it is also valid during an
initial era, well before
horizon entry and well before matter domination, giving
\be
\rfrac{aH}{k}^2 \delta\sk =\frac49 {{\cal R}}\sk (\mbox{initial})
\label{125} \ee

\subsection{Relation to the synchronous gauge}

So far we have worked exclusively in the comoving gauge. Many authors
work in the synchronous gauge described in Section \ref{DENSPER}.1. As we now
explain, the relation between the gauges is very simple during the initial and
final eras (Lyth \& Stewart 1990b).

As discussed in Section 2.1, each gauge corresponds to a slicing of
spacetime into
hypersurfaces, the density perturbation in that gauge being defined by
\eq{110}.
Choosing the averages to be the same, which corresponds to evaluating the
perturbations at the same coordinate time $t$,
one learns that the synchronous gauge
perturbation $\delta\rho_s$ is related to the comoving perturbation
$\delta\rho$ by
\be \delta\rho_s=\delta\rho +\dot{\rho} \, \delta t \ee
where $\delta t$ is the time displacement going from the comoving to the
synchronous hypersurface. From \eq{111} it is given by
\be \delta t=\int^t_0\frac{\delta p}{\rho+p} dt \ee
where $\delta p$, like $\delta\rho$ is to be evaluated in the comoving
gauge.

During the initial era, $\rho\propto p\propto a\mfour$ and $\delta p\propto
\delta\rho\propto a\mtwo$ with $a\propto t\half$, leading to
$\delta\rho_s=\frac43 \delta\rho$. Thus the synchronous and comoving density
perturbations are identical except for a {\em scale-independent} factor. After
matter domination $\delta p=0$ so that $\delta t$ remains constant, and
therefore quickly becomes negligible compared with the timescale $t$. Thus
{\em the synchronous and comoving gauges become identical after matter
domination}. We shall need these important facts in what follows.

\subsection{The transfer function}

For scales entering the horizon well after matter domination
($k\mone\gg k\mone\sub{eq} =80\Mpc$) the initial and final eras overlap so that
${{\cal R}}\sk (\mbox{initial})={{\cal R}}\sk (\mbox{final})$. Then,
\eqs{125}{126} determine the final density contrast in terms of the initial
one. On smaller scales, there is a linear transfer function $T(k)$,
which may be  defined by
\be {{\cal R}}\sk (\mbox{final})= T(k) {{\cal R}}\sk (\mbox{initial})
\label{127} \ee
An equivalent, and more usual, definition is
\be a\mone \delta\sk |\sub{final}
	=A T(k) \delta\sk|\sub{initial} \label{127a} \ee
where the (time dependent) right hand side is evaluated at an
arbitrarily chosen time during the initial era, and the constant $A$ is
chosen so that $T$ becomes equal to 1 on large scales.
{}From the results of the last section, this second definition can be used also
to define a synchronous gauge transfer function, which is {\em identical} with
the comoving one.

To calculate the transfer function, one needs an initial condition specifying
the relative abundance of the neutrinos, photons, baryons and cold dark matter
long before horizon entry. The most natural condition, which we adopt
in this subsection, is that
the abundances of all particle species are uniform on hypersurfaces of
constant total energy density. This is called the {\em adiabatic} condition,
for a reason that will be clear in a moment. It is indeed a reasonable one,
because at very early times a given comoving length scale far exceeds the
Hubble distance, over which causal processes can operate on the Hubble
timescale. On the other hand, as we discuss in Section \ref{DENSPER}.7, it is
not absolutely mandatory.

Adopting the adiabatic initial condition, consider a hypersurface displaced
from a hypersurface of constant density by time $\delta t(\bfx)$. The density
perturbation of any species $X$ is given by $\delta\rho\sub{X}=\dot\rho\sub{X}
\delta t $, so the perturbations of two species $X$ and $Y$ are related by
\be \frac{\delta\rho\sub{X}}{\dot\rho\sub{X}}
	-\frac{\delta\rho\sub{Y}}{\dot
	\rho\sub{Y}} =0\label{128} \ee
To first order in the perturbations, $\rho\sub{X}$ and $\rho\sub{Y}$ may be
taken to have their unperturbed values (their averages over a comoving
hypersurface). They each therefore satisfy the energy conservation condition
\eq{12},
\be \dot\rho\sub{X}=-3H(\rho\sub{X}+p\sub{X}) \label{129} \ee
It follows that each species of radiation has a common density contrast
$\delta_r$, and each species of matter has a common density contrast
$\delta_m$, with the relation
\be \delta_m-\frac34 \delta_r =0 \label{130} \ee
This relation holds in both the comoving and synchronous gauges.

Given the adiabatic initial condition, the transfer function is determined by
the physical processes occurring between horizon entry and matter domination.
If the radiation behaves as a perfect fluid, its density perturbation
oscillates during this era, with decreasing amplitude. The matter density
contrast living in this background does not grow appreciably before matter
domination because it has negligible self gravity. The transfer function is
therefore given roughly by
\bea T(k)&\!\!\!\!\!\!\!\!\!\!=1 \hspace{5mm}&(k<k\sub{eq}=80\Mpc\mone)
 \nonumber \\
T(k)&= (k\sub{eq}/k)^2 &(k>k\sub{eq}) \label{132}
\eea
An accurate description of the evolution under this perfect fluid assumption
is provided by \eq{118}, together with \eq{142} of Section \ref{DENSPER}.7
which describes the evolution of the entropy perturbation. Equivalent
equations exist in the synchronous gauge, and they were integrated numerically
by Peebles (1982a) (without including the neutrinos) to give the first
estimate of the transfer function.

The perfect fluid description of the radiation is far from correct after
horizon entry, because roughly half of the radiation consists of neutrinos
whose perturbation rapidly disappears through free-streaming. The photons are
also not a perfect fluid because they diffuse significantly, except between
the Silk scale $k\mone\sim 1\Mpc$ and the horizon scale at decoupling
$k\mone=180\Mpc$. One might therefore consider the opposite assumption, which
is that the radiation has zero perturbation after horizon entry. Then the
matter density perturbation evolves according to \eq{118}, with $\delta$ and
$\rho$ now referring to the matter alone, \be \ddot \delta + 2 H \dot \delta
-4\pi G \rho \delta =0 \label{133} \ee (This equation follows also from
Newtonian physics, as is clear from the fact that the fluid flow equations
\eqs{12}{13} coincide with the Newtonian ones for $p=0$.) The growing solution
is
\be \delta=A(1+1.5 y) \label{134} \ee
and the decaying solution is
\be \delta=B\left\{
(1+1.5y)\left[\ln\rfrac{(1+y)\half+1}{(1+y)\half-1}
\right] -3(1+y)\half \right\} \label{135} \ee
where $A$ and $b$ are constants and $y=a/a\sub{eq}$. Even the growing solution
increases only by a factor of order 2 between horizon entry (when \eq{133}
first becomes valid) and the epoch $a\sub{eq}$ when matter starts to dominate.
The transfer function is therefore again given roughly by \eq{132}.

A more sophisticated procedure (Blumenthal \& Primack 1984; Primack \&
Blumenthal 1984) is to keep both modes, matching to the (approximately
correct) time dependence $\delta\propto(aH)\mtwo$ at horizon entry, and one
then finds a somewhat bigger transfer function on small scales,
with \eq{132} replaced by $k\mtwo\ln k$.

Since the radiation consists of roughly half neutrinos, which free
stream, and half photons which either form a perfect fluid or
only diffuse, neither the perfect fluid nor the free-streaming approximation
looks very sensible. A proper calculation
should presumably take quantitative account of the following effects
\begin{itemize}
\item Neutrino free-streaming around the epoch of horizon entry.
\item The diffusion/free-streaming of photons around the time of horizon entry
(except on scales well above the Silk scale $k\mone\sim 1\Mpc$ and well below
the horizon scale at decoupling $k\mone=180\Mpc$).
\item The diffusion of the baryons along with the photons.
\item The establishment after matter domination of a common matter density
contrast, as the baryons fall into the potential wells created by the cold
dark matter.
\end{itemize}

To first order in the perturbations, all of these effects apply separately to
each Fourier component, so that a linear transfer function is indeed implied.
Excellent accounts of the standard (synchronous gauge) formalism for
calculating them have been given by Peebles (1980) and Efstathiou (1990).

To date, all numerical calculations of the cold dark matter transfer function
have been done in the synchronous gauge. They include some, but not
necessarily all of the effects mentioned above. Unfortunately, none of them is
fully documented, and the parametrisations which are supposed to fit them do
not agree very well with each other. We use one of the parametrisations given
by Bond and Efstathiou (1984), which has been widely used in the literature
(Efstathiou 1990)
\begin{equation}
T(k) = \left[ 1+\left( ak+\left(bk\right)^{3/2} + \left(ck\right)^2
	\right)^{\nu}\right]^{-1/\nu} \label{tran}
\end{equation}
where $\nu=1.13$ and
\bea
a\eqa 6.4 \left(\Omega_m h\right)^{-1} {\rm h\mone Mpc} \\
b\eqa3\left(\Omega_m h\right)^{-1} {\rm h\mone Mpc} \\
c\eqa1.7\left(\Omega_m h\right)^{-1} {\rm h\mone Mpc} \label{trana}
\eea
We postpone until Section \ref{MDM} a discussion of the dependence of this
expression on $h$ and
the present matter density $\Omega_m$, taking until then the
canonical values $\Omega_m=1$ and $h=0.5$. The transfer function is
shown in Figure 1a, where one sees that the behaviour estimated in \eq{132} is
roughly correct, though the break
at $k\mone\sub{eq}$ is not at all sharp. The transfer function is shown right
down to the scale $k\mone =10\mtwo\Mpc$, corresponding to the mass $M\sim
10^6\msun$ enclosed by a dwarf galaxy, but as we shall see it has so far been
compared with observation only for $k\mone\gsim 1\Mpc$, corresponding to the
mass $M\sim 10^{12}\msun$ enclosed by a bright galaxy.

Another widely quoted parametrisation is that given by Bardeen {\it et al}
(1986), based on a calculation of Bardeen (1986). It is compared in Figure 1b
with the one we are using, along with four other parametrisations. Not all
the parametrisations attempt an accurate reproduction down to the smallest
scales. One sees that even on the presently relevant scales $k\mone \gsim
1\Mpc$ the fits range more than 10\% each side of the one we are using. Adams
{\it et al} (1993) claim that on the scale $k\mone\simeq 10\Mpc$ the
parametrisation of Bardeen {\it et al} (1986) is extremely accurate for
$\Omega_B=0$, and that increasing $\Omega_B$ to the value $\Omega_B\simeq.06$
favoured by nucleosynthesis decreases it by about 15\%. One sees from Figure
1a that this would take it about 5\% below the parametrisation that we are
using.

In view of the crucial importance of the cold dark matter transfer function a
fully documented calculation seems to be called for, together with a
parametrisation which is accurate to $1\%$ or so for all $\Omega_B$ in the
range $0$ to $.1$, and all scales of cosmological interest.

The transfer function, which encodes the solution of linear equations, ceases
to be valid when the density contrast becomes of order 1. After that, the
highly non-linear phenomenon of gravitational collapse takes place, as
discussed in Section \ref{NONLIN}.

\subsection{Isocurvature density perturbations?}

Instead of the adiabatic initial condition, one could consider what is called
the isocurvature initial condition. According to this condition there is no
perturbation in the total energy density, or in the expansion rates of the
separate components (and therefore none in the overall expansion rate $H$).
There are however perturbations in the energy densities of two or more of the
components, which add up to zero.

Since there is no change in the energy density or the expansion rate the
Friedmann equation is unaffected, and the comoving hypersurfaces do not receive
a curvature perturbation. This is the reason for the term
`isocurvature'.\footnote{The specific reference to comoving hypersurfaces is
not usually spelled out in the literature, and one might form the impression
that it is the curvature of space-time which is supposed to be unaffected. Of
course that is not so, because in general isocurvature perturbations affect
the pressure which is related through the field equation to the space-time
geometry. In particular, the curvature scalar of space-time is $R=-8\pi
G(\rho-3p)$. }

The most general isocurvature density perturbation may be specified by giving
the perturbation $S_X$ in the ratio of the the number density of each species
$X$, to (say) the number density of the photons. (An equivalent definition is
that $-S_X$ is the perturbation in the photon entropy per particle.) Except
for particular epochs, each species constitutes to a good approximation either
radiation, $p=\rho/3$, or matter, $p=0$. Its energy density in these cases
evolves with time like its number density to the powers $4/3$ and $1$
respectively, so one has
\bea S_X&=\frac34 \delta_X-\frac34\delta_\gamma
\hspace{10mm} & \mbox{radiation} \label{138} \\
 S_X&\hspace{-10mm}=\delta_X-\frac34\delta_\gamma &
\mbox{matter} \label{139} \eea
Any initial energy density perturbation may be expressed as the sum of an
adiabatic perturbation and an isocurvature perturbation.

We shall focus exclusively on a perturbation of the cold dark matter  density.
Thus we assume that initially there is a common radiation density contrast
$\delta_r$, a baryon density contrast $\delta_B=\frac34\delta_r$ and a cold
dark matter density contrast specified by
\be S=\delta_c-\frac34 \delta_r \ee
To our knowledge, an isocurvature perturbation of the baryon density has not
been discussed in the context of cold dark matter, and an
isocurvature perturbation of the neutrino density has not been discussed at
all. In the context of cold dark matter this list is exhaustive, since we have
agreed to define the perturbations relative to that of the photon density.

Before seeing how to handle an isocurvature cold dark matter perturbation, we
briefly discuss its theoretical motivation. In the context of inflation, an
adiabatic perturbation of some magnitude is mandatory, because it corresponds
to a quantum fluctuation of the inflaton field. An isocurvature perturbation,
on the other hand, can originate during inflation only through the quantum
fluctuation of some other field. The resulting inhomogeneity in the field has
to survive after inflation, until some mechanism converts it into an
isocurvature density perturbation.

Until recently it seemed likely that the axion field, if it exists, could
easily generate an isocurvature perturbation\footnote{Since the axion
inhomogeneity generated through this mechanism is not necessarily small, a
significant isocurvature cold dark matter perturbation can be generated for
any axion density within the cosmologically allowed regime $10\mfour \lsim
\Omega_a \lsim 1$, even if something else is responsible for the bulk of the
cold dark matter. The perturbation is however non-Gaussian if the axion
inhomogeneity is big.} through this mechanism.\footnote{The status of the
subject in 1989 is described in the texts of Linde (1990) and Kolb and Turner
(1990). Developments since then can be traced from the papers of Lyth (1992),
Lyth and Stewart (1992b) and Lyth (1993a).} Indeed, if a quantum fluctuation of
the axion field survives to the end of inflation, and if in addition the
reheat temperature after inflation is too low to restore the Peccei--Quinn
symmetry, then the fluctuation will survive until the axion acquires a mass at
the epoch $T\sim\Lambda\sub{QCD}\sim 100\MeV$, to become an isocurvature axion
density perturbation. The condition that symmetry is not restored after
inflation is that the reheat temperature is less than the axion decay constant
$f_a\sim 10^{10}$ to $10^{13}\GeV$, which might well be satisfied. However, it
has recently been realised (Lyth \& Stewart 1992b) that the fluctuation may
not survive until the end of inflation, essentially because symmetry can be
restored by the Hawking temperature. The condition for this to happen turns
out to be $H_1\gsim f_a$, where $H_1$ is the Hubble parameter when the
observable universe leaves the horizon during inflation. According to
\eq{vqua} this condition is met in many models, so that an isocurvature
density perturbation arising from the axion is far from inevitable. Nor is
there any other well motivated way of generating an isocurvature cold dark
matter perturbation.

Despite the lack of motivation for isocurvature initial conditions, they are
worth discussing if only to see whether they can provide a viable explanation
of the observed large scale structure, without violating the microwave
background anisotropy measurements. Accordingly, the evolution of an
isocurvature cold dark matter density perturbation is described now, and its
effect on the microwave background is described in Section \ref{MWB}. The
conclusion is that such a perturbation cannot alone be the cause of large
scale structure.

\subsubsection*{The two fluid formalism}

The evolution of the density perturbation before horizon entry can be described
by ignoring the baryons, and treating the radiation and cold dark matter as a
pair of uncoupled fluids. Using the metric perturbation formalism in the
synchronous gauge, the resulting rather complicated equations have been given
by several authors, including Efstathiou and Bond (1986) who solved them
numerically. We shall work with the fluid flow formalism in the comoving gauge,
where the equations are simply Bardeen's equation \eq{118}, together with an
equation tracking the evolution of the entropy perturbation which was first
given by Kodama and Sasaki (1984, 1987). The equation is \be H\mtwo \ddot S\sk
+ (2-3c_z^2) H\mtwo \dot S\sk
	= \rfrac{k}{aH}^2\left[-c_z^2 S\sk
	+\frac13 (1+w)\mone \delta \sk  \right]
	\label{142} \ee
where $c_z^2=[{3+4(a\sub{eq}/a)}]\mone$. Evaluating $\delta P$ in terms of
$\delta$ and $S$, Bardeen's equation becomes
\bea H\mtwo\ddot{\delta}\sk  +
[2-3(2w-c_s^2)]H\mone \dot{\delta}\sk
-\frac32(1-6c_s^2+8w-3w^2)\delta\sk =\nonumber\\
-\left(\frac{k}{aH}\right)^2 \left[
c_s^2 \delta\sk + 3\frac
{1+(a/a\sub{eq})}{1+(4/3)(a/a\sub{eq})\mone }S\sk\right]
\label{142a}
\eea

\eqs{142}{142a} completely describe the evolution of the two fluids.
They have four independent solutions, which as we now see correspond at
earlier times to growing and decaying modes, each with either
adiabatic or isocurvature initial conditions.

\subsubsection*{The initial era}

At an initial epoch, well before horizon entry and well before matter
domination, there are the following independent modes (Kodama \& Sasaki 1984,
1987). (The subscripts $\bfk$ are dropped for clarity)

Adiabatic growing mode
\bea\delta= A a^2  \hspace{5em}&& S/\delta=O
\left[\frac{a}{a\sub{eq}}\rfrac{a}{kH}^2\right]\eea

Adiabatic decaying mode
\bea \delta= B a\mone  &&S/\delta=O
\left[\frac{a}{a\sub{eq}}\rfrac{a}{kH}^2\right]\nonumber\eea

Isocurvature growing mode
\bea S=C  &&\delta/S=O
\left[\rfrac{a}{a\sub{eq}}^2\rfrac{a}{kH}^2\right]\eea

Isocurvature decaying mode
\bea
S=D\ln(a/a\sub{eq}) &&\delta/S=O
\left[\rfrac{a}{a\sub{eq}}^2\rfrac{a}{kH}^2\right]\eea

The growing adiabatic mode corresponds to \eq{123}, the constant $A$ being
related to the constant initial value of ${{\cal R}}$.
The decaying adiabatic mode
corresponds to \eq{124}, with ${{\cal R}}=0$. We dropped it and will do the
same
with the decaying isocurvature mode. Thus, we are interested in the growing
isocurvature mode.\footnote{By considering the separate fluid flow equations,
one can show that the decaying modes correspond to relative motion between the
fluids, which is necessarily absent well before horizon entry (Lyth \& Stewart
1992b). In the case of an isocurvature mode this is a non-trivial result,
because it justifies the standard assumption that all of the initial $S$ value
feeds into the growing mode. It is this initial value which is related to the
inflationary potential, so for a given potential the predictions would be
altered if a decaying mode were initially present, even though it would become
negligible long before the present.}

\subsubsection*{The large scale radiation density perturbation}

After the initial epoch the evolution of the matter and radiation densities
can be followed by solving \eqs{142}{142a}, provided that diffusion and free
streaming are ignored. This is certainly permissible before horizon entry, and
in fact an analytic solution of the equations is available for that case
(Starobinsky \& Sahni 1984; Kodama \& Sasaki 1987). It shows that the entropy
perturbation $S$ is practically constant, corresponding to the fact that
particles do not have time to flow on a given scale before horizon entry. It
also shows that $|\delta/S|$ remains small. Substituting
$\delta\rho_c\simeq-\delta\rho_r$ into the definition of $S$, this implies
that during radiation domination $\delta_r\ll\delta_c$ and $S\simeq \delta_c$,
whereas during matter domination $\delta_c\ll\delta_r$ and
$S\simeq-\frac34\delta_r$. Since $\rho_\gamma\propto T^4$, this implies that
after matter domination the photon temperature has a perturbation \be
\frac{\delta T}{T}=-\frac13 S_X \label{151} \ee We shall discuss the
resulting microwave background anisotropy in Section \ref{MWB}.

\subsubsection*{The isocurvature transfer function}

Well after matter domination the pressure perturbation is negligible, so that
the density contrast has the behaviour \eq{126}. One can then define a
transfer function,
\be \left[\rfrac{aH}{k}^2
\frac{\delta\rho\sk }{\rho} \right]\sub{final}
=T\sub{iso}(k) S\sk (\mbox{initial}) \label{141} \ee
The subscript `initial' refers to the era well before horizon entry, when $S$
is practically constant, and the subscript `final' refers to the era well
after matter domination. This transfer function is the same in the
comoving and synchronous gauges, because $S$ is the same in both gauges
and so is the left hand side.

For scales entering the horizon, free streaming and diffusion are
negligible so that the transfer function from the
analytic solution of \eqs{142}{143} which has already mentioned.
This gives (Kodama \& Sasaki 1987)
\be T\sub{iso}(k)=2/15 \hspace{10mm} (k\mone\gg k\mone\sub{eq})
\label{143}\ee
Notice that according to this expression $|\delta\sk/S\sk |$ remains
small until horizon entry, in accordance with what was said earlier.

The  situation on smaller scales is similar to the one that we described
before for the adiabatic case. To estimate the transfer function roughly, one
notes that $\delta_m\sim S$ at horizon entry and that there is little
subsequent growth of $\delta_m$ until matter domination. As a result the
transfer function is again given roughly by \eq{132}. For a more accurate
calculation one could assume that the radiation free streams away promptly, or
go to the other extreme of taking it to be a perfect fluid. Neither
approximation looks very sensible, since the radiation consists of roughly
half neutrinos which free stream, and half photons which either are a perfect
fluid or else only diffuse. Efstathiou and Bond (1986) find however that the
perfect fluid approximation is valid to 5\%. We are not aware of any simple
reason why the effect of free-streaming and diffusion should be so small;
although the radiation density contrast relative to the matter density
contrast is far smaller, at horizon entry, in the isocurvature case than in
the adiabatic case (about the same instead of a factor $\rho_r/\rho_c$
bigger), it is still the thing that determines the shape of the transfer
function in the vital region around the break at $k=k\sub{eq}$. One would like
to know if the perfect fluid approximation is also accurate for the adiabatic
case.

\subsection{The peculiar velocity field}

After matter domination it is very useful to introduce the concept of a
peculiar velocity field. Since the comoving and synchronous gauges
coincide in this regime, the following treatment is valid in both of
them.

In any small region around a comoving observer, the
receding fluid defines a three-velocity field $u^i$. Its components are to be
defined in terms of locally inertial coordinates, in which the observer is
instantaneously at rest. By a `small region' is meant one which is small
compared with the Hubble distance, so that the curvature of space-time is
negligible. Ignoring the perturbations,
\be u^i=H y^i \label{154} \ee
where the observer is taken to be
at the origin of the spatial coordinates $y^i$.
 Including the perturbations, one defines a peculiar velocity
field $v^i$ by
\be u^i= \overline H y^i+ v^i \label{145} \ee
where $\overline H$ is the average defined in \eq{110b}.

The three-velocity field $u^i$ cannot be defined over an extended region
because of space-time curvature, but one might still hope to define a {\it
peculiar} velocity field, such that \eq{145} is satisfied in the small region
around each comoving observer. The field would be a four-vector, everywhere
orthogonal to the comoving four-velocity $u^\mu$ and with magnitude  much less
than $1$. As discussed in Section \ref{MWB}.3, such a globally defined peculiar
velocity field does not strictly speaking exist because of gravitational
waves. The gravitational waves can however be handled separately and it is
legitimate to ignore them in what follows.

Like any perturbation, the peculiar velocity field `lives' in flat space,
described by comoving coordinates $ x^i$ related to Cartesian coordinates by
$y^i=a x^i$. It can be decomposed into a longitudinal (irrotational) part and
a transverse (rotational) part,
\be \bfv=\bfv\sub{tran} + \bfv\sub{long}
	\label{146} \ee
where the transverse part satisfies $\del. {\bfv}\sub {tran}=0$ and the
longitudinal part is the gradient of some velocity potential. For a Fourier
mode with $\bfk$ along the $z$ direction, the longitudinal part has only a
$z$ component and the transverse part has only $x$ and $y$ components.

The transverse part contributes to the vorticity $\omega _{ij}$, and is the
quantity that we agreed earlier to subtract off so that comoving hypersurfaces
can be defined. It decays with time like $a\mfour (\rho+P)\mone$ and can be
ignored (Weinberg 1972; Peebles 1980; Lyth 1993b).

The longitudinal part of the peculiar velocity is directly related to
the density perturbation. After matter domination, which is the only
regime of interest in the present context, the relation reduces to a
well known Newtonian expression. The same expression follows from the present
analysis, because the fluid flow equations on which it is based reduce to the
Newtonian ones during matter domination. To derive it, start with the result
$\del. {\bfv} = 3\delta H$ which follows from \eq{14}. From
\eqsss{19}{120}{122}{123}, it follows that
\be \del. {\bfv}=(4\pi G\delta\rho ) t \label{147} \ee
The solution of this equation is
\be {\bfv}=-t\del\psi
\label{147a} \ee
or
\be v_i=- (t/a)\pdif{\psi(\bfx,t) }{x^i} \ee
where
\be \psi({\bfx},t) = - G a\mtwo \int\frac{\delta\rho(\bfx^\prime ,t)}
	{|\bfx^\prime -\bfx|} d^3x^\prime  \label{148} \ee
The factor $a\mtwo$ converts coordinate distances into physical distances.
Since it is related to the density perturbation by the Newtonian expression,
$\psi$ is called the peculiar gravitational potential. It is independent of
$t$ because $\delta \rho\propto a^2$, and from \eq{126} we see that it is
related to the spatial curvature perturbation by
\be \psi= - \frac35 {{\cal R}}(\mbox{final})  \label{149} \ee
The peculiar velocity on the other hand is proportional to $a\half$.

{}From \eqs{147a}{148} the Fourier components of $\bfv$, $\psi$ and $\delta$
are related by
\bea {\bfv}\sk \eqa i\frac{{\bfk}}{k}\rfrac{aH}{k} \delta\sk
\label{159}\\
\psi\sk \eqa -\frac32 \rfrac{aH}{k}^2 \delta\sk
\label{159a}\eea
It follows that on scales much bigger than the horizon they satisfy
\be |\delta\sk|\ll|\bfv\sk|
\ll|\psi\sk| \hspace{10mm} (aH/k\gg 1) \ee
For an isocurvature perturbation, it follows from \eqss{141}{143}{159a} that
\be \psi\sk=-\frac15 S\sk(\mbox{initial})
\hspace{10mm} (aH/k\gg 1) \label{161} \ee

\section{The spectrum of the density perturbation}
\label{SPECTRUM}
\setcounter{equation}{0}
\renewcommand\theequation{\thesection.\arabic{equation}}

During the linear regime one expects that the perturbations will  be random
fields with simple stochastic properties. For a Gaussian field these are
specified completely by the spectrum.

\subsection{The spectrum and stochastic properties}

We are interested in the density contrast $\delta$ and the quantity ${{\cal
R}}$ which measures the associated spatial curvature perturbation, and after
matter domination we are also interested in the components $v_i$ of the
peculiar velocity field, and the peculiar gravitational potential $\psi$. Let
$f$ denote any one of these perturbations. Its most important statistic is its
spectrum, which is essentially the smoothed modulus-squared of its  Fourier
coefficient. To be precise, the spectrum may be defined as the quantity
\be {{\cal P}}_f\equiv \rfrac{Lk}{2\pi}^3 4\pi \langle|f\sk|^2\rangle
	\label{21} \ee
where $L$ is the size of the box for the  Fourier expansion \eq{116}, and the
bracket denotes the average over a small region of $k$-space. The
normalisation is chosen to give a simple formula for the dispersion (root mean
square) of $f$, which we shall denote by $\sigma_f$. From the Fourier
expansion  one has $\sigma_f^2=\sum |f\sk^2|$, and since the possible
values of $\bfk$ form a cubic lattice with spacing $2\pi/L$ the transition
from sum to integral is
\be \rfrac{2\pi}{L}^3\sum\sk \longrightarrow 4\pi \int k^2  dk \ee
The dispersion $\sigma_f$ is therefore given by
\be
\sigma_f^2 \equiv\langle f^2(\bfx) \rangle
= \int^\infty_0 {{\cal P}}_f(k) \frac{dk}{k}
\label{DISP}
\ee
with the brackets now denoting the spatial average.
For the density perturbation $f=\delta$ it is useful to define the
correlation function $\xi(r)$ by
\be \xi(r)=\langle f(\bfr+{\bfx})f(\bfr)\rangle = \int^\infty_0 {{\cal P}}_f(k)
	\frac{\sin(kr)}{kr} \frac{dk}{k}
\label{corr} \ee
The analogous
quantity is useful for
other perturbations like the peculiar velocity components, though it is not
then called the correlation function.
For $r=0$ it clearly reduces to $\sigma_f^2$.

If the phases of the Fourier coefficients are random, $f$ is said to be
Gaussian, and then all of its stochastic properties are determined by its
spectrum. In particular the probability distribution of $f$, evaluated at
randomly chosen points, has a Gaussian profile. As long as the perturbations
are evolving linearly this Gaussian property is implied by the inflationary
models to be discussed in Section \ref{INFL}, and we take it for granted from
now on.

In linear theory the spectra of all of the perturbations are
($k$ dependent) multiples of that of the spectrum ${{\cal P}}_\delta$ of the
density contrast. In the literature,
our ${{\cal P}}_\delta$ is denoted variously
by ${{\cal P}}_\rho$ (Salopek, Bond \& Bardeen 1989), $P_\rho$ (Lyth \& Stewart
1990b), $\Delta^2$ (Kolb \& Turner 1990; Taylor \& Rowan-Robinson 1992)),
$\delta\rho_k/\rho$ (Linde 1990) and $d\sigma^2_\rho/d\ln k$ (Bond \&
Efstathiou 1991). An
alternative definition, used especially in the older literature and denoted by
$P$, is equal to $Nk\mthree {{\cal P}}_\delta$, where $N$ is an
author-dependent
normalisation factor which is often left undefined.

{}From \eqs{126}{127}, the spectrum of the density contrast after matter
domination may be written
\begin{equation}
{{\cal P}}_\delta(k) =\rfrac{k}{aH}^4 T^2(k) \delta_H^2(k)
\label{23} \end{equation}
The quantity $\delta_H$ specifies the initial spectrum.\footnote
{For $n=1$ it is equal to the $\eta_H$ of Scaramella and
Vittorio (1990) and to $4\pi$ times the $\epsilon_H$ of Abbott and Wise
(1984c).}
 In fact, from \eq{125}
it is related to the spectrum of the initial curvature perturbation
${{\cal R}}$ by
\be \delta_H^2(k)=\frac{4}{25} {{\cal P}}_{{\cal R}} (k)
\label{24} \ee
The  subscript $H$ has been used because $\delta_H^2$ is exactly equal to the
value of ${{\cal P}}_\delta$
on horizon entry on scales $k\mone\gg k\mone\sub{eq}$,
and approximately equal to it on smaller scales. As we shall see in Section
\ref{SPECTRUM}.2, this means that $\delta_H(k)$ is {\it roughly} equal to the
mean square value of $\delta$ at horizon entry, for a density enhancement with
comoving size of order $k\mone$.

The standard assumption is that $\delta_H^2$ is independent of $k$, which with
the older definition of the spectrum corresponds to $P\propto k$. A more
general possibility is to consider a spectrum
\be \delta_H^2\propto k^{n-1} \ee
corresponding to $P\propto k^n$. The exponent $n$ is called the {\it spectral
index}. The standard choice of $n=1$ was first advocated by Harrison (1970)
and Zel'dovich (1970) on the ground that it is the only one making the
perturbation small on all scales, at the epoch of horizon entry. On the other
hand, the power-law dependence on $n$ need only be an approximation, valid
over a limited range of scales. From this viewpoint the value $n=1$ is not so
special, and the possibility has recently been explored that $n$ is less than
1, tilting the Harrison--Zel'dovich spectrum to give more power on large
scales. Of course there is no reason in principle why the spectrum should have
a power-law dependence at all; the effective value of $n$ could change with
scale as one goes from the regime $k\mone\sim1\Mpc$ explored by the small
scale galaxy correlations to the regime $k\mone\sim 10^4\Mpc$ explored by the
large scale cmb anisotropy.

In Section \ref{INFL} we shall see how inflationary models of the early
universe inevitably give an adiabatic density perturbation at some level. For
a given inflationary model its spectrum can be calculated in terms of the
parameters specifying the potential. In typical inflation models there
is a quite accurate power-law dependence over the above range of scales,
the spectral index being determined by two small parameters
$\epsilon_1$ and $\eta_1$, which we shall define in Section \ref{INFL},
\be n=1+2\eta_1 -6\epsilon_1 \ee
The precise values of $\epsilon_1$ and $\eta_1$ depend on the inflationary
model. In many models they are extremely small,
making $n$ very close to 1. In others this is not so, and we shall
describe in Section \ref{INFL}.4 models which span the range $0\lsim n
\lsim 1.2$. The prejudice towards negative values is a reflection of the fact
that $\epsilon_1$ is positive by definition.

\subsubsection*{The isocurvature spectrum}

In the case of an isocurvature initial condition, \eq{23} is replaced by \be
{{\cal P}}_\delta(k)=\rfrac{k}{aH}^4 T^2\sub{iso}(k) {{\cal P}}_S(k) \ee At
least over a limited range one can hopefully define an isocurvature spectral
index by ${{\cal P}}_S(k)\propto k^{n\subsub{iso}}$. Since the adiabatic and
isocurvature transfer functions have a similar shape, a spectrum ${{\cal
P}}_\delta$ of given shape can arise either from an adiabatic perturbation
with spectral index $n$ or an isocurvature one with spectral index
$n\sub{iso}=n-1$.

If the isocurvature perturbation originates as a vacuum fluctuation
during inflation its spectral index is given by
\be n\sub{iso} =-2\epsilon_1 \ee
Many of the usually considered inflationary models require $\epsilon_1$ to be
very small, leading to a very flat spectrum, but others allow it to be
significant, tilting the spectrum to give more large scale power. Since
$\epsilon$ is positive by definition it is impossible (not just difficult, as
in the case of an adiabatic perturbation) to go the other way, at least within
the almost universally accepted slow-roll paradigm. (This restriction applies,
of course, only to an isocurvature perturbation generated in the manner we
have described. An isocurvature perturbation produced at a phase transition
will on the contrary have most of its power on the current horizon scale, and
so could form very small scale early structure. An example might be the
inhomogeneity of the axion density associated with domain walls present at the
epoch $T\sim1\GeV$ (Hogan \& Rees 1988).)

\subsection{The filtered density contrast}

The density contrast $\delta(\bfx)$ will evolve linearly as $\sigma\ll 1$,
except in those rare regions where it becomes $\gsim 1$ and gravitational
collapse takes place. Following the usual practice, we assume when necessary
that the linear evolution is at least roughly valid right up to the epoch
$\sigma\simeq 1$. Soon after that epoch, a large fraction of the matter
collapses into gravitationally bound objects, and linear evolution becomes
completely invalid. For definiteness let us define the epoch of non-linearity
as precisely the epoch $\sigma=1$. Since $\delta \propto a$ after matter
domination, it corresponds to a redshift $z\sub{nl}$ given by
$(1+z\sub{nl})=1/\sigma_0$, where $\sigma_0 $ is the {\em linearly evolved}
quantity evaluated at the present time.

A `filtered' density contrast $\delta(R_f,\bfx)$, which has a smaller
dispersion $\sigma(R_f)$, can be constructed by cutting off
the Fourier expansion of $\delta(\bfx)$
above some minimum wavenumber $\simeq 1/R_f$, or
equivalently by smearing it over a region with size $\simeq R_f$. The
filtered quantity will evolve linearly until a later epoch $\sigma
(R_f)=1$.

A precise definition of the filtered density contrast is made by means of a
`window function' $W(R_f,r)$, which is equal to 1 at $r=0$ and which falls off
rapidly beyond some radius $R_f$ (Peebles 1980; Kolb \& Turner 1990).
The filtered density contrast is
\be \delta(R_f,{\bfx})=\int W(R_f,|{\bfx}^\prime -{\bfx}|)
	\delta({\bfx}^\prime ) d^3 {\bfx}^\prime  \ee
and its spectrum is
\be {{\cal P}}_\delta(R_f,k)= \left[\widetilde W(R_f,k)/V_f\right]^2
	{{\cal P}}_\delta(k) \label{312} \ee
where
\be \widetilde W(R_f,k)=\int e^{-i \bf k\cdot x} W(R_f,r) d^3x \ee
and
\be V_f=\int W(R_f,r) d^3x \ee
The filtered dispersion is
\begin{equation}
\sigma^2(R_f)
= \int_0^{\infty} \left[\widetilde W(R_f,k)/V_f\right]^2
{{\cal P}}_\delta(k) \frac{{\rm d}k}{k}
\end{equation}

The quantity $V_f$ is the volume `enclosed' by the filter. It is convenient to
define the associated mass $M=\rho_0 V_f$, where $\rho_0$ is the
present mass density.
One normally uses $M$ instead of $R_f$ to specify the scale,
writing $\delta(M,\bfx)$ and $\sigma(M)$.

The two popular choices are the Gaussian filter
\bea W(R_f,r) \eqa \exp(-r^2/2 R_f^2) \label{gone} \\
V_f \eqa (2\pi)\threehalf R_f^3 \\
\widetilde W(R_f,k)/V_f \eqa \exp(-k R_f) \\
M \eqa 4.36\times 10^{12} h^2 (R_f/1\Mpc)^3 \msun
\label{gtwo} \eea
and the top hat filter which smears uniformly over a sphere of radius $R_f$
\bea W(R_f,r)\eqa\theta(r-R_f) \\
V_f \eqa 4\pi R_f^3/3\\
\widetilde W(R_f,k)/V_f \eqa 3 \left( \frac{\sin(k R_f)}{(k R_f)^3}
-\frac{\cos(k R_f)}{(k R_f)^2} \right) \label{322} \\
M \eqa 1.16\times 10^{12} h^2 (R_f/1\Mpc)^3 \msun
\eea
The Gaussian filter is the most convenient for
theoretical calculations, but the top hat filter is widely used to as
a means of presenting data.

Provided that the spectrum is increasing, say like $k^m$ with $m$ not too
different from 1, the filtered dispersion will be given roughly by
\be \sigma^2(R_f)\simeq {{\cal P}}_\rho(k=R_f\mone) \ee
Furthermore, as we shall see in Section \ref{NONLIN}, the filtered dispersion
is roughly the rms density contrast of structures of comoving size $R_f$. Thus
{\it the square root of the spectrum of the density contrast at wavenumber $k$
is typically a measure of the average density contrast within an overdense
region of comoving size $R_f\sim k\mone$}. This rule holds in the regime where
the spectrum is increasing reasonably strongly, which corresponds to
$k\mone\gsim .1\Mpc$.

We have focussed on the density perturbation, but any other perturbation may
be filtered in the same way. Thus, the entire cosmological perturbation
theory can be applied to perturbations on any chosen scale, and remains valid
for longer the larger that scale. For filtering scales $R_f\gg 10\Mpc$, it is
still valid at present.

\subsection{The conventional normalisation}

For a given spectral index, the shape of the spectrum after matter domination
is determined by the transfer function. It remains to specify its
normalisation. In the past this has usually been done by comparing the
linearly evolved theory with observations of the galaxy correlation function
on a scale of order $10 h^{-1}$ Mpc.
Two standard prescriptions exist (Davis \& Peebles 1983). The first requires
that the dispersion $\sigma_g(r)$ of galaxy counts in spheres of radius $r$
has the observed value unity at $r=8h\mone\Mpc$. The second requires that
$J_3(r)$, the integral of the second moment of the galaxy correlation function
up to distance $r$, has the observed value 270 $h^{-3}$ Mpc$^{-3}$ at
$r=10h^{-1}$ Mpc.

These normalisation schemes both refer to the statistics of the clustering of
optical galaxies, not to the clustering of mass as described by the spectrum
${{\cal P}}_\delta$.
It is an important ingredient of the CDM cosmogony that these
are not necessarily taken to be the same --- so-called biased galaxy
formation (Bardeen {\it et al} 1986). To be specific, the mass density
correlation function $\xi$ given by \eq{corr}
is assumed to be related to the correlation function $\xi_{gg}$ of luminous
galaxies by
\be \xi_{gg}(r) \simeq b^2 \xi(r) \label{bias} \ee
with a bias parameter $b$ which is roughly independent of the scale $r$.
Including the bias parameter, the normalisations implied by the two schemes
are
\bea \sigma_g^2(r)/b^2
\eqa 9 \int_0^{\infty} {{\cal P}}_0(k)
	\left[ \frac{\sin kr}{(kr)^3} - \frac{\cos kr}{(kr)^2}
	\right]^2 \frac{{\rm d}k}{k}\nonumber\\
\eqa (1/b)^2 \mbox{\hspace*{1cm} at\ }r=8h\mone\Mpc
\label{sig} \eea
and
\bea
2r\mthree J_3(r)/b^2 \eqa \int_0^{\infty} 2 {{\cal P}}_0(k)
\rfrac{1}{k r}^2
\left[\frac{\sin kr}{kr}-
	\cos kr \right] \frac{ {\rm d} k}{k}\nonumber\\
\eqa (.73/b)^2 \mbox{\hspace*{1cm} at\ }r=10h\mone\Mpc
\eea
In each case, ${{\cal P}}_0$ is the {\it linearly evolved} spectrum,
evaluated at the present epoch.

One sees that $\sigma_g(r)/b$ is the dispersion of the linearly evolved
density contrast with top hat filtering on the scale $r$, whereas $(2r\mthree
J_3(r))\half/b$ is its dispersion on the same scale with a different choice of
filter. In what follows we always adopt the $\sigma_g$ normalisation scheme,
in the sense that normalisation of the spectrum is specified by quoting the
quantity $b$ {\em defined} by \eq{sig}, which will be denoted by $b_8$. In
other words, the normalisation is defined by giving the quantity
\be 1/ b_8\equiv \sigma_8\equiv \sigma\sub{TH} (8h\mone\Mpc) \ee
where $\sigma\sub{TH}$ is the present value of the {\em linearly evolved}
dispersion, with top hat filtering. Only occasionally do we invoke the much
stronger assumptions that the galaxy correlation function is given by
$\xi_{gg}(r)\simeq b^2\xi(r)$, with $b$ scale-independent and with $\xi(r)$
given by linear theory on scales $r\gsim10\Mpc$.

As noted earlier, the dispersion of a filtered quantity evolves
linearly only as long as it is $\lsim1$, and this condition is only marginally
satisfied by the quantities encountered in the two conventional normalisation
schemes. $N$-body simulations show that in
practice though one doesn't do too badly as the nonlinear corrections shrink
the comoving size of the fluctuation so that it doesn't contribute fully to
the filtered variance. Evaluating the integrals, one finds that the $J_3$
normalisation is typically 10\% lower than the normalisation using $\sigma_g$,
across the range of $n$ in which we shall be interested. These are shown in
Figure 2. As remarked above, different choices of transfer function
parametrisation can give a correction of around 10\% as well.

Figure 3a shows ${{\cal P}}_0(k)/\sigma_8$
for a selection of $n$ values, where as
always ${{\cal P}}_0$ is the
{\it linearly evolved} value of ${{\cal P}}_\delta$ at the
present epoch. The spectra for different $n$ typically cross at a scale
$k^{-1}\sim 10$ Mpc. Figure 3b shows the corresponding dispersion
$\sigma(M)/\sigma_8$ for $n=1$ and $n=0.7$, for both Gaussian and top hat
filtering. The mass scale $M$ runs from $M=10^6\msun$, the Jeans mass at
decoupling (Peebles 1980; Kolb \& Turner 1990) to $M=10^{18}\msun $, the mass
of the universe on the scale of the CfA survey. The comoving wavenumber
$k\mone$ runs over the corresponding range of filtering radius $R_f$.

For $n=1$ the spectrum
${{\cal P}}_\delta(k)$ is practically flat on small scales.
For $n<1$ the power on small scales is reduced (though even for $n=0$ not by
as much as in the standard hot dark matter model (White, Frenk \& Davis 1983)
or with string-seeded hot dark matter (Albrecht \& Stebbins 1992b)). As a
result the spectrum starts to fall on scales $k\mone\lsim .1\Mpc$. The
implication of this change for small scale structure is discussed in Section
\ref{NONLIN}.

\section{The cosmic microwave background anisotropy}
\label{MWB}
\setcounter{equation}{0}
\renewcommand\theequation{\thesection.\arabic{equation}}

So far we have focussed on the matter perturbation, which can be observed
through the distribution and motion of the galaxies. In addition there are the
perturbations in the cosmological neutrinos and photons,
where `cosmological' denotes particles originating in
the early universe as opposed to particles emitted later from astrophysical
sources. The cosmological neutrinos are completely unobservable, but the
cosmological  photons are observed as the cosmic microwave background
radiation, variously abbreviated as the cmb, mbr or cbr.

\subsection{The spectrum of the cmb}

Discounting the possibility of early re-ionisation,\footnote{Very early
structure could cause early re-ionisation, which moves the surface of last
scattering closer. On large angular scales this has little effect on the
anisotropy, but it could be significant for scales less than a few degrees. As
we shall see in Section \ref{NONLIN}, if the spectrum of the density
perturbation is even roughly flat the upper limit on its normalisation implied
by the large scale anisotropy prevents such early structure from forming.
Reionisation is more likely in the case where there are nongaussian seeds,
which can favour sharp overdensities. An example of this is the texture
scenario (Turok \& Spergel 1990), though it appears to be in trouble primarily
from the COBE observation, comparison with which is effectively unaffected by
the possibility of reionisation (Pen, Spergel \& Turok 1992).} the cmb last
scattered at the epoch of decoupling. Its surface of last scattering therefore
lies practically at the particle horizon whose comoving distance is
$x=2H_0^{-1}=1.2\times10^4\Mpc$. At this surface an angle $\theta$ degrees
subtends a comoving distance
\be x \simeq 200\, \theta \Mpc \label{41} \ee

The intensity of the cmb as a function of frequency is a very accurate
blackbody distribution (Mather {\it et al} 1990), with a temperature which is
almost independent of direction, $T=2.736\pm.017K$. Here we want to discuss
the small anisotropy of the cmb. What one will actually calculate and measure
is the variation in intensity at fixed frequency, but it can be expressed as
an equivalent variation $\Delta T$ in the temperature\footnote{As we shall
see, $\Delta T/T$ is less than $10^{-5}$; by specifying such a number one
does not imply that the blackbody form is so accurate, still less
that such accurate measurements of $T$ exist.}.

The temperature as a function of direction $\bfe$ may be expanded in
multipoles,
\be T(\bfe)=\overline T+\bfv.\bfe+\Delta T(\bfe) \ee
where
\be \frac{\Delta T}{T}=\sum_{l=2}^{\infty}
	\sum_{m=-l}^{+l} a_l^m Y_l^m (\bfe) \label{mult} \ee
The dipole term $\bfv.\bfe$ is well measured, and as we shall see $\bfv$ can
be ascribed to our motion at least if the isocurvature perturbation is
negligible.

The mean square anisotropy over the whole sky is
\be \left\langle \rfrac{\Delta T}{T}^2 \right\rangle
	=\frac1{4\pi} \sum_{l=2}^\infty\sum_{m=-l}^l |a_m^l|^2 \ee
The multipoles, and therefore the mean square anisotropy, depend of course on
the position of the observer. For a randomly placed observer
\be \left\langle\left\langle \rfrac{\Delta T}{T}^2 \right\rangle
	\right\rangle\sub{position}
	=\frac1{4\pi}\sum_{l=2}^\infty (2l+1) \Sigma_l^2 \ee
where
\be \Sigma_l^2=\left\langle |a_l^m|^2 \right\rangle\sub{position} \ee

As we shall see in a moment, the anisotropy can be calculated in terms of the
density perturbations of the matter and radiation (together with any
cosmological gravitational waves), and as a consequence the mean square
multipoles $\Sigma_l^2$ are smooth functions of $l$. One can
therefore replace the sum over $l$ by an integral,
\be \left\langle\left\langle \rfrac{\Delta T}{T}^2 \right\rangle
	\right\rangle\sub{position}
	\simeq\frac1{4\pi}\int^\infty_2
l(2l+1)\Sigma_l^2 \frac{dl}{l} \label{4last} \ee
Another nice feature of large $l$ is the relation $\theta \sim 1/l$ between
the angular size of a feature in the sky (in radians) and the order $l$ of the
multipoles that dominate it. (This is, of course, analogous to the relation
$r\sim 1/k$ between the linear size of a feature and the wavenumbers that
dominate its Fourier expansion.) Translating to degrees we have the following
relation between $l$ and the angular scale
\be \frac\theta{1^0}\simeq \frac{60}{l} \ee

{}From \eq{4last}, one sees that $(2l+1)l\Sigma_l^2/4\pi$
gives the contribution to
the mean square anisotropy per unit logarithmic interval of $l$, and therefore
the expected temperature contrast of structures in the sky with angular size
$\theta=60/l$ degrees. It is commonly referred to as `the cmb anisotropy on
scale
$\theta$'. As yet it has been observed only in the COBE observation which we
shall discuss in Section \ref{COBE}, but the theoretically expected value in
the CDM model is shown in Figure 4. Before discussing this result in detail
for large angular scales, let us give a brief overview starting from the
smallest scales.

On scales less than a few arcminutes, the distance \eq{41} subtended at the
last scattering surface is less than the thickness of this surface, which is
about $14\Mpc$ (Hogan, Kaiser \& Rees 1982). As a result, any initial
anisotropy is wiped out, and it turns out that no significant additional
anisotropy is acquired by the cmb on its journey towards us.

On bigger scales the cmb has significant anisotropy when it starts out, which
can be calculated in terms of the cosmological perturbations of the matter and
radiation. An scales less than a degree or so it acquires negligible
additional anisotropy on its journey towards us.\footnote{We here discount the
Sunyaev-Zel'dovich effect by which the cmb acquires anisotropy through photon
up-scattering off dust on its way through a galaxy or cluster. It occurs on
scales of order one arcminute and is easily identified for nearby clusters.
The cumulative effect of distant clusters is expected to be negligible
(Makino \& Suto 1993).} From \eq{41} this arcminute regime explores distance
scales of order 10 to 100 Mpc, the same as the one explored by large scale
galaxy surveys.

As the scale is increased towards one degree, the cmb begins to acquire
significant additional anisotropy on its journey towards us, which dominates
on scales in excess of a degree or so. This anisotropy, called the Sachs-Wolfe
effect, is caused by the density perturbation, and (if they exist with
significant amplitude) by cosmological gravitational waves. \eq{41} still
holds roughly, so we see that scales in excess of a few degrees explore scales
in excess of $10^3\Mpc$.

The expected anisotropy from these sources is shown in Figure 4. In this
report we deal only with the large scale anisotropy, corresponding to angular
scales $\theta$ in excess of a degree or so, and multipoles with $l\lsim60$ We
have three reasons for making this restriction. First, the COBE detection
which is the only one reported at the time of writing lies in this regime.
Second, the theoretical description in terms of the Sachs-Wolfe effect is
extremely simple. Finally, this is the regime which explores very large linear
scales, which cannot be explored using galaxy surveys. For a discussion of
smaller scales the reader is referred to the papers of Bond and Efstathiou
(1987), Linder (1988), Bond {\it et al} (1991a), Vittorio {\it et al} (1991)
and Dodelson and Jubas (1992).

\subsection{The contribution of the density perturbation}

We now calculate the CMB anisotropy associated with the density perturbation.
First we calculate the anisotropy of the cmb from this source when it first
set out on its journey towards us, and then we calculate the Sachs-Wolfe
effect describing the additional distortion which it experiences on the way.
In calculating this latter effect we do not employ the usual description in
terms of the metric perturbation. Instead, we work with the peculiar velocity
field, imagining that the cmb passes a sequence of comoving observers. Peebles
(1980) has given a similar treatment, though he does not explicitly work with
the peculiar velocity.

In the small region around a comoving observer, use a locally inertial
coordinate system in which he is instantaneously at rest. Denote these
coordinates by $y^i$ with $y^i=0$ the observer's position, and consider
light received by the observer which was emitted by a comoving source with
coordinates $dy^i=e^i dy$, with $e^i$ a unit vector. Its wavelength is
redshifted by an amount $d\lambda$, given in terms of the recession
velocity $u^i$ by
\bea d\lambda/\lambda \eqa e^i du_i\\
\eqa  e^i \pdif {u_j}{y^i} dy^j \\
\eqa e^i e^j \pdif {u_j}{y^i} dy
\eea

For a perfectly homogeneous and isotropic universe, \eq{154} gives Hubble's
law
\be d\lambda/\lambda= H(t) dt = da/a \ee
To calculate the redshift of light from a distant comoving source, we may
suppose that it is observed by a sequence of nearby comoving observers in its
path, and integrate the above expression. This leads to the familiar
result
\be 1+z\equiv\frac{\lambda_0}{\lambda\sub{em} }
=\frac{a_0}{a\sub{em} }\ee
where the subscript $0$ denotes the observer's position and the subscript em
denotes the position of the comoving source which emitted the radiation.

Including the perturbation, \eq{145} gives
\be \frac{d\lambda}{\lambda}=\frac{da}{a}
	+e^i e^j \pdif{v_j(\bfx,t)}{x^i} dx \ee
Here, as always, the scale factor $a$ is defined with respect to the average
Hubble parameter on a comoving hypersurface,  $\overline H=\dot a /a$, and the
comoving coordinates $x^i$ are related to inertial ones at each point by
$dy^i=a dx^i$. Since we are dealing with a photon trajectory, $dt=-dy =-a dx$.
To first order in the perturbations the anisotropy $\delta T/T= -\delta
\lambda/\lambda$ is given by
\be \frac{\Delta T(\bfe)}{T} = \rfrac{\Delta T}{T}\sub{em}-
	\int^{x\subsub{em}}_0 e^i e^j \pdif{v_j (\bfx,t)}{x^i} dx
	\label{415} \ee
In this expression, $x\sub{em}=2H_0\mone$ is the coordinate distance of
the last scattering surface.

The first term is the anisotropy of the cmb on a {\it comoving} hypersurface,
evaluated just after it leaves the last scattering surface. On scales entering
the horizon after decoupling it is easily calculated, because the initial
adiabatic condition still holds. Since $\rho_r \propto T^4$, $\Delta
T(\bfe)/T= \frac14\delta_r=\frac13\delta_c$, the last equality following from
the adiabatic condition. Since the cmb starts out during matter domination,
one has $\delta\rho=\rho_c\delta_c+\rho_r\delta_r \simeq \rho\delta_c$. Thus
\be \rfrac{\Delta T(\bfe)}{T}\sub{em}=\frac13 \delta(\bfx) \ee
where $\bfx=\bfe x\sub{em}$ is the point of origin of the
cmb arriving from the direction $\bfe$.

Now consider the second term, which is the Sachs--Wolfe contribution.
According to \eq{147a} it may be written
\bea \rfrac{\Delta T(\bfe)}{T} \sub{SW}
\eqa \int^{x\subsub{em}}_0 \frac ta \frac{d^2\psi}{dx^2} dx \\
\eqa -\int^{x\subsub{em}}_0 \left[\frac{d}{dx}\rfrac ta
\right] \frac {d\psi}{dx}
dx + \left[ \frac ta \frac{d\psi}{dx} \right] ^{x\subsub{em}}_0
\eea
To evaluate the integral, we need the $x$ dependence of $(t/a)$, which using
$a\propto t^{2/3}$ is given by
\be x(t)=\int^{t_0}_t \frac{dt}{a} = 3
\left[ \rfrac ta _0 - \frac ta \right] \ee
The other term may be expressed in terms of peculiar velocities using
\eq{147a}. The final result is
\be \frac {\Delta T(\bfe)}{T} =
\frac13\delta(\bfx)+
\frac 13 [\psi(\bfx)-\psi_0]-\bfe. \bfv(\bfx) +\bfe . \bfv_0 \ee
In this expression, the subscript $0$ denotes our position.

Of these terms, $\bfe . \bfv_0 $ is the part of the dipole due to our peculiar
velocity\footnote{It may be shown (Lyth 1993b) that the total dipole
contribution from the above expression is $\bfe. (\bfv_0-\overline{\bfv})$
where the bar denotes the average over all of space inside the surface of last
scattering. This average is completely negligible compared with our own
peculiar velocity, thus verifying the usual statement that the dipole measures
our peculiar velocity for an adiabatic perturbation. An isocurvature
perturbation can in contrast give a significant dipole (Turner 1991a).} and
$-\psi_0/3$ has no angular dependence. Since we are dealing with scales far
outside the horizon the other terms are related by \eqs{159}{159a}, and one
sees that the potential term dominates. Dropping the whole of the dipole as in
\eq{mult}, the final result is therefore
\be \frac{\Delta T(\bfe)}{T}=\frac13 \psi(\bfx)
\label{424} \ee
This is often called the Sachs--Wolfe effect.

Inserting the Fourier expansion of $\psi$ and projecting out a
multipole leads to the expression
\be a_l^m=-2\pi i^l \sum \sk  \rfrac{aH}{k}^2
\delta\sk  j_l(kx) Y^m_l(\Omega\sk) \label{417} \ee
where $\Omega\sk$ is the solid angle in $\bfk$ space.

This expression applies to any observer at the origin $\bfx=0$ of the
coordinates. When the position of this observer, and therefore that of the
origin, is chosen randomly the coefficients $\delta\sk$ have random phases,
and this implies that the multipoles $a_l^m$ have a Gaussian distribution. The
variance of the distribution is (Peebles 1982b)
\begin{equation}
\Sigma_l^2 = \pi \int_0^{\infty} \frac{{\rm d}k}{k} \, j_l^2
	\left( 2k/aH \right) \; \delta^2_H(k)
\end{equation}
where $j_l$ is the spherical Bessel function. With
$\delta_H^2(k)\propto k^{n-1}$ this becomes
\be \Sigma_l^2=\frac{\pi}{2} \left[\frac{\sqrt\pi}{2}l(l+1)
\frac{\Gamma((3-n)/2)}{\Gamma((4-n)/2)}
\frac{\Gamma(l+(n-1)/2)}{\Gamma(l+(5-n)/2)}
\right]
\frac{\delta^2_H(H_0/2)}{l(l+1)} \label{422} \ee
The square bracket is equal to 1 for $n=1$. For $l\gg1$ and
$l\gg|n|$ it can be replaced by 1, if $\delta_H$
is evaluated on the scale $k\simeq lH_0/2 $ which dominates the
integral.

\subsubsection*{The isocurvature contribution}

The Sachs--Wolfe effect \eq{424} gives the anisotropy acquired by the
radiation on its way towards us, whether the initial condition is adiabatic or
isocurvature. For the isocurvature case, \eq{161} allows one to write it as
\be \rfrac{\Delta T(\bfe)}{T}\sub{SW}=-S(\bfx) /15
\label{156}
\ee
where $S$ is the intial entropy perturbation. In that case, however, one must
add the anisotropy which the cmb started out with, and according to \eq{151}
it is five times as big. The total anisotropy is therefore six times bigger
than in the adiabatic case, for a given density perturbation (Starobinsky \&
Sahni 1984; Efstathiou \& Bond 1986; Kodama \& Sasaki 1987).

In the second half of this report we will find, from studies of the
distribution and motion of the galaxies, that for a spectral index $n=1$ the
spectrum of the density perturbation has more or less the right normalisation
to account for the observed cmb anisotropy if it is adiabatic. Since the
isocurvature and adiabatic transfer functions have more or less the same
shape, it follows that the cmb anisotropy would be much too big if the density
perturbation were isocurvature. In other words the isocurvature density
perturbation, if
it exists, can be only be a small fraction of the total
density perturbation. Lacking theoretical
guidance about the size of this fraction, we ignore the isocurvature
contribution from now on, including its contribution to the cmb anisotropy
which could in principle be significant.

\subsection{The contribution of gravitational waves}

\eq{415} for the anisotropy of the cmb involves the gradient $\pa_i u_j$ of
the four-velocity of comoving worldlines. To the extent that gravitational
waves are significant, this gradient cannot be written as the gradient of a
globally defined peculiar velocity. Rather one must write (Peebles 1980; Lyth
1993a)
\be
\pdif{u_j}{y^i} = \overline H\delta_{ij}+ a\mone(t) \pdif {v_j(\bfx,t)}{x^i} +
\frac12 \pdif{h_{ij}(\bfx,t)}{t} \label{428} \ee

The second term is the one we dealt with already, and the third is the
contribution of the gravitational waves.
It involves that part of the metric perturbation in \eq{10} which
is traceless, $\delta^{ij} h_{ij}=0$, and transverse, $\pa_i h_{ij}=0$.
This means that each Fourier component is of the form
$h_{ij}=h_+ e^+_{ij} + h_\times e^\times_{ij}$, where
in a coordinate system where $\bfk$ points along
the $z$-axis
the nonzero
components of the polarisation
tensors are defined
by $e^+_{xx}=-e^+_{yy}=1$ and $e^\times_{xy}=e^\times_{yx}=1$.
The spectrum ${{\cal P}}_g$ of the
gravitational wave amplitude may be defined
by summing \eq{21} over all four components,
\be {{\cal P}}_g=2\rfrac{Lk}{2\pi}^3 \left( \langle|h_+(\bfk)|^2\rangle
+\langle|h_\times (\bfk)|^2\rangle \right)
\ee
Each Fourier component satisfies the massless wave equation, which in comoving
coordinates is
\be \ddot h_{ij} +3H \dot h_{ij} +(k/a)^2 h_{ij} =0 \ee
Well before horizon entry it has constant initial value.
For
scales entering the horizon after matter domination
its subsequent evolution is
\be h_{ij}(t)
=\left[3\sqrt\frac{\pi}{2} \frac{J_{3/2}(x)}{x\threehalf}
\right] h_{ij}(\mbox{initial}) \ee
where $x=2k/(aH)$. Well after horizon entry
one has redshifting radiation.

By substituting this expression into \eq{428} one can calculate the cmb
multipoles in terms of the initial amplitude (Fabbri \& Pollock 1983;
Starobinsky 1985), and hence calculate $\Sigma_l^2$ in terms of the spectrum
${{\cal P}}_g(k)$ of the initial amplitude. Each gravitational wave gives its
dominant contribution as it enters the horizon, since its amplitude is
practically constant before that and redshifts away afterwards. As a result
the gravitational wave contribution to the cmb anisotropy cuts off below the
angular scale $\simeq 1^0$, corresponding to the smallest linear scale which
enters the horizon after decoupling $k\mone\sub{dec}=180\Mpc$. The
corresponding multipole cutoff is at $l\simeq k\mone\sub{dec} H_0/2\simeq 70$.
For $l$ well below this cutoff the gravitational wave contribution
is given in the case of
a flat spectrum (${{\cal P}}_g(k)$ independent of $k$) by (Starobinsky 1985)
\be l(l+1)\Sigma_l^2=
\frac{\pi}{36}\left( 1+\frac{48\pi^2}{385}\right)
{{\cal P}}_g C_l \label{427} \ee
If one ignored the cutoff due to the redshift, the coefficient $C_l$ would
become $1$ in the limit $l\gg1$. Starobinsky gives the values $C_2=1.118$,
$C_3=0.878$  and $C_4=0.819$. Numerical calculation, including the effect of
the cutoff, shows that a value $C_l$ close to $1$ is indeed achieved for
$l\sim10$, before the cutoff takes effect (Figure 4).

For $l\gg1$ the above result is good also if ${{\cal P}}_g(k)$ has moderate
scale-dependence, provided that it is evaluated at the scale $k\simeq l H_0/2$
which dominates the $l$th multipole.

Within a given inflationary model one can calculate ${{\cal P}}_g(k)$.
Defining the spectrum $n_g$ of the gravitational waves by ${{\cal P}}_g\propto
k^{n_g}$, one finds \be n_g=-2\epsilon_1 \ee where $\epsilon_1$ is the small
positive parameter mentioned earlier and defined in Section \ref{INFL}.

Setting the coefficient $C_l$ of \eq{427} equal to 1, the ratio of the
gravitational and density contributions is given in terms of this same
parameter by
\be R\equiv \frac{\Sigma_l^2\mbox{(grav)}}
	{\Sigma_l^2(\mbox{density})}=12.4\epsilon_1\label{430} \ee
In most of the usually considered models of inflation, $\epsilon_1$ is very
small, corresponding to a very small gravitational wave contribution with a
very flat spectrum. The important exceptions are power-law and extended
inflation, where $\epsilon_1$ need not be very small, so that the spectral
index can be much less than one (more power on large scales) and the ratio
$R$ can be significantly bigger than 1. It is impossible to have a positive
spectral index for the gravitational waves, without going outside the almost
universally accepted slow-roll paradigm.

As we noted earlier the gravitational wave amplitude is strongly reduced by
redshifting on scales $\lsim100\Mpc$, so that it gives a negligible
contribution to the cmb anisotropy on arcminute scales. Nevertheless,
estimates  with a flat spectrum $n_g=0$ show that the waves could give a
significant contribution to the planned LIGO detector (Linder 1988; Sahni
1990; White 1992; Souradeep \& Sahni 1992). Since the spectrum is tilted, if
at all, towards more large scale power, these estimates are upper limits. They
ensure by many orders of magnitude that the gravitational waves do not
contribute significantly to the locally defined peculiar velocity field, which
we observe within $100\Mpc$ or so of our position.

\section{Inflation}
\label{INFL}
\setcounter{equation}{0}
\renewcommand\theequation{\thesection.\arabic{equation}}

It is widely supposed that the very early universe experienced an era of
inflation. By `inflation' one means that the scale factor has positive
acceleration, $\ddot a>0$, corresponding to repulsive gravity and $3p<-\rho$.
During inflation $aH=\dot a$ is increasing, so that comoving scales are
leaving the horizon (Hubble distance) rather than entering it, and it is
supposed that at the beginning of inflation the observable universe was well
within the horizon.

The inflationary hypothesis is attractive because it holds out the possibility
of calculating cosmological quantities, given the Lagrangian describing the
fundamental interactions. The Standard Model, describing the interactions up
to energies of order $1\TeV$, is not viable in this context because it does
not permit inflation, but this should not be regarded as a serious setback
because it is universally agreed that the Standard Model will require
modification at higher energy scales, for reasons that have nothing to do with
cosmology. The nature of the required extension is not yet known, though it is
conceivable that it could become known in the forseeable future (Weinberg
1993). But even without a specific model of the interactions (ie., a specific
Lagrangian), the inflationary hypothesis can still offer guidance about what
to expect in cosmology. More dramatically, one can turn around the
theory-to-observation sequence, to rule out otherwise reasonable models.

In this section we explain how these things work, with particular reference to
the density perturbations and to cosmological gravitational waves. Both of
these could originate during inflation as quantum fluctuations, which become
classical as they leave the horizon and remain so on re-entry. The original
quantum fluctuations are of exactly the same type as those of the
electromagnetic field, which give rise to the experimentally observed Casimir
effect.

\subsection{The basic features of inflation}

The implications of an inflationary era were worked out by many people,
following the seminal paper by Guth (1981), and they have been described in
several texts (Kolb \& Turner 1990; Linde 1990). One of the most dramatic and
simple effects is that there is no fine-tuning of the initial value of the
density parameter $\Omega=8\pi\rho/3 m_{Pl}^2 H^2$. From the
Friedmann equation, $\Omega$ is given by
\be \Omega-1=\rfrac{K}{aH}^2 \label{51} \ee
Its present value $\Omega_0$ is certainly within an order of magnitude of 1,
and in the absence of an inflationary era $\Omega$ becomes ever smaller as one
goes back in time, implying an initial fine tuning. In contrast, if there is
an inflationary era beginning when the observable universe is within the
horizon, \eq{51} implies that $\Omega_0$ will be of order 1, provided only
that the same is true of $\Omega$ at the beginning of inflation. A value of
$\Omega_0$ extremely close to 1 is the most natural, though it is not
mandatory (Lyth \& Stewart 1990a; Ellis, Lyth \& Mijic 1991).\footnote {An
argument has been given for $\Omega_0$ very close to 1 on the basis of effects
on the cmb anisotropy from regions far outside the observable universe (Turner
1991a), but it is not valid as it stands because it ignores spatial curvature.}

Another effect of inflation is that it can eliminate particles and topological
defects which would otherwise be present. Anything produced before inflation
is diluted away, and after inflation there is a maximum temperature (the
`reheat' temperature) which is not high enough to produce all the particles
and defects that might otherwise be present. As we shall remark later, this
mechanism can remove desirable, as well as undesirable, objects.

The most dramatic effect of inflation is that it may offer a way of
understanding the homogeneity and isotropy of the universe, or at any rate of
significant regions of it. We have nothing to say about this complex issue in
its full generality,  but a more modest version of it is our central concern.
In this version, one begins the discussion at some early stage of inflation,
when the universe is supposed already to be {\it approximately} homogeneous
and isotropic (Section 5.1).
 One then argues that in that case, scales far inside the
horizon must be {\it absolutely} homogeneous and isotropic, except for the
effect of vacuum fluctuations in the fields. Finally, one shows that after
they leave the horizon, such fluctuations can become the classical
perturbations that one deals with in cosmological perturbation theory. This
possibility was first pointed out for gravitational waves by Starobinsky
(1979) and for density perturbations by several people (Guth \& Pi 1982;
Hawking 1982; Starobinsky 1982) As we shall go to some trouble to demonstrate,
the vacuum fluctuations can be evaluated unambiguously once an inflationary
model is specified.

To have a density perturbation of the right magnitude, the energy density when
the observable universe leaves the horizon must typically correspond to an
energy scale $\rho\quarter\sim 10^{16}\GeV$, not far below the Planck scale.
One is thus making a big extrapolation above the scales $\rho\quarter\sim
T\lsim 10^3\GeV$ at which the standard model has been tested (and an even
bigger one above the scale $T\sim 10^{-6}\GeV$, corresponding to the epoch a
few Hubble times before the smallest cosmologically interesting scales
enter
the horizon, which are all that we need to consider in the rest of the present
report). Mercifully, a density perturbation
leaving the horizon during
inflation can be followed to the epoch of horizon entry without any
knowledge of the intervening era, as
we saw in Section \ref{DENSPER}. But one would be very lucky to find a well
motivated model of inflation without understanding that era. At present such
an understanding requires that one believes in some model of the GUT type
which has elementary Higgs fields, as opposed to one of the technicolour type
where the Higgs fields are composite.

\subsubsection*{Scalar field inflation}

Two mechanisms for inflation have been proposed. The simplest one
(Guth 1981) invokes a scalar field, termed the inflaton field.
An alternative (Starobinsky 1980) is to invoke a modification of
Einstein gravity, and
combinations of the two mechanisms have also been proposed. During inflation
however, the proposed modifications of gravity
can be abolished by redefining the spacetime metric tensor, so that one
recovers the scalar field case. We focus on it for the moment,
but modified gravity models
will be included later in our survey of
specific models.

In comoving coordinates a homogeneous scalar field
 $\phi$ with minimal coupling to gravity has the
equation of motion
\be \ddot \phi+3 H\dot \phi +V^\prime (\phi) =0 \label{52} \ee
Its energy density and pressure are
\bea \rho\eqa V+\frac12\dot\phi^2\\
p\eqa-V +\frac12\dot\phi^2 \eea
If such a field dominates $\rho$ and $p$, the inflationary condition
$3p<\rho$ is achieved provided that the field rolls sufficiently slowly,
\be \dot\phi^2<V \ee

Practically all of the usually considered models of inflation satisfy three
conditions. First, the motion of the field is overdamped, so that the `force'
$V^\prime $ balances the `friction term' $3H\dot\phi$,
\be
\dot{\phi} \simeq -\frac{1}{3H} V' \label{56}
\ee
Second,
\be
\epsilon \equiv \frac{m_{Pl}^2}{16\pi}
\left( \frac{V'}{V} \right)^2 \ll 1 \label{57} \ee
which means that the inflationary requirement $\dot\phi^2<V$ is well
satisfied and
\be H^2\simeq \frac13 \frac{8\pi}{\mpl^2} V \label{57a} \ee
These two conditions imply that
$H$ is slowly varying, and that the scale factor increases more or less
exponentially,
\be a\propto e^{Ht} \label{58b} \ee
The third condition that is usually satisfied is
\be |\eta|\ll1 \label{58} \ee
where
\be \eta \equiv \frac{m_{Pl}^2}{8\pi} \frac{V''}{V} \ee
It can be `derived' from the other two by differentiating
the approximation \eq{56} for $\dot\phi$ and noting that consistency with the
exact expression  \eq{52} requires $\ddot \phi\ll V^\prime $ is satisfied.
However
there is no logical necessity for the derivative of an approximation to be
itself a valid approximation, so this third condition is logically independent
of the others. Conditions involving higher derivatives of $V$ could be
`derived' by further differentiation, with the same caveat, but the two that
we have given, involving only the first and second derivatives, are the ones
needed to obtain the usual predictions about inflationary perturbations. The
term `slow-roll inflation' is generally taken to denote a model in which they
are satisfied and we are adopting that nomenclature here. Practically all of
the usually considered models of inflation satisfy the slow-roll conditions
more or less well, and we adopt them in this section, showing how to handle
the general situation in Appendix B.

It should be noted that the first slow-roll condition is on a quite different
footing from the other two, being a statement about the {\em solution} of the
field equation as opposed to a statement about the potential that
defines this equation. What we are saying is that in the usually
considered models one can show
that the first condition is an attractor solution, in a regime typically
characterised by the other two conditions, and that moreover reasonable initial
conditions on $\phi$ will ensure that this solution is achieved well before
the observable universe leaves the horizon.

\subsubsection*{Begining and ending inflation}

Given a parametrisation of the potential, one can calculate the spectrum of
the density perturbation and of the gravitational waves, and see if they are
consistent with observation in a theoretically acceptable regime of parameter
space. A parametrisation of the potential,  suggested more or less loosely by
some theoretical consideration, is referred to as a model (Linde 1990; Olive
1990; Kolb 1991).

Models of inflation may usefully be classified according to the way in which
they {\em enter} the inflationary era, and the way in which they {\em exit}
that era.
Two modes of entry have been discussed, which one may term {\em thermal} and
{\em chaotic}. In the thermal mode, one supposes that inflation is preceded by
an epoch of radiation domination, with at least some of the components in
thermal equilibrium. Specific attempts to realise the thermal case, which will
be discussed in more detail below, are `old inflation', `new inflation' and
`extended inflation'. In the thermal case, one does not usually discuss what
happened before thermal equilibrium set in. In the chaotic mode  it is
supposed that the universe, or at least our part of it, contains only smoothly
varying fields from the moment of its creation until the end of inflation.
Subsequently, most of the fields quickly adjust themselves to minimise the
effective potential, leaving behind the slowly rolling inflaton field and
perhaps some other fields.

Two modes of exit are commonly discussed, which one may term {\it first-order}
and {\em oscillatory}. In the case of first-order exit, the universe is
supposed
to be sitting in a false vacuum during inflation. Exit from inflation occurs
via a first-order phase transition, in which bubbles of the true vacuum form
by quantum tunnelling. The original (scalar field) inflationary model, usually
called `old inflation', was of this kind, and was quickly shown not to be
viable because the bubbles never coalesce. More recently, first-order
inflation models termed `extended inflation' have been formulated, which rely
on a modification of gravity to coalesce the bubbles. As we shall show, the
COBE data when combined with large scale galaxy surveys puts these models in
jeopardy too (Liddle \& Lyth 1992, 1993). Different first-order inflation
models discussed below, employing two scalar fields, may still be viable.

In the case of an inflation is supposed to end when the inflaton field moves
sufficiently close to its minimum that it starts to oscillate, without the
intervention of any phase transition. For the commonly considered potentials,
this epoch is marked by the failure of one of the conditions \eqs{57}{58}, so
denoting it by the subscript `end' we have
\be \max\{\epsilon\sub{end},|\eta\sub{end}|\}\simeq 1 \label{510a} \ee
The oscillations correspond to a gas of `inflaton' particles with negligible
pressure, so that there ensues a matter dominated era which ends only when the
inflatons decay into radiation. Of the thermal-entry models mentioned earlier,
`new inflation' invokes an oscillatory exit. Chaotic inflation is normally
also thought of as invoking an oscillatory exit, though this is not mandatory.

In addition to these two commonly considered cases, there exists the
possibility that inflation occurs in a false vacuum, which is however exited
through a {\em second order} phase transition as opposed to a first order one.
That is, the local minimum of the effective potential which represents the
false vacuum turns smoothly into a local maximum, after which the fields
quickly adjust themselves to the true vacuum values. By definition, bubble
production is negligible in a second order phase transition, but as with a
first order transition there may be production of topological defects such as
cosmic strings or domain walls. A simple two field model which seems to give
rise to a second order exit has been given by Linde (1991b); it may be termed
{\em two-scale inflation} because the fields have very different masses $m$
and $M$, related roughly be $m m_{Pl} =M^2$. We will describe the two-scale
model in Section \ref{INFL}.4, along with a selection of better known ones.

In any model, the universe (or at least the constituents dominating the energy
density) must achieve thermal equilibrium at some point after inflation.
Because the first models of inflation invoked a thermal entry, this event is
usually called `reheating', even in the chaotic scenario where no previous
thermalisation has occurred. The corresponding `reheat temperature' is the
biggest temperature ever achieved after inflation and plays a crucial role in
cosmology. Unfortunately, no reliable estimate of it is known at present.
Reheating will occur promptly if inflation ends in a phase transition
(certainly if it is a first-order one), but it may be long delayed if
inflation ends with oscillations of the inflaton field. The reason is that the
couplings of the inflaton field to other fields is typically quite weak. A
discussion of reheating, with references to earlier work, has recently been
given by Kofman, Linde and Starobinsky (1993).

\subsubsection*{The epoch of horizon exit}

A first requirement for an inflationary model is that the observable
universe should be within the horizon at the beginning of inflation.
We therefore need to know the epoch when a given scale
leaves the horizon, something which is necessary also to calculate the
inflationary perturbations. Denoting it by a star, this epoch is given
by
\be a_* H_*=k \label{ashs} \ee
The epoch of horizon exit is therefore related to the
present magnitude of the scale in Hubble units by
\be
\frac{a_0H_0}{k} =\frac{a_0 H_0}{a_* H_*} \ee
Let us denote the end of inflation a subscript `end' and the epoch of
reheating by `reh', assuming a matter domination during the era
(if any) between these two epochs. Let us also assume that after
`reh' there is radiation domination, until the epoch `eq'
at which the dark matter density becomes equal to that of the radiation.
Throughout the history of the universe the
radiation energy density is proportional to $a\mfour$, that of
the matter is proportional $a\mthree$, and the total is proportional to
$H^2$. It follows that
\bea \frac{k}{a_0 H_0}\eqa\frac{a_*}{a\sub{end}}
\frac{a\sub{end}}{a\sub{reh}}
\frac{a\sub{reh}}{a_0}
\frac{H_*}{H_0}\\
\eqa e^{-N_*} \rfrac{\rho\sub{reh}}{\rho\sub{end}}\third
\rfrac{\rho_{0r}}{\rho\sub{reh}}\quarter
\rfrac{\rho_*}{\rho_0}\half
\eea
where $\rho_{0r}=(a\sub{eq}/a_0)\rho_0$ is the present radiation energy
density and $N_*$ is the number of Hubble times between horizon exit and the
end of inflation,
\bea N_*\equiva \ln(a_*/a\sub{end})\\
\simeqa \int^{t\sub{end}}_{t_*} H dt \label{nsta}
\eea
It follows that
\be N_* = 62- \ln \frac k{a_0 H_0}
- \ln \frac{10^{16}\GeV}{V_*^{1/4}}
+ \ln \frac{V_*^{1/4}}{V\sub{end}^{1/4}} -\frac13 \ln\frac{V\sub{end}
\quarter}{\rho\quarter\sub{reh}}
\ee

This equation relates the three energy scales $V_*\quarter$,
$V\sub{end}\quarter$ and $\rho\sub{reh}\quarter$. The first two scales are
related by another expression for $N_*$, which follows from
\eqsss{nsta}{56}{57}{57a}
\be
N_*=\frac{8\pi}{m_{Pl}^2} \int_{\phi\sub{end}}^{\phi_*} \frac{V}{V^\prime}
 d\phi =
\sqrt{\frac{4\pi}{m_{Pl}^2}} \; \left|\int_{\phi\sub{end}}^{\phi_*}
	\epsilon\mhalf d\phi \right| \label{nhub} \ee

The biggest scale that can be explored is roughly the present Hubble distance,
$a_0/k =H_0\mone=6000\Mpc$. We shall refer to the epoch of horizon exit for
this scale as `the epoch when the observable universe leaves the horizon', and
denote it by the subscript 1. As displayed in \eq{vqua}, the COBE observations
require that $V_1\quarter\lsim 10^{16}\GeV$, with the equality holding in most
models of inflation. Also, \eq{nhub} gives in most models of inflation
$V_1\quarter\simeq V\sub{end}\quarter$. If reheating is prompt, we learn that
in most models the observable universe leaves the horizon about 62 e-folds
before the end of inflation. If reheating is long delayed $N_1$ could be
considerably reduced, being equal to 32 for the most extreme possibility of
$V\quarter\sub{reh}\sim 1000\GeV$ (corresponding to reheating just before the
electroweak transition which is presumably the last opportunity for
baryogenesis). For most purposes, however, one needs only the order of
magnitude of $N_1$.

The smallest scale on which the primeval perturbation can be probed at present
is around $1\Mpc$, and one sees that this scale leaves the horizon about 9
Hubble times after the observable universe. Discounting the case of very late
reheating, we conclude that in the usually considered inflationary models,
scales of cosmological interest leave the horizon 50 to 60 Hubble times before
the end of inflation.

\subsection{Calculating the density perturbation}

Soon after inflation was proposed, it was realised that the vacuum fluctuation
of the inflaton field would inevitably give rise to an adiabatic density
perturbation. Its spectrum, specified say by the quantity $\delta_H(k)$ that
we introduced earlier, can be calculated in terms of the inflationary
potential. The first calculations (Guth \& Pi 1982; Hawking 1982; Starobinsky
1982; Bardeen, Steinhardt \& Turner 1983) quoted only order of magnitude
results, largely because they did not specify the normalisation of the
spectrum that the result was supposed to apply to, but partly also because the
calculations themselves were not very precise. The precise result, valid
subject to the three slow-roll conditions listed above, was first given by
Lyth (1985),
\be \delta_H^2(k)= \frac{32}{75}\frac{V_*}{\mpl^4}\epsilon_*\mone
\label{512} \ee
As always, a star denotes the epoch of horizon exit $aH=k$.

We now derive this result, using essentially the original method with a
simplification due to Sasaki (1986). Here in the text we ignore the
back-reaction of the metric perturbation on the inflaton field, explaining in
Appendix B why it is negligible within the slow-roll paradigm. We also explain
there how departures from slow-roll, including the back-reaction, can easily
be evaluated if desired (Stewart \& Lyth 1993). A somewhat related formalism
was developed by Mukhanov (1985, 1989), and yet another by Salopek, Bond and
Bardeen (1989), but neither of them is as simple as the one advocated here,
especially in the context of the slow-roll paradigm.

To define the perturbation $\delta \phi$ of the inflaton field, one has to
choose a slicing of space-time into spacelike hypersurfaces, just as was the
case for the perturbations $\delta\rho$, $\delta p$ and $\delta H$ that we
studied earlier. Since the inflaton field is supposed to dominate the
energy-momentum tensor, the momentum density vanishes if its spatial gradients
vanish (Bardeen, Steinhardt \& Turner 1983).  In other words, $\delta \phi$
vanishes if the hypersurfaces are chosen to be comoving!

As we explain in Appendix B, the comoving choice of hypersurfaces becomes
singular in the limit $\epsilon\to 0$ of exponential inflation. If instead we
make a choice which is non-singular, it becomes very easy to handle the
inflaton field perturbation. Its field equation, given the slow-roll
conditions, is the one that would na\"{\i}vely be obtained by perturbing
\eq{52}, ignoring the perturbation of the metric,
\be (\delta\phi\sk)\,\ddot{}+3H(\delta\phi\sk)\dot{}
+\left[\rfrac ka ^2 +V^{\prime \prime} \right] \delta
\phi\sk =0 \label{510} \ee

Until a few Hubble times after horizon exit \eq{58}
ensures that $V^{\prime \prime}$ can be dropped, so that the equation becomes
\be (\delta\phi\sk)\,\ddot{}+3H(\delta\phi\sk)\dot{}
+\rfrac ka ^2 \delta
\phi\sk =0 \label{511} \ee
Well before horizon entry $\delta\phi\sk $ is a massless
field living in
flat space-time, since its wavenumber $k/a$ is much
bigger than
the Hubble parameter $H$.
It can be quantised in the usual way so that its
quantum state  is labelled by the number of inflaton particles
 present with momentum $\bfk$.

The fundamental assumption leading to \eq{512} is that on cosmologically
relevant scales {\it the inflaton field is in the vacuum state, corresponding
to no inflaton particles}. Let us pause to ask about the status of this
assumption.\footnote{The following arguments, which we have not seen written
down before, were developed in collaboration with Ewan Stewart.} The idea
behind it is that at any epoch during inflation there is a cutoff $k\sub
{max}$ in
the energy spectrum of inflaton particles. Such a cutoff is obviously present
in order to avoid infinite energy density, but the point at issue is its
magnitude, in Hubble units. If inflation is preceded by an epoch of thermal
equilibrium, the equilibrium occupation number
(Kolb \& Turner 1990) $2(\exp(k/aT)\pm1)\mone$
implies $k\sub{max}/a\sim T$. In that case inflation begins when the
radiation energy $\sim T^4$ falls below the inflaton field energy $\sim V$, so
the cutoff is independent of the epoch and is given by $(k\sub{max}/a)^4 \sim
V\sub{begin}$. But since $V$ decreases during inflation,
and $a$ increases rapidly,
this implies that
once inflation is underway
\be \rfrac{k\sub{max}}{a}^4 \ll V \label{513}  \ee
Irrespective of what happens before inflation, this condition is in any
case necessary for consistency, just to ensure that
the radiation energy density due to the inflaton particles
is much less than the
potential $V$ which is supposed to be dominant. Indeed, since the density of
states in $\bfk$ space is $(2\pi)\mthree$, one inflaton per state up to
$k\sub{max}$ implies a radiation energy density given by
\be \rho\sub{rad}=\frac{1}{2\pi^2}\int ^{k\subsub{max}}
_0 \rfrac ka^4 \frac{dk}{k} =\frac1{8\pi^2}
\rfrac{k\sub{max}}{a}^4 \ee
Apart from the numerical factor this leads to the desired condition, \eq{513}.

The condition is enough to justify the vacuum assumption. To see this, rewrite
it in the form
\be \rfrac{k\sub{max}}{aH}\ll \rfrac{\mpl}{H}\half \label{516} \ee
We shall see that typically the right hand side is of order $10^3$, so assume
this value for definiteness. Now suppose that inflation begins some number
$N\sub{before}$ of Hubble times before the observable universe leaves the
horizon. Since the scale factor
$a$ grows like $e^{Ht}$, with $H$ is slowly varying,
\eq{516} applied at the beginning of
inflation leads to
\be \frac{k\sub{max}}{k_1}\lsim 10^3 e^{-N\subsub{before}} \ee
The vacuum assumption is therefore ensured if inflation begins at least 7 or so
Hubble times before the observable universe leaves the horizon. Clearly, the
result is not very sensitive to the assumed value of $\mpl/H$.

Now let us see how the spectrum \eq{512} is implied by the vacuum assumption.
The first step is to calculate the vacuum expectation value of
$|\delta\phi\sk|^2$. Well before horizon entry we need only the flat
space-time version of quantum field theory, rewritten in terms of the comoving
coordinate $x^i$ and the comoving wave vector $\bfk$. They are related to the
Cartesian coordinates and physical wave vectors by the scale factor $a(t)$,
whose time dependence is negligible on the timescale $a/k$. Working in the
Heisenberg representation, $\phi\sk$ is associated with an operator
\be \hat\phi\sk(t)
=w_k(t) \hat a\sk+w^*_k(t)\hat a^\dagger_{-{\mbox{\scriptsize\bf k}}}
\label{518}  \ee
The annihilation operator $\hat a\sk$ satisfies the commutation
relation
\be [\hat a_{\mbox{\scriptsize \bf k }_1},
\hat a^\dagger_{\mbox{\scriptsize \bf k }_2} ]= \delta _{
\mbox{\scriptsize \bf k}_1, \mbox{\scriptsize \bf k}_2} \label{520} \ee
and the field satisfies the
commutation
relation,
\bea[\hat \phi(\bfy _1,t),\pdif{}{t}
\hat\phi(\bfy _2,t)]
\eqa  i \delta^3(a \bfy_1-a \bfy _2)  \\
\eqa ia^3\delta(\bfy_1-\bfy_2)
\eea
As a result the functions $w_k$ are given by
\be w_k(t)=a\mthreehalf(2k/a)\mhalf
e^{-i(\chi+(kt/a))} \label{519} \ee
The phase factor $\chi$ is arbitrary, and like $a$ it must have negligible
variation on the timescale $a/k$. The vacuum state is the one annihilated by
$\hat a\sk$, so the vacuum expectation value of the field perturbation is
\be \langle|\delta\phi\sk|^2\rangle=|w_k|^2
\label{521} \ee

To extend these results to the epoch of horizon exit and beyond, we have to
accept the validity of free field theory in curved space-time. It seems to
have first been formulated for a general Robertson--Walker metric by Parker
(1969), and an extended account of the subject in this and other space-times
is given in the text of Birrell and Davies (1982). Hardly any of this general
theory is in fact necessary for the present application; all we need to assume
is that there is a Heisenberg picture, in which operators satisfy the
classical equations of motion and state vectors are time independent. Then
$\hat\phi\sk$ continues to satisfy the field equation \eq{510}, and
\eqss{518}{520}{521} still hold, where $w_k$ is the solution of the field
equation reducing to \eq{519} well before horizon entry. As one easily checks,
the required solution is
\be w_k(t)=\frac{H}{(2k^3)\half }
\left(i+\frac{k}{aH}\right) e^{ik/aH} \label{523} \ee
A few Hubble times after horizon exit,
the vacuum expectation value is therefore
\be \langle|\delta\phi\sk|^2 \rangle=\frac{H^2}{2k^3}
\label{524} \ee

A measurement of the $\phi\sk$'s
will yield random phases, and a distribution of moduli whose dispersion is
given by \eq{524}. Accordingly the spectrum of the inflaton field, defined by
\eq{21}, is given a few Hubble times after horizon exit by (Vilenkin \& Ford
1982; Linde 1982; Starobinsky 1982)
\be {{\cal P}}\half_\phi(k)=\frac{H}{2\pi} \label{529} \ee
This equation does not generally hold more than a few Hubble times after
horizon exit, for two separate reasons. First, \eq{523} fails to be a solution
of the massless field equation \eq{511} unless $H$ is practically constant, as
opposed to just slowly varying. Indeed, after horizon exit the third term of
\eq{511} can be dropped, and its solution remains constant whether or not $H$
varies. Second, the massless equation will not in general be correct, because
the term $V^{\prime \prime}$ in
\eq{510} will become significant. In order to derive our
desired result however, it will be enough to know that \eq{529} holds a few
Hubble times after horizon exit. Since $H$ is slowly varying on the Hubble
timescale it can be written
\be {{\cal P}}\half_\phi(k)=\frac{H_*}{2\pi} \label{526} \ee
where the star denotes the epoch of horizon exit, and
${{\cal P}}_\phi$ is evaluated a few Hubble times after horizon exit.

\subsubsection*{Massless field inhomogeneity}

Before continuing the derivation of the spectrum of the density perturbation,
we briefly digress to consider the case of a massless scalar field $\psi$.
Provided that the field has minimal coupling to gravity it satisfies
\eq{511}, so one arrives at \eq{526} with $\phi$ replaced by $\psi$,
\be {{\cal P}}_\psi\half(k)=\frac{H_*}{2\pi} \ee
In contrast with \eq{526} though, this equation remains valid
arbitrarily long
after horizon exit, because as we noted already the massless field equation
ensures that $\psi\sk$ remains constant. Defining a spectral index by
${{\cal P}}_\psi\propto k^{n_\psi}$ or equivalently
\be n_\psi=\frac{d\ln {{\cal P}}_\psi}{d\ln k} \ee
one concludes from \eqssss{56}{57}{57a}{ashs}{58b} that (Liddle
\& Lyth 1992)
\be n_\psi=-2\epsilon_1 \label{527} \ee
To the extent that the spectral index is well defined $\epsilon$ does not vary
significantly while interesting scales leave the horizon, but for definiteness
we may evaluate it when the observable universe leaves the horizon and this is
indicated by the subscript 1.

If the inhomogeneity of $\psi$ is converted later into an isocurvature density
perturbation $S$ as discussed in Section \ref{DENSPER}.7, the spectrum
${{\cal P}}_S$ has the same shape, so \eq{527} gives its spectral index.

\subsubsection*{The curvature perturbation}

Returning to the derivation of \eq{512}, we have so far dealt with the
inflaton field perturbation $\delta\phi$, defined on hypersurfaces which are
non-singular in the limit $\epsilon\to 0$. It is shown in Appendix A that the
curvature perturbation ${{\cal R}}$
of comoving hypersurfaces is given in terms of
this quantity by
\be {{\cal R}}=\frac{H}{\dot\phi} \delta\phi \label{532} \ee
A relation equivalent to this one was derived by Lyth (1985), but its simple
form in terms of the ${{\cal R}}$
variable was first appreciated by Sasaki (1986),
who pointed out that it followed directly from a formula given by Bardeen
(1980). A simple derivation of it is given in Appendix A.

The spectrum of ${{\cal R}}$ is therefore given by
\be {{\cal P}}_{{\cal R}}\half=\frac{H}{\dot\phi}{{\cal P}}_\phi\half \ee
In Section \ref{DENSPER}.4
we learned that ${{\cal R}}$ is constant after horizon
exit, so ${{\cal P}}_{{\cal R}}$ remains constant even though $H/\dot\phi$ and
${{\cal P}}_\phi$ might vary separately. It follows that as long as the scale
is
far outside the horizon,
\be {{\cal P}}_{{\cal R}}\half= \frac{H_*^2}{2\pi \dot\phi_*} \ee
where the star denotes the epoch of horizon exit.
Using \eq{56}, this
leads to \eq{512}, concluding our proof of that equation.

\subsubsection*{The spectral index}

The spectral index $n$ of the density perturbation is defined
by $\delta_H^2\propto k^{n-1}$, or equivalently
\be n-1\equiv\frac{d\ln \delta_H^2}{d\ln k} \ee
The power-law dependence on $k$ is not supposed to be exact, so that the
latter expression should be taken as the proper definition of the spectral
index, being useful if $n$ does not vary much over a cosmologically
interesting range of scales. Differentiating \eq{512} with the aid of
\eqssss{56}{57}{57a}{ashs}{58b} leads to the result (Liddle \& Lyth
1992)
\be n=1+2\eta_1-6 \epsilon_1 \label{532a}
\end{equation}
To the extent that $n$ is scale-independent the right hand side can be
evaluated at any epoch, which for definiteness we have taken to be the epoch
when the observable universe leaves the horizon.

In order to have significant deviations from the flat $n=1$ spectrum one must
come close to violating at least one of the slow-roll conditions $\epsilon\ll1
$ and $|\eta|\ll1$. Nevertheless it turns out that three of the half dozen or
so currently favoured models can naturally give a value of $n$ significantly
below 1. It is more difficult to have $n$ significantly above 1 because
$\epsilon$ is positive by definition;
none of the usually considered models has $n>1$ but we shall see that
small positive values are possible in
two-scale models.

\subsubsection*{The inflationary energy scale}

As we see shall see later, the cmb anisotropy observed by COBE determines the
value of $\delta_H$ on the scale corresponding roughly to the size  of the
observable universe, if the density perturbation is adiabatic and dominates
the anisotropy. The value is $\delta_H=1.7\times 10^{-5}$
with a one-sigma
uncertainty of about 20\%. From \eq{512} this determines the potential and the
Hubble parameter in terms of $\epsilon_1$, at the epoch when the observable
universe leaves the horizon
\bea V\quarter_1\eqa 6.2\epsilon_1\quarter
\times 10^{16}\GeV \label{vqua} \\
H_1\eqa 9.1\epsilon_1\half\times 10^{14}\GeV \label{hone}
\eea

If there are other contributions to the cmb anisotropy these estimates are
reduced. One therefore has the important conclusion (Lyth 1984) that {\em the
inflaton potential when the observable universe leaves the horizon satisfies
$V_1\quarter\lsim 10^{16}\GeV$.} This is also an upper bound on the potential
$V\sub{end}$ at the end of inflation, and in fact with very weak assumptions
(Lyth 1990) one can say more precisely that $V\sub{end}\quarter<1.6 \times
10^{16}\GeV$.

\subsubsection*{The classical nature of the predicted perturbation}

To end this long section, we want to explain in what sense the predicted
density perturbation is a classical quantity. Similar considerations
apply to the predicted fluctuation of a massless field, and to the
predicted gravitational waves to be discussed in a moment.

In arriving at \eq{526} for the spectrum well after horizon exit, we supposed
that a measurement of the field had been made, so that the squared Fourier
components $|\phi\sk|^2$ took on values drawn from the probability
distribution predicted by the theory. The expected value
$\langle|\phi\sk|^2\rangle$ is given by the theory, and this is all that is
required to evaluate the spectrum. Nevertheless, what one observes (through
the density perturbation which is linearly related to it) is the field itself,
as a function of spacetime position. The question is, whether the quantum
uncertainty is small enough to permit this field to have a sharply defined
value over an extended period of time. The answer turns out to be yes, but for
this positive answer it is crucial that the field is measured only well after
horizon exit.

Our fundamental hypothesis is that each perturbation starts out, well before
horizon entry, as a vacuum fluctuation of a massless field in flat spacetime.
The state of the system is the vacuum state, in which the expectation value of
the field vanishes. A measurement of the field made well before horizon entry
would produce a new state in which it has a nonzero expectation value, but of
course we have to suppose that no such measurement is made or our fundamental
hypothesis would be violated. The probability of each possible outcome of the
measurement would be given by quantum theory; to be precise, the real and the
imaginary part of each Fourier component is uncorrelated, with a Gaussian
probability distribution whose variance is given by (half of) \eq{521}.

The flat spacetime vacuum fluctuation is an essentially quantum object because
it is not possible to put the field into a condition where it remains sharply
defined over an extended period of time, unless the field strength is far out
on the tail of the probability distribution. This is clear from the fact that
the real and imaginary parts of each Fourier component have the same dynamics
as a quantum oscillator, the vacuum state being the ground state of the
oscillator; it is well known that for such an oscillator it is not possible to
make a `wave packet' which does not spread significantly, unless its
displacement from the origin is far in excess of the root mean square value
for the ground state.

The situation is quite different well after horizon exit. Then, one can check
that it is possible to give the field a sharply defined value, whose strength
as measured by the modulus squared of the Fourier coefficients is around the
quantum expectation value (Lyth 1985; Guth \& Pi 1985; Grishchuk 1993;
Albrecht {\em et al} 1993). The underlying reason for the classical behaviour
seems to be the fact that $w_k$ becomes real after horizon exit (Starobinsky
1982, 1986) so that $\hat\phi\sk$ has only trivial time dependence
\be \hat\phi\sk\simeq w_k(t)[ \hat a\sk+\hat a^\dagger_{-
{\mbox{\scriptsize \bf k}} }] \ee

This quantum-to-classical behaviour is a tremendous success for the theory,
which does not seem to have received adequate publicity. If it had failed, the
prediction for spectrum of the density perturbation would have had nothing to
do with reality.

Of course there remains the usual interpretation problem, arising whenever one
talks about measurement in quantum theory. Even after horizon exit one is not
allowed to say that the $\phi\sk$'s have particular values before they are
measured, because in principle one is supposed to be able to measure
observables whose operators do not commute with the $\hat\phi\sk$'s. A similar
situation arises when one observes an electron created in, say, beta decay.
Even if its observed position is a metre away from the source, one is not
allowed to say that it arrived at that position on the classical trajectory.
The reason is that one might in principle have chosen to measure, say, its
angular momentum, which involves the full outgoing spherical wave function.
The problem with the inflaton field is, however, more severe because as we
shall see it is essentially equal to the density perturbation. Whatever one
might think about an electron, one is certainly unwilling to say that the
density perturbation of the universe is not there until it is measured. The
problem has been discussed by many authors using the concept of decoherence
(Sakagami 1988; Halliwell 1989; Padmanabhan 1989),  but it seems far from a
satisfactory solution.

\subsection{Primordial gravitational waves?}

We stated earlier that the gravitational wave amplitude $h_{ij}$ satisfies the
massless wave equation. To derive this equation, one can evaluate the
contribution of $h_{ij}$ to the Einstein Lagrangian
\be {\cal L}=-\frac1{16\pi}R \ee
where $R$ is the curvature scalar of spacetime. As shown in for example the
review of Mukhanov, Feldman \& Brandenberger (1992) one finds that up to a
total
derivative each of the four non-zero Fourier components discussed in Section
\ref{MWB}.3has the same Lagrangian as a Fourier component of a massless scalar
field $\psi=(\mpl^2/16\pi)\half h_{ij}$.

This result not only proves the field equation, but also allows us to quantise
the gravitational wave amplitude. Each non-zero Fourier component has the same
vacuum fluctuation as a massless field, with the above conversion factor. Thus
the spectrum of the gravitational waves is
\be {{\cal P}}_g(k)= 4\times \frac{16\pi}{\mpl^2} \rfrac{H_*}{2\pi}^2 \ee
Putting this expression into \eq{427} and dividing it by \eq{422} gives the
ratio \eq{430} of the gravitational wave and density contributions to the cmb
anisotropy.

Just as in \eq{527}, the spectral index of the gravitational waves
is (Liddle \& Lyth 1992)
\be n_g=-2\epsilon_1 \ee

\subsection{A survey of inflationary models}

We now apply the formalism that has been developed to some popular models of
inflation, as well as to the `two-scale' model mentioned earlier. The survey
is by no means exhaustive, and does not consider `designer' inflation (Kofman
\& Linde 1987; Kofman \& Pogosyan 1988;
Salopek, Bond \& Bardeen 1989; Linde 1992), where one tunes the
shape of the potential, typically with more than one inflaton field, to obtain
more or less any desired spectrum. Somewhat similar surveys have been given by
Salopek (1992a, 1992b), Davis {\em et al} (1992b), Liddle \& Lyth (1992) and
Turner (1993). The outcome of our discussion is summarised in Table 2.

\subsubsection*{Polynomial Chaotic inflation}

In chaotic inflation (Linde 1983, 1990) the inflaton potential is supposed to
emerge, at the Planck scale, with a value $V\sim\mpl^4$.
If the potential is $V\propto\phi^{\alpha}$ inflation can begin at the
Planck scale, with an oscillatory exit at the
epoch determined by \eq{510a}, $\phi\sub{end}\simeq \alpha (\mpl^2/8\pi)\half$.
{}From \eq{nhub}
with $N_1=60$, $\phi_1^2\simeq120\alpha m_{Pl}^2/8\pi$. This leads to
$1-n=(2+\alpha)/120$  and $R=.05\alpha$, leading to the following
relation between the gravitational wave contribution and the spectral
index
\be R=6(1-n)-0.1 \label{ntor} \ee
For $\alpha=2$, 4, 6 and 10 one has
$1-n=.033$, $.05$, $.067$ and $.1$, and
$R=.10$, $.20$, $.30$ and $.50$.
Shafi (1993) has shown that a potential of this kind follows from a class of
superstring-inspired gauge theories, with an
index $\alpha$ varying between $6$ and $10$.

\subsubsection*{$R^2$ inflation}

Instead of using a scalar field to drive inflation one can modify gravity with
the addition of an $R^2$ term,
\be {\cal L}=-\frac{\mpl^2}{16\pi}[R+R^2/(6M^2)]+ {\cal L}\sub{nongr} \ee
The first term is the Lagrangian for Einstein gravity, and the third term is
the Lagrangian of the non-gravitational sector (standard model, GUTS or
whatever), and the second term modifies Einstein gravity. A more complicated
version of this Lagrangian was the first model of inflation (Starobinsky
1980). The simple version given here has recently been justified in the
context of supergravity (Cardoso \& Ovrut 1993), and because of its simplicity
it has been analysed by several authors (Starobinsky 1983; Barrow \& Ottewill
1983; Whitt 1984; Duruisseau \& Kerner 1986; Mijic, Morris \& Suen 1986;
Starobinsky \& Schmidt 1987; Suen \& Anderson 1987; Kofman, Mukhanov \&
Pogosyan 1987). The $R^2$ term can be eliminated by a conformal transformation
of the metric (Whitt 1984) to give a new Lagrangian
\be {\cal L}=-\frac{\mpl^2}{16\pi}\widetilde R
+\frac12 \tilde g ^{\mu\nu}\pa_\mu\phi \pa_\nu \phi
-V(\phi) + \widetilde{\cal L}\sub{nongr} \ee
where
\be V(\phi)=\frac{3\mpl^2 M^2}{32\pi}
\left[ 1-\exp\left(-\rfrac{16\pi}{3\mpl^2}\half \phi\right) \right]^2
\ee
The new Lagrangian has standard gravity, and a scalar field $\phi$ with
canonical kinetic terms which did not appear at all in the original
Lagrangian, as well as  the original non-gravitational sector which now
appears in {\it modified form}. In the regime $\phi\gsim m_{Pl}$, the
potential $V(\phi)$ satisfies the slow-roll conditions so that inflation
occurs. The non-gravitational sector is irrelevant during inflation, and if
the $R^2$ term quickly becomes negligible afterwards we have a model in which
inflation is driven by a scalar field, without any modification of gravity.
What has been achieved is to motivate the otherwise bizarre form of the
inflationary potential.

In the regime $\phi\gsim\mpl$ where they are small, the parameters appearing
in the slow-roll conditions are
\bea
\eta\eqa - \frac43
\exp\left( -\sqrt\frac23 \frac{\sqrt{8\pi}}{\mpl}\phi\right)\\
\epsilon\eqa\frac34 \eta^2
\eea
There is an oscillatory exit to inflation at the epoch given by \eq{510a},
$\phi\sub{end}\sim (\mpl^2/8\pi)\half $. From \eq{nhub} with $N_1=60$, \be
\phi_1\simeq 5 \frac{\mpl}{\sqrt{8\pi}} \ee leading to $\eta_1\simeq -.02$ and
$\epsilon_1\sim 10\mfour$. Thus the spectral index is $n=0.96$,
but the gravitational wave contribution is completely
negligible.

Salopek, Bond and Bardeen (1989) have considered different modifications
of gravity, termed `induced gravity' and `variable Planck mass' models,
which after a conformal transformation lead to inflation with the same
potential as with $R^2$ inflation, in the relevant regime of parameter space.
These models therefore give the same spectral index, and again negligible
gravitational waves.

\subsubsection*{Power-law inflation}

Power-law inflation $a \propto t^p$ (Abbott \& Wise 1984b; Lucchin \&
Matarrese 1985; Barrow 1987; Liddle 1989) corresponds to an exponential
potential
\begin{equation}
V(\phi) = V_0 \exp \left( \sqrt{\frac{16\pi}{p\, m_{Pl}^2}} \, \phi \right)
\end{equation}
With the required value $p>1$ such a potential does not have any known
theoretical motivation, except in the context of extended inflation which we
consider in a moment. In contrast with the previous two models, the
featureless exponential potential cannot provide an oscillatory end to
inflation, but the calculation of the spectra of the density perturbation
and the gravitational waves does not depend on what kind of exit occurs.
The slow-roll conditions are satisfied for all $\phi$ if $p\gg 1$.
The slow-roll parameters
are $\epsilon=\eta/2=1/p$ independent  of time.
The spectral indices are given by $-n_g=1-n=2/p$,
and $R=12/p$ leading to the relation
\be R=6(1-n) \ee
The chaotic result \eq{ntor} becomes the same as this result for $1-n\gsim.1$.

Power-law inflation is of particular interest to researchers in inflation
because exact analytic solutions exist both for the dynamics of inflation and
for the density perturbations generated. As a result one can check that the
slow-roll approximation remains reasonable, even when the condition $p
\gg 1$
is not very well satisfied. The exact gravitational wave spectrum was
calculated by Fabbri, Lucchin and Matarrese (1986), and the exact density
spectrum by Lyth \& Stewart (1992a). The exact spectral indices are
\begin{equation}
-n_g=1-n= 2/(p-1)
\end{equation}
The normalisations of the spectra do not deviate much from the slow-roll
expressions. From \eq{422} it follows that the same is true of the density
contribution to the microwave anisotropy. Full results have not been given for
the gravitational contribution, but for $\ell=2$ the value of $Rp/12.5$ can be
read from Figure 2 of Fabbri, Lucchin and Matarrese (1986), and is within
$10\%$ or so of unity for $.6<n<1$.

\subsubsection*{Extended Inflation.}

Extended inflation is based on a modification to the gravitational sector of
the Lagrangian, which allows a first-order inflationary phase transition to
complete satisfactorily (La \& Steinhardt 1989; Kolb 1991). The modification
can be eliminated by a conformal transformation of the metric. In general this
is not a very useful thing to do because it modifies the Lagrangian of the
non-gravitational fields, but it is useful during inflation. In fact, after the
transformation one has power-law inflation, with unmodified gravity (Kolb,
Salopek \& Turner 1990).

The original model (La \& Steinhardt 1989) was based on a Brans--Dicke theory
with parameter $\omega$, and although this proved insufficient to allow present
day tests of general relativity to be satisfied, it has remained the paradigm
around which more complicated working models are based. After the conformal
transformation, one has power-law inflation with $2p = \omega + 3/2$.

A crucial difference between power-law and extended inflation is that extended
inflation suffers an additional constraint, as one must avoid the large
bubbles generated as inflation ends from being so profuse as to unacceptably
distort the microwave background. This constrains $\omega$ as a function of
the inflaton energy scale $M$ as (Liddle \& Wands 1991; Liddle \& Lyth 1992)
\begin{equation}
\omega < 20 + 0.7 \log_{10} \left( M/m_{Pl} \right)
\end{equation}
Bounding $M$ using the microwave background limits on the fluctuation amplitude
gives $\omega \lsim 17$, corresponding to $n \lsim 0.75$. We have discussed the
significance of this bound in earlier papers (Liddle \& Lyth 1992, 1993) and
the prognosis for the extended inflation model is not good, as we shall recap
in the Conclusion.

\subsubsection*{Natural Inflation.}

The natural inflation model (Freese, Frieman \& Olinto 1990; Adams {\it et al}
1993) is based on a pseudo--Nambu--Goldstone boson evolving in a potential
\begin{equation}
V(\phi) = \Lambda^4 \left( 1 \pm \cos (\phi/f) \right)
\label{562} \end{equation}
where $\Lambda$ and $f$ are mass scales.
In this model
\bea \epsilon\eqa r \mone\tan^2\frac{\phi}{2f} \\
\eta\eqa -r \mone\left( 1-\tan^2\frac{\phi}{2f} \right)
\eea
where $r\equiv (16\pi f^2)/\mpl^2$. Since $\epsilon-\eta=r\mone$ one must have
$r\gg 1$ to be in the slow-roll regime anywhere on this potential. On the
other hand, one would like to have $f\lsim (\mpl^2/8\pi)\half$ so that quantum
gravity does not spoil the field theory on which \eq{562} is based. To this
extent `natural' inflation is somewhat unnatural, and one will not
wish to have $1/r$ smaller than is necessary to obtain a viable model.

Inflation ends with an oscillatory exit at the epoch given by \eq{510a},
$\tan(\phi\sub{end} /2f)\simeq r\half$ which is of order 1. Using this result,
\eq{nhub} with $N_1=60$ gives $\phi_1/2f\simeq\exp({-60/r})$, leading to
\bea \epsilon_1\eqa \frac 1{r} e^{-120/r} \\
\eta_1\simeqa -1/r\\
1-n\simeqa 2/r
\eea
Thus, natural inflation makes the gravitational waves negligible but tends to
give a significantly tilted spectrum.

A similar analysis is valid for any model which has the inflaton field near
the top of an inverted harmonic oscillator potential, $V(\phi) = V_0 -
\frac{1}{2} m^2 \phi^2$.

That natural inflation gives the same spectrum as power-law inflation may seem
surprising in the light of claims that one can reconstruct inflationary
potentials from a given spectrum (Hodges {\em et al} 1989;
Hodges \& Blumenthal 1990). In fact, these
two models can be regarded as different regimes of an all-encompassing
potential (which one can calculate in the manner of Hodges and Blumenthal)
which is essentially $1/\cosh^2 \phi$ with various factors thrown in (Copeland
{\it et al} 1993). In the $\phi \simeq 0$ region we have the inverted harmonic
oscillator, while at large $\phi$ we have the exponential region. In a sense,
this is the unique potential from which all inflationary models giving
power-law spectra arise. With this potential, the scalar field can roll from
the top down to the exponential region, while in a slow-roll approximation
generating an exact power-law spectrum. However, one must also note that as
discussed above, the fact that the spectrum has been tilted implies that the
slow-roll approximations are at best only just satisfied, so there will be
corrections to the slow-roll spectrum in all regions of this potential. In the
exponential region these corrections have long been known to affect only the
amplitude and not the slope; Stewart and Lyth (1993) have shown that the
corrections are also not very significant near the top of the potential.

\subsubsection*{Two-scale inflation}

The two-scale model written down by Linde (1991b) corresponds to the
potential
\be V(\psi,\phi)=\frac14\lambda (\psi^2-M^2)^2+\frac12m^2\phi^2+
\frac12\lambda^{\prime}\phi^2\psi^2 \ee
The couplings
$\lambda$ and $\lambda^{\prime }$ are supposed to be somewhat less
than unity,
and in the numerical examples we set $\lambda=\lambda^\prime =0.1$.
For $\phi^2 >
\phi\sub{end}^2 = \lambda M^2/\lambda^\prime $, the potential for
the $\psi$ field has a local minimum at $\psi = 0$, in which the field is
assumed to sit (except for the quantum fluctuation). Inflation can occur as
$\phi$ slowly rolls down its potential
\be V(\phi)=\frac14 \lambda M^4 +\frac 12 m^2 \phi^2
\label{twsc} \ee
When $\phi$ falls below $\phi\sub{end}$ this local minimum becomes a local
maximum, so that there is a phase transition.
Provided that bubble formation is negligible the transition will be of
second order, the fields rolling quickly down to their vacuum values
$\psi=M$ and $\phi=0$. (Otherwise we have a first order exit
model akin to the two-field first-order
models of for instance Adams and Freese (1991) and
Linde (1991a), which are briefly discussed in the following subsection.)

We will suppose that while observable scales leave the horizon the first term
of \eq{twsc} dominates, since in the opposite case we recover the $\phi^2$
potential already considered. This means that
\be \frac2\lambda \frac{m^2\phi^2_1}{M^4} \ll 1 \label{domi} \ee
Of the parameters $\epsilon$ and $\eta$ which are required to be small,
the second is independent of $\phi$,
\be \eta=\frac4\lambda X^2 \ee
where
\be X^2\equiv \frac{m_{Pl}^2}{8\pi} \frac{m^2}{M^4} \ee
The ratio $X$ must therefore be significantly less than 1. From
\eq{nhub},
\be \phi_1=\sqrt{\lambda/\lambda^\prime } \, M e^{N_1\eta} \ee
Consistency with \eq{domi} requires roughly $\eta\lsim .1$.
The other small quantity $\epsilon_1$ is given by
\be \epsilon_1=\frac12 \frac\lambda{\lambda^\prime }
\frac{8\pi}{m_{Pl}^2}M^2 \eta^2 e^{2N_1\eta} \ee
and \eq{domi} requires $\epsilon_1\ll \eta$.
It therefore follows that the spectral index $n$ is {\em bigger} than
$1$ in the two-scale model.

Setting $\lambda=\lambda^\prime =.1$ and imposing the COBE normalisation
$\delta_H=1.7 \times 10^{-5}$
determines all of the parameters in terms of
$m$. For $m=100\GeV$ one has $M=4\times10^{11}\GeV$ leading to
$\eta=10\mfour$ and $\epsilon_1
=10^{-23}$. The gravitational waves are
absolutely negligible, and
the spectral index is extremely close to 1.
The maximum value of $m$ permitted by \eq{domi} is roughly
$m=10^{13}\GeV$, giving
$M=2\times 10^{16}\GeV$, $\eta=.07$ and $\epsilon_1
=10\mthree$. The gravitational waves are still negligible, but
$n=1.14$, significantly {\em bigger} than 1.

The low end of the permitted mass range,
$m=10^2$ to $10^4\GeV$, is particularly
interesting because it corresponds to the Higgs field(s) of the standard
model. Moreover, the other mass scale $M$ is very roughly of order
$ m m_{Pl}/(8\pi)\half$, which is the `intermediate mass scale'
invoked in supergravity models (Nilles 1984).

So far we have considered a single scalar field $\psi$, possessing the discrete
symmetry $\psi\to-\psi$. When inflation ends, domain walls will form along the
surfaces in space where $\psi$ is exactly zero, so to make the model
cosmologically viable one would have to get rid of the walls by slightly
breaking the symmetry. However, it is clear that one can easily consider more
fields, with continuous global or gauge symmetries. In particular, if $\psi$
is complex one can use the same potential with the replacement
$\psi^2\to|\psi|^2$. This leads to exactly the same inflationary model, but
now global strings are formed instead of domain walls. A particular
realisation of this case might be to identify $\psi$ with the Peccei-Quinn
field.

Although the two-scale model seems to be quite attractive, one should bear in
mind that unlike the other models listed it has not yet been the subject of
detailed public scrutiny.

\subsubsection*{Double field inflation}

The double field inflation model (Adams \& Freese 1991; Linde 1991a) is very
similar to the two-scale model just discussed, whereby one scalar field (say
$\phi$) is utilised to trap a second (say $\psi$) in a metastable state for a
finite amount of time, terminated due to some physical effect of the evolution
of $\phi$. The difference is that in this model the transition is to be a
first-order one proceeding through bubble nucleation, so the role of the
second scalar field is to alter the nucleation probability without allowing
the barrier to disappear entirely. As the nucleation probability is typically
exponentially sensitive to the details of the barrier, it is certainly
possible for this to be achieved for very modest evolution of $\phi$.

This model is interesting in that it does appear to permit inflation to end
via a first-order phase transition, which variants on the extended inflation
paradigm seem unable to do (Liddle \& Lyth 1993) under current constraints
(working extended models all rely on ending inflation by some dynamical means
other than bubble nucleation, or by erasing the effects of bubbles by a
subsequent inflationary epoch). This is because of the considerable
flexibility in modifying the nucleation rate permitting one to avoid the
danger of any excessively large bubbles. However, in practice, the
interactions of the two fields must be exceptionally finely balanced (Adams \&
Freese 1991), and so these models should be regarded as rather unnatural.

\subsubsection*{Intermediate Inflation}

Intermediate inflation (Barrow 1990; Barrow \& Saich 1990) is a model in which
the scale factor expands according to
\begin{equation}
a = \exp \left( At^f \right)
\end{equation}
where $A$ and $f$ are constants with $0<f<1$. It arises as an exact solution
for the rather complicated potential
\begin{equation}
V(\phi) = \frac{8A^2}{(\beta+4)^2} \left[ \frac{\phi}{(2A\beta)^{1/2}}
	\right]^{-\beta} \, \left[ 6 - \frac{\beta^2}{\phi^2} \right]
\end{equation}
with $\beta = 4(f^{-1}-1)$, and rather more interestingly as a slow-roll
solution for a potential falling off asymptotically as a power-law $V(\phi)
\propto \phi^{-\beta}$.

Although rather contrived, this potential has some interesting properties
differing from those we have already seen (Barrow \& Liddle 1993). Most
interestingly, in the case $f=2/3$ the density perturbation spectrum is (in
the slow-roll approximation) exactly the Harrison-Zel'dovich case. Indeed,
because this is brought about by trading off the $\epsilon$ and $\eta$
slow-roll parameters, there is no automatic requirement that the gravitational
wave production be small {\em despite} the flatness of the spectrum. The
second interesting aspect is that this potential provides a second example of
$n>1$, which arises with $2/3 < f < 1$.

In practice, a substantial gravitational wave contribution (or $n$
significantly greater than one) does prove hard to achieve (Barrow \& Liddle
1993). The drawback of the model is that (as with pure power-law inflation)
there is no natural exit from the model --- one needs a mechanism to break the
power-law form of the potential. Indeed, in this model the slow-roll
parameters {\em decrease} as the field rolls down the potential, giving
smaller and smaller deviations from flatness with negligible gravitational
waves as it proceeds. Only by fine-tuning the timing of the mechanism which
ends inflation can one arrange for presently observable scales to leave the
horizon when the field is placed so as to give large deviations from the
standard predictions.

\section{Large Scales: Normalisation from the COBE data}
\label{COBE}
\setcounter{equation}{0}
\renewcommand\theequation{\thesection.\arabic{equation}}

For the rest of this report we are mostly concerned with the comparison
between theory and observation. We want to know what observation can tell us
about the spectrum $\delta_H$, on various scales, and start here with the
largest scales which are explored by the cmb anisotropy. We remind the reader
that we are expressing our spectral normalisations via $\sigma_8$, using the
CDM spectrum to translate between scales.

\subsection{Calculating the observed quantities}

As we write, the only positive detection of anisotropy is the COBE result,
announced in April 1992 (Smoot {\it et al} 1992).\footnote{As we were revising
for the final version, a Princeton--MIT balloon experiment operating on
similar angular scales produced a positive result with amplitude similar to
the COBE $1$-sigma lower limit.}. However, there are many interesting
microwave results expected in the near future. Before focussing on the COBE
data, we give here some generally applicable formalism, following closely
Scaramella and Vittorio (1990). [See also Abbott \& Wise 1984a;
Bond \& Efstathiou 1987; Scaramella \& Vittorio 1988; Efstathiou 1990, 1991.]

By analogy with \eq{corr}, one can define a temperature correlation
function,
\be C(\alpha)=\left\langle \frac{\Delta T(\bfe)}{T}
\frac{\Delta T(\bfe^\prime )}{T} \right\rangle \ee
Here $\bfe$ and $\bfe^\prime $ specify the directions in which the
anisotropy is observed, and the
average goes over directions separated by
an angle $\alpha$. It is given in terms of the multipoles by
\be C(\alpha)=\sum_{l=2}^\infty Q_l^2 P_l(\cos\alpha) \ee
where
\begin{equation}
Q_l^2 \equiv \frac{1}{4\pi} \, \sum_{m=-l}^{+l} |a_l^m|^2
\label{ql2} \end{equation}
Note that our $Q_l^2$ has a different normalisation to that of Scaramella and
Vittorio (1990).

We introduced earlier the notion of a filtered density contrast $\delta(R_f,
{\bfx})$, obtained by smearing over a region whose radius is of order
$R_f$, or equivalently dropping Fourier modes with $k\gsim R_f\mone$. The same
notion is useful for the microwave background anisotropy, so that one can
smear $\Delta(T)/T$ over a region of angular size $\theta_f$ ($\ll 1$
radian),
which is equivalent to dropping multipoles with $l\gsim \theta_f\mone$. For a
Gaussian filter with dispersion $\theta_f$ (equal to .425 times the full width
at half maximum) the precise statement is that each multipole is reduced
according to the formula
\be (a_l^m)^2\to \exp \left[ -((2l+1)\theta_f/2)^2\right] (a_l^m)^2\ee
Associated with the smeared quantity is a correlation function
\be C(\theta_f,\alpha)=\sum_{l=2}^\infty
\exp \left[ -((2l+1)\sigma/2)^2\right]  Q_l^2 P_l(\cos\alpha)
\label{corf} \ee

A given
experimental setup typically measures something which can be directly related
to $C$. The simplest one is a single beam whose resolution can be represented
by a Gaussian profile. Averaged over the sky the anisotropy measured by such a
beam is
\be \left\langle \left[ \frac{\Delta T(\theta_f,\bfe)}{T} \right]^2
\right\rangle=C(\theta_f,0) \label{single} \ee
For more complicated setups involving two or three beam switching,
still with Gaussian profiles, the measured anisotropy is
\be \left\langle \left[ \frac{\Delta T(\theta_f,\bfe)}{T}
-\frac{\Delta T(\theta_f,\bfe^\prime )}{T} \right]^2 \right\rangle
= 2\left[ C(\theta_f,0)-C(\theta_f,\alpha) \right]
\label{double} \ee
and
\bea
&&\left\langle \left[ \frac{\Delta T(\theta_f,\bfe)}{T} -\frac12
	\frac{\Delta T(\theta_f,\bfe^\prime )}{T} -\frac12 \frac{\Delta
	T(\theta_f,\bfe^{\prime \prime })}{T} \right] ^2\right\rangle
\nonumber\\
\eqa\frac32
	C(\theta_f,0)-2C(\theta_f,\alpha)+\frac12 C(\theta_f,2\alpha)
\label{triple} \eea
In the second expression,
$\bfe^\prime$ and $\bfe^{\prime \prime }$ lie on
opposite sides of $\bfe$, aligned on a great circle and each at an angular
distance $\alpha$. In a typical setup the `beam throw' $\alpha$ is of the same
order of magnitude as the `antenna resolution' $\theta_f$.

If we denote the left hand side of Eq.~(\ref{single}), (\ref{double}) or
(\ref{triple}) generically by $(\Delta T/T)^2$, the prediction in terms of
multipoles may be written
\be \rfrac{\Delta T}{T}^2 =\sum_{l=2}^\infty
2 F_l Q_l^2 \label{filter} \ee
where $F_l$ is a filter function (Bond \& Efstathiou 1987; Efstathiou 1991).
The filter function is normalised to $1/2$, rather than $1$,
in the limit of no filtering,
but we have retained this normalisation for compatability with earlier
work.

These expressions involving Gaussian beam profiles are accurate
representations of some experimental setups, including that of COBE,
but require modification in some other cases.\footnote
{We are indebted to George Efstathiou for a clarifying conversation on
this issue.} Accurate filter functions for some currently mounted
observations are shown in Figure 4, which is reproduced from the
paper of Crittenden {\em et al} (1993).

\subsubsection*{The cosmic variance}

The multipoles $a_l^m$ which appear in the above expressions depend on the
position of the observer. We saw in Section \ref{MWB} that the probability
distribution of each multipole as a function of the observer's position is
Gaussian with zero mean, and that cosmological perturbation theory can
predict the variance $\Sigma_l^2$ of this distribution, which is the expected
value of $|a_l^m|^2$. It thus predicts the expected values of the observed
anisotropies defined by \eqssss{ql2}{corf}{single}{double}{triple}.

In comparing theory with observation, it is important to appreciate that what
one can actually measure is only the temperature anisotropy at a single
observer point, our own. Thus in practice one only gets a single realisation
of the Gaussian probability distribution. This leads to statistical
uncertainties, as an individual realisation may not reflect the properties of
the ensemble averaged system. This effect, normally called the {\em cosmic
variance}, is significant for observables which depend only on a limited
number of the $a_l^m$, and in particular for the quadrupole $Q_2$ which only
depends on the five $a_2^m$ (two of which are redundant rotational
information). This is the fundamental reason why the quadrupole measurement is
not particularly useful for constraining theories. On the other hand, the
anisotropy on a scale of a few degrees depends on a significant number of the
$a_l^m$ and is far less susceptible to statistical vagaries.

\subsection{The COBE quadrupole anisotropy}

The COBE observations yield enough data to reconstruct the temperature
fluctuation field across the entire sky. We focus first on the quadrupole,
though we shall find that it is not particularly useful for constraining the
models, and we shall ultimately drop it in favour of the $10^0$ result.
Nevertheless, it is worth examining the predictions of the model to see why we
come to this conclusion.

The above equations readily allow us to calculate the mean of the predicted
distribution function for $Q^2_2$. It is given in Figure 5, again as a function
of spectral slope, for both types of inflationary model. Because the spectral
normalisation is uncertain by the bias factor, we have plotted the quantity
$\langle Q_2^2 /\sigma_8^2\rangle$. The COBE result of $2.5 \times 10^{-11}$
corresponds almost exactly to the mean prediction for the flat spectrum when
the bias is one, whereas at $n = 0.6$ the predicted mean is nearly 12 times the
COBE result even for the natural inflation scenario, seemingly requiring a bias
of over three. However, that conclusion neglects the statistical nature of the
quadrupole prediction.

As discussed above, theory does not predict a unique values for the
quadrupoles $Q_l$, but rather a probability distribution. Each quadrupole is a
sum \eq{ql2} of the square of $2l+1$ quantities with Gaussian probability
distributions, and its probability distribution is called a $\chi^2_{(2l+1)}$
distribution. For a given observer, the prediction is for a given realisation
from that distribution. Let us discuss the quadrupole predictions in terms of
a quantity $q \equiv 10^{10} \, b^2 \, Q_l^2$. Then the mean value $\bar{q} =
10^{10} \, b^2 \, \langle Q_2^2 \rangle$, and the $\chi_5^2(q)$ {\em pdf} with
that mean is
\begin{equation}
\chi_5^2 (q) = \frac{1}{3 \sqrt{2\pi} \left(\bar{q}/5 \right)^{5/2}} \,
	q^{3/2} \exp \left( - \frac{5q}{2\bar{q}} \right)
\end{equation}
This distribution has a very broad spread (the variance being $2/5$ of the
mean squared), and thus realisations of it may differ considerably from the
mean. The spread is indicated in Figure 5 by the vertical bars through the
mean predictions.

Let us for the time being assume the COBE measurement to be perfect ({\em ie}
not subject to observational errors). Then one can only exclude a given theory
on the basis of some exclusion level, where the experimental result is far
along the tail of the distribution. A conventional choice would be that 95\% of
the distribution should predict values above (or below) the experimental
measurement. Thus one can exclude at 95\% confidence only those theories which
predict a mean sufficiently high that 95\% of the distribution is above the
COBE value of $q_{\rm exp} = 0.25 \, b^2$. This one can readily calculate via
error functions, to discover that one can only say with 95\% confidence that
$\bar{q} < 1.1 b^2$, corresponding to $\sqrt{\langle Q_2^2 \rangle} < 1.05
\times 10^{-5} b$.

One can loosen this constraint even further by incorporating the fact that the
COBE results possess experimental errors. One can model the COBE data with a
probability distribution function describing the expected values were the
experiment to be repeated --- a sensible choice might be to assume $Q_2^{{\rm
exp}}$ to be gaussian distributed with mean $5 \times 10^{-6}$ and width $1.5
\times 10^{-6}$ --- and use this to construct a {\em pdf}~~$p_{\rm exp}
(q_{\exp})$ for the experimental measurement $q_{\rm exp}$. One then tests
each member of the experimental distribution for exclusion and takes a
weighted mean, thus constructing a rejection functional on the predicted
theoretical {\em pdf}s as
\begin{equation}
{\cal R} = \int_0^{\infty} p_{\rm exp} (q_{\rm exp}) \left| 1-2 \int_0^{q_{\rm
	exp}} \chi_5^2(q) dq \right| dq_{\rm exp}
\end{equation}
where $|$ signals the modulus. Defined on one side of the probability
distribution like this, a value ${\cal R} > 0.9$ signals a 95\% rejection of
the theoretical {\em pdf} in the light of the modelled experimental data.
With the experimental modelling as suggested above, predicted mean values of
$\bar{q}$ up to $2.3b^2$ are allowed, corresponding to 95\% confidence that
$\sqrt{\langle Q_2^2 \rangle} < 1.5 \times 10^{-5} b$.

Hence one sees that the statistical uncertainties in the theoretical
quadrupole prediction make it of little use in constraining theories. For
example, with this last constraint $n = 0.6$ is allowed with a mild bias of
1.15.

\subsection{The variance at $10^0$}

When one carries out an experiment based on a particular beam configuration,
the experimental configuration typically involves averaging over fluctuations
on scales below the beam resolution. This cuts off the contribution from
higher multipoles. From \eq{filter}, the  expected value of the anisotropy seen
in any particular experiment is
\begin{equation}
\left\langle \left( \frac{\Delta T}{T} \right)^2\right\rangle = \frac{1}{2\pi}
	\sum_l (2l+1) \Sigma_l^2 F_l \label{tenvar}
\end{equation}

For ground-based experiments which typically feature two and three beam
configurations, this filter function can be rather complex, as
we saw earlier. For a single beam experiment like COBE it is much
simpler.  The original COBE beam is well approximated by a
gaussian with Full Width Half Maximum of $7^0$. However, before calculating
the variance they smooth again by convolving with a further $7^0$ FWHM
gaussian, a procedure equivalent to an original smoothing by a $10^0$ FWHM
gaussian. Such a gaussian has a variance $\sigma^2 = (4.25^0)^2$. Thus the
appropriate form of the filter function is
\begin{equation}
F_l = \frac{1}{2} \exp \left( - \left(4.25 \pi (l+1/2)/180 \right)^2 \right)
\end{equation}

It is now trivial to calculate the predicted $10^0$ variance as a function of
$n$, and in Figure 6 we plot its square root, henceforth denoted $\left.
\frac{\Delta T}{T}\right|_{10^0}$, multiplied by the bias, for both power-law
and natural inflation. In the latter case, one sees that the results are
essentially exactly linear. We have been unable to show why this should be
analytically. The appropriate fitting function for natural inflation, which
has no gravitational waves, is
\begin{equation}
\left. \frac{\Delta T}{T} \right|_{10^0}(n) = \exp \left(2.62 (1-n) \right)
	\; \left. \frac{\Delta T}{T} \right|_{10^0} (n=1)
\end{equation}
while that for power-law inflation which does have gravitational waves is
\begin{equation}
\left. \frac{\Delta T}{T} \right|_{10^0} (n) = \sqrt{\frac{15-13n}{3-n}} \exp
	\left(2.62 (1-n) \right) \; \left. \frac{\Delta T}{T} \right|_{10^0}
	(n=1)
\end{equation}
Remember that we are assuming the same normalisation at $8h^{-1}$ Mpc. If
instead one makes the same normalisation to the COBE $10^0$ result, then the
$n$-dependent factor is transferred to the normalisation $\sigma_8$.

One can also readily calculate the cosmic variance, giving the spread in
values about these means that would be measured by differently positioned
observers. The variance of the $Q_l^2$ is $2\langle Q_l^2 \rangle^2/(2l+1)$.
For the $10^0$ result, this is 10\% at $n=1$, rising to 12\% at n=0.6 (this
result remains true whether or not there is a gravitational wave contribution
to the anisotropies). This is a negligible correction to the larger COBE error
in the present observations, but is ultimately a limiting obstacle to the
conclusions one can draw on this large angular scale.

Readers may find useful a relation between the $10^0$ variance and the
expectation of the quad\-rupole as $n$ is varied. For a power-law spectrum one
has
\begin{equation}
\frac{\Sigma_l^2}{\Sigma_2^2} = \frac{\Gamma(l+(n-1)/2)}{\Gamma(l+(5-n)/2)} \,
	\frac{\Gamma((9-n)/2)}{\Gamma((3+n)/2)}
\end{equation}
For the cases we examine, this is also true of the gravitational wave
contribution. The factor $\Sigma_2^2$ can now be pulled out of
Eq.~(\ref{tenvar}) to give an analytic expression for the ratio as a function
of $n$ alone, but in a rather awkward form. For the COBE result, a reasonable
approximation in the range $n \in [0.5,1]$ is just a linear fit
\begin{equation}
\frac{\left. \frac{\Delta T}{T} \right|_{10^0}}{\sqrt{Q_2^2}} =
	2.03 + 0.68 \, (n-1)
\end{equation}

\subsection{Microwave background constraints}

In establishing constraints on the normalisation $\sigma_8=1/b_8$
at fixed spectral index $n$, it is the $10^0$ data
which are of primary interest. Some particular values worthy of note are that
for the flat spectrum of $\left. \frac{\Delta T}{T}\right|_{10^0} = 1.05
\times 10^{-5}1/b_8$, and that for $n=0.6$ of $\left. \frac{\Delta T}{T}
\right|_{10^0} = 5.17 \times 10^{-5}1/b_8$ for power-law
inflation and $\left.
\frac{\Delta T}{T} \right|_{10^0} = 2.99 \times 10^{-5}1/b_8$ for natural
inflation. These of course are to be compared with the COBE observations of
$\left. \frac{\Delta T}{T}\right|_{10^0} = (1.1 \pm 0.2) \times 10^{-5}$.

The mean quadrupole prediction is also very similar to the flat spectrum
prediction with bias one, but the statistical uncertainties make this
comparison rather less relevant. It is worth recalling that an extrapolation
of the quadrupole from the COBE data at smaller angles, assuming a power-law
spectrum, gives a somewhat larger prediction for the {\em mean} quadrupole of
$16 \pm 4 \mu K$, corresponding to $(6 \pm 1.5) \times 10^{-6}$ (Smoot {\it
et al} 1992). Note that this extrapolation estimates the mean quadrupole, not
a specific realisation. Hence this again favours values of the bias not much
exceeding $1$, unless one inserts large-scale power to boost the theoretical
prediction for the mean.

Lacking full access to the COBE data, one must make an operational choice as
to what to take as the observational limits. The COBE team provide fits of
their data to power-law spectra, where both the slope and amplitude are
treated as fitting parameters. The two pertinent pieces of information are
that the observed (root of the) variance at $10^0$ is $(1.1 \pm 0.2) \times
10^{-5}$, and the allowed slopes in the fit carried out by the COBE team are
$n = 1.1 \pm 0.5$, where both errors are $1$-sigma. These are of course not
independent, as $10^0$ is at the lower end of the fitting range. Nevertheless,
if one takes the upper $2$-sigma value of the extrapolated quadrupole ($24
\mu$ K), then this actually gives a prediction for the $10^0$ variance which
is well above the $2$-sigma limit on the $10^0$ data even for the central $n$
value. This conclusion (also implicit in Efstathiou, Bond and White (1992)),
argues that the fits are not rigid enough to be of much use in constraining
theories. Because of the cosmic variance and also experimental uncertainties,
the quadrupole itself also appears of little use in this type of analysis.
Consequently, we choose to adopt simply the $10^0$ result as it stands,
without further incorporation of the COBE data. We relax their error bars to
$2$-sigma, and for many purposes we are only interested in the upper limit
thus given of $1.5 \times 10^{-5}$. This is a particularly useful way of
utilising the results, as it seems likely that this number can only go
downwards if it is to avoid conflict with other experiments and so any limits
quoted on its basis are likely only to become stronger with improved
observations.

The full limits on the bias as a function of $n$ from this criterion, for
both styles of inflationary model, are plotted in Figure 15 in
Section \ref{MDM} along with other constraints derived in later sections.
Some sample results are that for $n=1$ we have the obvious $.7< b_8 < 1.5$,
while for $n=0.6$ we require $3.4 < b_8 < 7.4$ (power-law inflation) and $2.0
< b_8 < 4.3$ (natural inflation). The following section will assess the
validity of different values for the bias parameter. Let us finally recall the
uncertainties of normalisation. The use of the Bardeen
{\it et al} (1986) transfer function rather than the one we use would reduce
the required bias by a further 10\%, and a higher baryonic content raise it by
a similar amount.

\subsection{Prospects for the near future}

It can be expected that COBE will release improved
results once the data from later years has been analysed. The ground based
Tenerife experiment (Watson {\it et al} 1992) operating on similarly large
scales should also soon be in a position to confirm the COBE detection. Much
interest has been focussed on the South Pole experiment operating on smaller
angular scales of around a degree, which has recently announced very strong
upper limits (Gaier {\it et al} 1992). This is particularly interesting as
those scales are below the horizon size at decoupling and so a gravitational
contribution to the anisotropies will be suppressed. Indeed, with a flat
spectrum the South Pole upper limit is alarming low, and may indeed indicate
that the COBE result has a significant gravitational wave contribution.
However, the situation remains to be clarified. In particular, one would like
to know whether a tilted spectrum without gravitational waves ({\it a la}
natural inflation) is compatible with COBE and the South Pole, or whether the
gravitational waves are necessary. Further, the COBE result could be towards
the $2$-sigma minimum of its range, in which case the conflict may disappear.
While the data remain controversial, we do not feel that this paper is an
appropriate place to investigate these issues.

\section{Medium scales: Galaxy Clustering and Bulk Velocities}
\label{CLUSTER}
\setcounter{equation}{0}
\renewcommand\theequation{\thesection.\arabic{equation}}

Now we consider scales which are big enough that the filtered density contrast
is still evolving linearly, yet small enough that there exist observations of
the galaxies and their motion with which one might hope to compare the linear
theory. Thus we are discussing scales from around $10h^{-1}$ Mpc up to perhaps
$100h^{-1}$ Mpc.

The cleanest observable to interpret theoretically is the galaxy peculiar
velocity field smeared on scales $\gsim 10\Mpc$, which is termed the bulk
flow. It is generally assumed that the galaxy bulk flow will be the same as
that of the underlying matter, which can be calculated from linear theory as
discussed in Section \ref{DENSPER}.8. The other quantities which might hope to
use are the number density perturbations  $\delta N/N$ of specific classes of
object, notably optical galaxies, infrared galaxies and galaxy clusters. On
the assumption that they are related to the underlying density perturbation
$\delta \rho$ by a scale-independent bias factor,
\be \frac{\delta N\sub{obj}}{N\sub{obj}}=b\sub{obj} \frac{\delta\rho}{\rho}
\label{bobj} \ee
they can be compared with the linear theory on scales
$r\gsim 10\Mpc$. Otherwise, their calculation takes us into the
non-linear regime discussed in Section \ref{NONLIN}, even on scales $r\gsim
10\Mpc$.

The peculiar velocity field, if it is available, determines both the
normalisation $\sigma_8$ and the spectral index $n$ of the density
perturbation. The number density perturbations, on the other hand, give
information only about $n$ even if the bias is indeed scale-independent. We
discuss these two types of information now, in reverse order.

\subsection{The galaxy correlation function}

There is ever increasing evidence that the amount of large-scale galaxy
clustering is greater than can be accommodated in the standard CDM cosmogony.
The most striking piece of evidence is provided by the measurement of the
galaxy angular correlation function $w(\theta)$ in the APM survey (Maddox {\it
et al} 1990, 1991), based on a sample of over two million galaxies. The
angular correlation function measures the clustering pattern as seen in
projection on the sky, and thus does not require redshift data, enabling such
large catalogues to be obtained by automated measurements on photographic
plates. With large redshift samples becoming available the further evidence
for excess large scale clustering is mounting. Conclusions in broad agreement
with APM are provided by the `counts in cells' of the QDOT survey (Saunders
{\it et al} 1991), and more recently by  surveys of redshifts of APM galaxies
(Dalton {\it et al} 1992; Loveday {\it et al} 1993),
 and in the power spectrum inferred from the CfA
survey (Vogeley {\it et al} 1992), that inferred from the Southern Sky
Redshift Survey (Park, Gott \& da Costa 1992) and that from the 1.2 Jansky
survey (Fisher {\it et al} 1993).

If the bias parameter is indeed scale-independent, all of these data indicate
that the CDM, $n=1$ spectrum normalised at $k\mone\simeq10\Mpc$ lacks power on
larger scales. To quantify the amount of extra power required, Wright {\it et
al} (1992) introduced a quantity which they called the excess power, which is
a functional of the power spectrum defined as
\begin{equation}
E[{\cal P_{\delta}}]= 3.4 \, \frac{\sigma(25h^{-1} {\rm Mpc})}{\sigma(8h^{-1}
	{\rm Mpc})}
\end{equation}
They chose the normalisation so that with their transfer function $E=1$ for
the CDM $n=1$ model; with our transfer function, $E = 0.95$ for that model.
Another way of defining the required extra power is to give the value of $n$
which would produce it (Liddle, Lyth \& Sutherland 1992). The connection
between these two specifications is
\begin{equation}
E[n] = 1.44 - n/2 \quad ; \quad 0.3 \leq n \leq 1
\end{equation}

To estimate the required value of $E$ or $n$, let us analyse the APM data
following Bond (1989) and Liddle, Lyth and Sutherland (1992).
On the large angular scales
where the discrepancy arises, the angular correlation function $w(\theta)$ can
readily be calculated in linear theory. It is given (Peebles 1980) by Limber's
equation as an integral over the galaxy correlation function $\xi_{gg}(r) =
b_g^2 \xi(r)$, where the mass correlation function $\xi(r)$ is defined
by \eq{corr}. A linear calculation is expected to be reliable on
scales above about $2^0$. The results appear in Figure 7, reproduced from
Liddle, Lyth and Sutherland (1992). The flat spectrum $n=1$ falls well below
the observational data, but one can see that as the spectrum is tilted, the
extra power does indeed make itself evident in the clustering statistics. It
is suspected that the APM survey may contain small residual systematics which
bias the observational estimates upwards (Will Sutherland, private
communication; Fong, Hale-Sutton \& Shanks 1992), so it seems reasonable to
regard values of $n$ from $0.3$ to $0.6$ as good fits to the excess clustering
data. The corresponding range for $E-1$ is $0.29$ to $0.14$. We postpone for
the moment the question whether the CDM model becomes viable in all respects,
with a value of $n$ in this range.

A different way of specifying the required power, used for instance by
Efstathiou, Bond and White (1992), is to specify the modification of the
transfer function which would generate it, through an alteration of the
parameter $\Gamma\equiv\Omega h$ appearing in, for instance,
\eqs{tran}{trana}. The equivalent value of $n$ is
\begin{equation}
\Gamma = \frac{1}{2} \left( \frac{1.88}{2.88-n}\right)^{10/3}
\end{equation}
The range of $\Gamma$ these authors consider a reasonable fit to the APM data
is $0.15 < \Gamma < 0.30$, corresponding to $0.15 < n < 0.67$, which we see is
actually rather looser than the range we took above. We postpone until Section
\ref{MDM} the question of whether this modification actually corresponds to
altering the physical value of the present dark matter density $\Omega$ (with
$h$ fixed at say $h=.5$), and whether such a change would lead to a viable
model.

All of this analysis assumes that linear evolution is valid on scales $k\mone
\gsim10\Mpc$. The assumption has been questioned by Couchman \& Carlberg
(1992), who claim that a highly evolved model with $\sigma_8=1$ as suggested
by the COBE data (though it is important to note their analysis preceeded
COBE) can explain the APM data. This apparently occurs via a combination of
non-linear effects altering the shape of the spectrum and the choice of a
rather unusual method of identifying the galaxies in their simulations. Their
analysis, which also gives (indeed requires) a significant velocity bias, has
been questioned by subsequent authors, and is not generally accepted at
present.

Although the galaxy clustering data is usually taken to indicate the need for
extra large scale power, the alternative explanation of a scale dependent bias
factor is sometimes mooted. A particularly elegant proposal in this direction
is to allow quasars to suppress local galaxy formation (Babul \& White 1991).
There is also `cooperative galaxy formation' (Bower {\it et al} 1993), a
phenomenological (at present) model in which galaxy formation is favoured in
the neighbourhood of other galaxies.

\subsection{The bulk flow}

There are many controversial aspects to the measurements of the bulk flow, but
nonetheless they provide a useful measure of the absolute magnitude of the
power spectrum on intermediate scales (for a recent review, see Kashlinsky \&
Jones 1991). It is well known (Kolb \& Turner 1990) that typical theories,
including standard hot and biased cold dark matter models and also models
seeded by topological defects, tend to predict bulk velocities rather lower
than those observed, particularly for standard CDM with high bias. Here we try
to quantify this effect for cold dark matter, bearing in mind the
uncertainties of theory and observation.

Just as one can filter the density contrast, one can also filter the peculiar
velocity field to obtain a bulk flow quantity, on any desired scale. As long
the density contrast on that scale is in the linear regime, the bulk flow can
be constructed from the linearly evolved Fourier components $\bfv\sk $, given
by \eq{159}, and compared with observation. There are several different ways
in which this comparison can be carried out.

For our conclusions, the most important is the one which we consider first ---
an analysis of the correlation of the bulk flow with the QDOT density field.
This provides strong constraints on the bias parameter for galaxies in the
IRAS survey, and combined with other results from the QDOT survey concerning
the clustering of IRAS galaxies leads directly to strong limits on the density
field power spectrum. We also discuss the reconstruction of the full three
dimensional density field via POTENT, and the cosmic Mach number.

\subsubsection*{Correlating QDOT and peculiar velocities}

One certainly does not at present have a catalogue of galaxy peculiar
velocities as large as one would like. This is due to the difficulties of
requiring independent distance estimators in addition to redshifts. The most
commonly used catalogue at present is that compiled by Burstein (from
measurements by a variety of different researchers) containing around 1000
galaxies, unfortunately distributed very anisotropically across the sky.

This velocity catalogue has been used to estimate the density contrast on
intermediate scales by the QDOT collaboration (Kaiser {\it et al} 1991). The
QDOT survey (Saunders {\it et al} 1991) consists of a redshifting of galaxies
sparsely (1 in 6) sampled from the catalogue of infrared emitting galaxies
compiled by the IRAS satellite. This survey has the advantage of even and
nearly complete sky coverage, and allows the galaxy density field to be
constructed.

In order to obtain a value for the bias, this density field is correlated with
the velocities from the Burstein sample. The redshift survey supplies the
position of the galaxies in redshift space, which amounts to assuming that
they have zero peculiar velocity along the line of sight. One attempts to
invert this into the true galaxy distribution in real space. To do this, one
constructs a smoothed galaxy number density contrast field, and appeals to a
constant bias to generate the density field from it. This density field allows
one to calculate the expected galaxy peculiar velocities, and use this to
correct their position. With luck, this generates an improved galaxy
distribution, on which the whole process can be repeated iteratively. This
success of this iteration will depend on the accuracy of the choice of bias
parameter\footnote{And in general on $\Omega$, though the inversion here
depends only on the combination $\Omega^{0.6}/b$ and we are anyway
presupposing $\Omega=1$.}, as an incorrect choice will systematically
incorrectly estimate the density field and hence the peculiar velocities.
Thus the correlation with the observed peculiar velocities constrains the
bias.

In general, the bias of the infra-red selected (and thus typically young) IRAS
galaxies will not be the same as that of the optically selected galaxies
discussed thus far, and so we denote this bias by $b_I$. It is often stated
that IRAS galaxies are somewhat less clustered than their optical
counterparts; for example Saunders, Rowan-Robinson and Lawrence (1992) suggest
$b_I = (0.69 \pm 0.09)b_g$ at the $1$-sigma level. Kaiser {\it et al} (1991)
obtain from their analysis $b_I/\Omega^{0.6} = 1.16 \pm 0.21$. This provides
the first dynamical evidence which is manifestly consistent with $\Omega = 1$,
as we are assuming here.

There are other ways of constraining $b_I$. According to Taylor and
Rowan-Robinson (1992), there are presently three independent dynamical
estimates of $b_I$ (one being that described above), giving respectively $b_I
= 1.23 \pm 0.23$, $1.16 \pm 0.21$ and $1.2 \pm 0.1$ where all errors are
$1$-sigma. The first is obtained from the convergence of the velocity dipole,
the second as discussed above and the third from the covariance tensor of
reconstructed flow fields. One cannot statistically combine errors from the
same data set, so we adopt as a $1$-sigma result $b_I = 1.2 \pm 0.2$.

This estimate of $b_I$ can be turned into an estimate of $\sigma_8$
(Efstathiou, Bond \& White 1992), because the
QDOT survey supplies the dispersion of counts of QDOT galaxies in $30
h^{-1}$ Mpc cubes, given as $0.46 \pm 0.07$ (Saunders, Rowan-Robinson \&
Lawrence 1992) (we have doubled the quoted error bars). Because we have a
good estimate of the bias, we can immediately calculate the mass dispersion in
these cubes, and so determine the normalisation of the spectrum as defined by
$\sigma_8$. The result of this calculation is shown in Figure 11.

\subsubsection*{POTENT reconstruction of the velocity field}

Another simple comparison which can be made is to the mean peculiar velocity
of the galaxies in a sphere of radius $R$ around us. For $R\gsim 10\Mpc$ this
should be equal to the linearly evolved filtered peculiar velocity field
$\bfv(R,\bfx)$, with a top hat filter of radius $R$ and evaluated at our
position. The expected value of its modulus squared is therefore the variance
$\sigma^2_v(R)$ of the smeared velocity field, which from
\eqss{159}{DISP}{312} is given by
\be \sigma_v(R)=H_0^2\int^\infty_0 W(R,k) \frac{{{\cal P}}_\delta(k)}{k^2}
\frac{dk}{k} \label{sigv} \ee
Here, $W(R,k)$ is the top hat filter $\widetilde W/V_f$ defined
by \eq{322}
Because of the $k\mtwo$ factor, the influence of large scales filters down
significantly to small scales.

Using this approach it is easy to consider the effect of $n$. Figure 8
illustrates the scaling of the dispersion with $R_f$ for a top hat filter.
Smaller values of $n$ give significantly higher predictions, but the scaling
with $R_f$ is roughly the same.

To compare this result with observation we use the velocity field
reconstruction results from the POTENT method, pioneered by Bertschinger,
Dekel and collaborators (Bert\-schinger \& Dekel 1989; Dekel, Bertschinger \&
Faber 1990; Bertschinger {\it et al} 1990; Dekel 1991). This is a very
powerful technique, at present rather frustrated by the inadequacies of
available data, which allows one to take the radial component of the peculiar
velocity, which is all one can measure, and reconstruct the full three
dimensional velocity flow. The procedure is to assume that the velocity field
is given as the gradient of a scalar (in practice this will be identified as
the peculiar gravitational potential of Section \ref{DENSPER}.8). This
scalar can be calculated as a line-of-sight integral requiring only the radial
peculiar velocity component; the full velocity field is then obtained by
taking the gradient. The power and potential of this method has been vividly
demonstrated by tests on simulated data, but on real data problems of
sparseness and uneven sampling present awkward obstacles to a reconstruction
of the true velocity field around us. However, although the detailed form of
the reconstruction is open to question, it certainly provides useful measures
such as the bulk flow in spheres about us.

It is
not trivial to compare theory with observations, because one predicts only the
{\em rms} velocity averaged over the entire universe and the observed
distribution may not be a fair sample (phrased another way, the comparison is
subject to a large cosmic variance due to the limited amount of data
available; in the language of Section \ref{COBE}, one is dealing with a
$\chi^2_3$ distribution since $v^2=v_x^2+v_y^2+v_z^2$). In particular, the
possible contaminating effects of the great attractor (if backside infall can
be unambiguously identified (Mathewson, Ford \& Buchhorn 1992)), may distort
observations. In the Monte Carlo work analysing bulk flows of Tormen {\it et
al} (1992), special criteria for choosing the observer points are employed
{\em before} statistical comparisons are made.

For this statistic, it is particularly notable that the influence of large
scales filters down significantly to small scales, so an unexpectedly large
fluctuation (and the great attractor would certainly be one in these CDM
models) can easily move the whole observed data set to well above the mean, as
predicted from the spectrum, that would be observed in a fair sample. That is,
although one might think that there were many independent samples on the scale
at which one is looking, the dependence on large scales correlates the
supposedly separate samples leading to a cosmic variance well in excess of
that one might have expected.

The POTENT results are obtained via a two-stage smoothing (Dekel 1991). First,
the original data is smoothed with a gaussian of radius $12h^{-1}$ Mpc to
remove short-scale sampling difficulties, and then a top hat of radius $R_f$
is used to provide the published data. In Figure 9, we apply this two-stage
smoothing to our spectra, which reduces the predictions, especially at short
scales. These are compared with the POTENT data at different normalisations.
It is seen that a normalisation $\sigma_8$ bigger than $.7$ is preferred, in
keeping with the conclusion that we reached earlier. This conclusion holds
more or less independently of the shape of the spectrum, say in the range
$.7<n<1$, because the scale probed by the bulk flow data is not all that
bigger than the normalisation scale.

Briefly we note that an alternative comparison framework, using the Burstein
sample of peculiar velocities and making a direct comparison of power-law
models with observations, has been made by Tormen {\it et al} (1992). The
utilise Monte Carlo simulations, and consider 18 models combining all possible
combinations of
$b_8\equiv1/\sigma_8 = 1$, $1.5$, $2$; $n=1$, $0.5$, $0$; $\Omega = 0.4$, $1$.
Of particular interest is the $\Omega =1$, $n = 0.5$, $b_8 = 1.5$ model.
They employ a maximum likelyhood test, and find that this model is twice as
likely as the model with flat spectrum and bias $1.5$, and is four times more
likely than the flat spectrum with bias $1$. The analysis is however
complicated by their requirement that the observer points in their simulations
must be chosen to match the properties of the local group.

\subsubsection*{The cosmic Mach number}

The cosmic Mach number test was devised by Ostriker and Suto (1990) in an
attempt to probe the shape of the spectrum as opposed to its normalisation.
They calculated a cosmic Mach number ${{\cal M}}\sub{OS}(R)$ defined by
\be ({{\cal M}}\sub{OS}(R))^2=\frac{\langle|\bfv(R,\bfx)|^2\rangle }
{\langle|\bfv(\bfx)-
\bfv(R,\bfx)|^2\rangle }
\ee
where $\bfv(R,\bfx)$ is the velocity field smeared on a scale $R$ and
$\langle\rangle $
denotes the average over all positions $\bfx$. A large value of ${{\cal M}}(R)$
(greater than one) indicates a fairly coherent (cold) flow on the given scale,
while a small value indicates a fairly random (hot) flow.

In order to be (marginally) within the linear regime they smeared $\bfv$ on
a scale $R_s=10\Mpc$ before applying the definition, and then used it only for
$R\gg R_s$. This gives, in terms of the window function $W(R,k)$ used in
\eq{sigv},
\be ({{\cal M}}\sub{OS}
(R))^2=\frac{\int^\infty_0 W^2(R_s)W^2(R){{\cal P}}_\delta(k) dk/k^3}{\int
^\infty_0 W^2(R_s)[1-\int W^2(R)] {{\cal P}}_\delta(k) dk/k^3} \label{mach}\ee
(They also multiplied $W^2(R)$ in the square bracket by $(1+k^2R^2/9)$
to allow for the fact that the spherically symmetric part of the
peculiar velocity field around an observer will not be detected by local
observations, but for simplicity we drop this small correction.)
Clearly, ${{\cal M}}\sub{OS}$ depends only on the shape of the spectrum, not
on its normalisation.

For a Gaussian filter
\be ({{\cal M}}\sub{OS}(R))^2=\frac{(\sigma_v(R+R_s))^2}
{(\sigma_v(R_s))^2-(\sigma_v(R+R_s))^2}\ee
and for any filter one has in the limit $R_s\ll R$
\be ({{\cal M}}\sub{OS}(R))^2 \simeq
	\frac{(\sigma_v(R))^2}{(\sigma_v(R_s))^2-(\sigma_v(R))^2}\ee
Suto and Ostriker calculate
${{\cal M}}\sub{OS}$,
and find that for CDM $n=1$ model it is $\simeq1$ on a scale
$R=20\Mpc$ falling to $\simeq.2$ at $R=100\Mpc$. Increasing the power on large
scales increases ${{\cal M}}\sub{OS}$, but the effect is not very strong with a
reduction to $n=.5$ increasing ${{\cal M}}\sub{OS}$ by at most a factor $1.5$.

{}From the data that we presented and discussed in Figure 9 it is clear that
${{\cal M}}\sub{OS}$ is difficult to estimate from published observations,
especially when one remembers the cosmic variance. Such a direct comparison
was not attempted by Ostriker and Suto. Instead, they essentially considered
the observational value of a different quantity, which corresponds to a more
conventional definition of the Mach number
\be {{\cal M}}(R)=v(R)/\sigma \ee
where $v(R)$ is the mean, and $\sigma$ is the dispersion,
of the peculiar velocity in a sphere of radius $R$ around us.
{}From optical
galaxy data they estimated (see also Groth, Juszkiewicz and Ostriker (1989))
${{\cal M}}(16\Mpc)=2.2\pm.5$
and ${{\cal M}}(36\Mpc)=1.3\pm .4$. If one were prepared
to regard these as estimates of ${{\cal M}}\sub{OS}$, a lot of extra large
scale
power would be indicated, corresponding to a value $n\simeq0$. Subsequent work
(Suto \& Fugita 1990; Suto, Gouda \& Sugiyama 1990; Suto, Cen \& Ostriker
1992) confirmed this, and also claimed on the basis of statistical
analysis that ${{\cal M}}$
and ${{\cal M}}\sub{OS}$ can be identified within a factor
better than 2 with reasonable confidence.

Our conclusion is that the Mach number test seems to point in the same
direction as the galaxy clustering data, in that more rather than less
large-scale power is indicated, but that the latter data are much more
decisive at present.

\section{The non-linear regime}
\label{NONLIN}
\setcounter{equation}{0}
\renewcommand\theequation{\thesection.\arabic{equation}}

In the last section we considered the galaxy bulk flow, and the correlation
functions of galaxies and galaxy clusters, on scales $\gsim 10\Mpc$. The bulk
flow can be compared directly with linear theory, and so can the correlation
functions on the assumption that there is a scale independent bias factor.
Most observational quantities cannot be calculated using only linear theory. A
partial list of them is the following.
\begin{itemize}
\item The redshift of formation of a given class of objects, such as
optical galaxies, infrared galaxies or galaxy clusters.
\item The correlation function $\xi\sub{obj}(r)$ of such a class,
and therefore the bias factor $b\sub{obj}$ defined by \eq{bobj}
to the extent that it is a valid concept.
\item The number density $n(z,>M)$ of all gravitationally bound systems with
mass bigger than $M$, which exist at redshift $z$.
\item The mean virial velocity of a given class of objects.
\item The number density $n(z,>v)$ of
all gravitationally bound systems with virial
velocity bigger than $v$, which exist at redshift $z$.
\item The dispersion $\sigma_{||}(r)$ of the relative
line-of sight velocity of a pair of
galaxies separated by distance $r$.
\end{itemize}

Although none of these quantities can be calculated using only linear
cosmological perturbation theory, some of them can be estimated if the linear
theory is combined with a spherical model of gravitational collapse. This
quasi-linear approach can be used in different ways. At the crudest level its
use is relatively uncontroversial; it allows one to show that structure forms
in a bottom-up manner, with light objects forming first, and it allows one to
estimate the epoch of formation for objects of given mass. These two
applications form the subject of the Sections 8.1 and 8.2 below.

If one is prepared to take the model more seriously, it is possible to make
more definite statements, by identifying the high peaks of the density
contrast in the linear regime with the sites of structure formation.  This
quasi-linear approach has a long history, a very influential early paper being
that of Press and Schechter (1974). Then, in another influential paper Bardeen
{\it et al} (1986, henceforth BBKS) developed it in considerable detail, at
the same time applying it to the CDM model. It's defect, as BBKS and many
other authors have pointed out, is that estimates made from it depend
sensitively on at least two parameters (the threshold density contrast, and
the ratio of the filtering mass to the mass of the objects studied which is
usually set arbitrarily equal to 1) whose values are quite uncertain; or to be
more precise, are ill-defined except under very idealised assumptions.

Because of its widespread use, we explore the quasi-linear approach in detail
in Sections 8.3--8.8. The basic assumptions are explained, then the essential
results of Press-Schechter and BBKS described, and finally the sensitivity of
the results to the unknown parameters will be quantified. The conclusion will
be that the quasi-linear approach is of somewhat limited use, its main value
being to exclude very low normalisations of the spectrum on scales
$k\mone\lsim1\Mpc$.

Finally in Section 8.9 we go on to consider the results of numerical
simulations, which avoid using the quasi-linear approach or even the spherical
collapse model, by modelling in various ways the non-linear evolution of the
dark matter and baryons. Such simulations appear to be capable of giving
reasonably definite information, and indicate that the observations cannot be
fitted by the CDM model if it is normalised to fit the COBE data.

\subsection{Gravitational collapse}

The quasi-linear approach makes essential use of the filtered density contrast
$\delta(M,\bfx)$. Following the usual practice for theoretical calculations,
we use the Gaussian filter, defined by \eqs{gone}{gtwo}.
The dispersion of the filtered density contrast is thus given by
\be \sigma^2(M)=\int^\infty_0 {{\cal P}}_\delta(k)\exp(-R_f^2 k^2)\frac{dk}{k}
\label{gfdis} \ee

At this point we need to distinguish between the density contrast
of the cold dark matter, and that of the baryonic matter. After decoupling the
baryons fall into the potential wells created by the cold dark matter, and
achieve the same density contrast, on scales in excess of the Jeans mass of
order $10^6\msun$ (Peebles 1980, 1984a). On smaller scales their pressure
prevents the baryons from collapsing, so that their density remains
practically uniform. Thus, luminous galaxies have mass in excess of
$10^6\msun$ according to the cold dark matter model, which is in rough
agreement with observation. (Note though that, as discussed in for example the
review of Blumenthal {\it et al} (1984), mass scales of order $10^6\msun$ to
$10^8\msun$ are also distinguished by another consideration; they are the
smallest mass scales on which the baryons can efficiently radiate energy, to
become the rather tightly bound structures that are observed.) From now on we
identify the density contrasts of the cold dark matter and the baryons.

As long as $\sigma(M)\lsim1$,
the filtered density contrast evolves linearly
except in those rare regions of space where it exceeds $1$ in magnitude. In
the 50\% of these regions where it is positive, gravitational collapse takes
place. At least initially, the collapsing regions have mass $\gsim M$, because
the filtered density contrast does not `see' structure on smaller scales. In
the approach that we are considering, gravitational collapse is modelled by
taking the collapsing region to
be spherically symmetric. One then finds (Peebles
1980, Eqs.~19.50 and 19.53) that when a given mass shell stops expanding, the
mean density inside it is a factor $9\pi^2/16$ times bigger than the mean
density of the universe. If the matter inside the shell had evolved linearly
(density contrast $\propto (1+z)\mone$), its density contrast at that time
would have been $\delta=(3/5)(3\pi/4)^{2/3} =1.06$. After it has stopped
expanding, the  shell collapses. If spherical symmetry continues to hold, and
one neglects pressure forces, the shell has collapsed to a point by the time
that the age of the universe has increased by precisely a factor 2,
corresponding to a mean density contrast within the linearly evolved shell of
$\delta=(3/5)(3\pi/2)^{2/3}=1.69$. Numerical studies (eg. Peebles 1970)
indicate that by about this time, pressure forces will in fact have virialised
the random motion of the constituents of the object. After this initial
virialisation, the object can lose energy (dissipation) which further
increases its virial velocity.

The use of the spherical collapse model in the present context is somewhat
problematical. First, filtering on a mass scale $M$ will distort the profile
of a peak with mass of order $M$, making it lower and broader. Thus the
linearly evolved \it filtered \rm density contrast will not have precisely the
behaviour described above, even for a spherical peak. Second, a peak of the
filtered density will not be spherically symmetric, and the corresponding peak
of the unfiltered quantity will be even less so. Finally, any initial
departure from sphericity is amplified during the collapse (Peebles 1980).
Nevertheless, one might hope that two features of the spherical collapse model
translate into reality.
\begin{itemize}
\item The regions with mass $\gsim M$ which have
undergone gravitational collapse can be at least approximately identified with
the regions where the linearly evolved density contrast $\delta(M,\bfx)$
exceeds some threshold $\delta_c$ which is roughly of order 1.\footnote
{Note that throughout this section the subscript c denotes the threshold
value of $\delta$, whereas in earlier sections it was used to distinguish the
cold dark matter density contrast from the radiation density contrast.}
\item A collapsing region does not fragment into a large number of separate
objects, which means that the mass of the resulting gravitationally bound
systems is also $\gsim M$.
\end{itemize}

These two assumptions form the basis of the quasi-linear approach, and they
will be adopted for the rest of the discussion.

\subsection{The epoch of structure formation}

The quasi-linear approach that we are going to discuss is valid only in the
regime in which the mass density is evolving linearly, which ends when the
filtered linearly evolved dispersion $\sigma(z,M)$ of the density contrast is
of order 1. With the usual $n=1$ spectrum it is known that this is the epoch
of structure formation for objects of mass $M$, so that one has a bottom-up
scenario in which low mass objects form first. In this section we derive this
result, and ask how far it remains valid if $n$ is less than 1.

For definiteness, let us take the epoch when non-linearity sets in to be
the epoch when the linearly evolved quantity  $\sigma(z,M)$ is
precisely equal to 1. Denoting this epoch by
$z\sub{nl}$ one has
\be 1+z\sub{nl}(M)=\sigma_0(M) \ee
where as always the subscript $0$ denotes the present value of the {\em
linearly evolved } quantity. At about this epoch, the regions where
gravitational collapse is taking place on scales $\gsim M$ become common,
whereas before they were quite rare. If $\sigma(M)$ is increasing
significantly as the mass decreases, then at this same epoch the density
contrast filtered on mass scales substantially bigger than $M$ is still
evolving linearly in most parts of the universe. The result is a bottom-up
scenario, in which gravitationally bound systems of successively bigger mass
$M$ form at the successively later epochs $z\sub{nl}(M)$. After a system has
formed, it may become part of a bigger gravitationally bound system, remaining
a discrete object, but it may also merge with other systems.

Let us ask if $\sigma(M)$ does, in fact, increase significantly as $M$
decreases. For $M>10^6\msun$  it has already been plotted in Figure 3b both for
the standard choice $n=1$ and for some smaller values. One sees that it
increases quite fast in the regime $M\gsim 10\msun$, but flattens off a lower
masses and in fact turns over for $n$ significantly less than 1. This is a
consequence of the small-scale behaviour ${{\cal P}}_\delta(k)\propto T^2(k)
k^{3+n}\propto k^{n-1}$ of the spectrum (up to a logarithmic factor). That
behaviour is expected to persist down to some very small scale, called the
coherence scale, below which the spectrum cuts off sharply and $\sigma(M)$
becomes practically equal to the unfiltered quantity $\sigma$. The coherence
scale depends on the nature of the cold dark matter. If the cold dark matter
consists of subnuclear particles, the coherence scale is determined by their
interactions. For instance, if it consists of the lightest supersymmetric
particle the coherence scale is the Hubble scale at the epoch when the
particle becomes non-relativistic. If it consists of the axion, the coherence
scale is the Hubble scale at the epoch when the axion acquires mass through
QCD effects.

On the basis of these considerations, we arrive at the following picture of
structure formation, which covers the case $n<1$ as well as the familiar case
$n=1$. The first structure forms at some epoch given by
\be 1+z\sub{nl}=\sigma_0 \ee
where $\sigma_0$ is the unfiltered quantity.
For $n=1$, this epoch is very early
and depends on the nature of
the cold dark matter. Decreasing $n$ makes the epoch  later,
and to an increasingly good approximation
allows it to be calculated without knowing the nature of the
cold dark matter.
After this epoch, structure forms according
to the bottom-up picture.

Let us quantify the
extent to which the bottom-up behaviour is satisfied,
by estimating the the maximum mass $M\sub{max}$ of structures forming
at the epoch $z\sub{nl}(M)$; in other words, by estimating the width of
the band $M$ to $M\sub{max}$ of the masses that are formed at that epoch.
It is the mass such that $\sigma_0(M\sub{max})$ is
significantly less than 1, so that the density contrast filtered
on the scale $M\sub{max}$ is still evolving linearly
almost everywhere
in the universe.
To be definite, let us require that it
is still evolving linearly in 90\%
of the volume of the universe. From the Gaussian distribution, this
corresponds to $\sigma_0(M\sub{max})= 1/1.63=.61$.
For $M=10^6\msun$ this criterion gives
$M\sub{max}=10^{9}\msun$ ($10^{10}\msun$) for $n=1$ ($.7$).
In the regime $M\sim10^6\msun$,
$z\sub{nl}(M)$ is a {\it slowly varying} function of $M$, so
the conclusion is that
a {\it broad band} of masses downwards of $10^{10}\msun$
or so
collapses at the the epoch $z=z\sub{nl}
(10^6\msun)$.
For $M=10^{12}\msun$, the criterion gives $M\sub{max}=
2\times 10^{13}\msun$, more or less independent of $n$.
In the regime $M\sim 10^{12}\msun$,
$z\sub{nl}(M)$ is a fairly {\it rapidly
varying} function, so the conclusion is that a fairly {\it narrow} band of
masses $M=10^{12}\msun$ to about $10^{13}\msun$ collapse
around the epoch $z=z\sub{nl}(10^{12}\msun)$.

We conclude that the bottom-up picture is more or less the same for
any $n$ in the range $.7<n<1$. A broad range of masses below about
$10^{10}\msun$ collapses at about the same time, but subsequent collapse
takes place in an increasingly narrow mass range.

Finally we estimate the actual value of the redshift of nonlinearity
$z\sub{nl}(M)$. It can be read from Figure 3b, for $n=1$ and $n=.7$ and one
sees
that it becomes more recent as $M$ or $n$ are reduced. Let us consider three
representative values of $M$. First take $M=10^6\msun$. Then $z\sub{nl}(M)$ is
quite early, $1+z\sub{nl}(M)=18\sigma_8$ ($9\sigma_8$)
for $n=1$ ($.7$). A broad range of
mass scales downwards of about $10^9 \msun$ ($10^{10}\msun$) collapse at
around this epoch. Second, take $M=10^{12}\msun$. Then $1+z\sub{nl}(M)=4.5
\sigma_8$
($3.7\sigma_8$)
for $n=1$ ($.7$). This is the epoch when a significant fraction of
the mass of the universe collapses into objects with mass of order
$10^{12}\msun$, the mass of large galaxies. However, the favoured explanation
of a bias factor $b_8>1$ is that \it luminous \rm galaxies originate from
exceptionally high peaks of the evolved density contrast, and therefore form
well before the epoch $z\sub{nl}(M)$. We shall discuss later to what
extent the quasi-linear approach is capable of estimating this epoch.
Finally consider
$M=10^{15}\msun$. Then, $1+z\sub{nl}(M)=.82\sigma_8$, almost independent of
$n$.
This means that a significant fraction of
the mass of the universe is only now beginning to
collapse into large galaxy clusters.
The prediction is therefore that the observed large clusters originated from
exceptionally high peaks of the density contrast, which again implies a bias
factor for these objects (Kaiser 1984). Again, we will discuss later
whether the quasi-linear approach can quantify this effect.

A different criterion is sometimes proposed for the epoch of structure
formation, namely that it is the epoch when the linearly evolved spectrum
${{\cal P}}_\delta(k)$
is equal to 1 on the corresponding scale $k\mone=R_f$. This
is practically equivalent to the
criterion $\sigma(M)=1$ if ${{\cal P}}_\delta(k)$
is significantly increasing, say like $k^m$ with $m\sim 1$, because then the
filtered spectrum is ${{\cal P}}_\delta(k)\exp(-k^2R_f^2)
\simeq{{\cal P}}_\delta(R_f\mone)\delta(\log(kR_f))$, where $\delta$ denotes
the
Dirac delta function. This condition
fails, in particular, when $k$ corresponds to the maximum of ${{\cal
P}}_\delta$
which occurs when $n<1$, so it is not true that in that case the first
structure forms at the epoch when $1+z$ is equal to the maximum value of
${{\cal P}}_\delta(k)$,
as was incorrectly stated in Liddle, Lyth and Sutherland
(1992).

\subsection{The statistics of the collapsed regions}

Now we study the stochastic properties of the collapsed regions, defined as
regions where the linearly evolved density contrast $\delta(M,\bfx)$ exceeds
a threshold $\delta_c$. Since they
represent exceptionally
large fluctuations of a Gaussian random field, there are powerful mathematical
results concerning their stochastic properties. We shall use some of them in
what follows, drawing extensively on the work of BBKS.

The correct choice of the
threshold $\delta_c$, insofar as it is well defined,
is a matter of debate. Many authors take the value $1.69$
inspired by the spherical collapse model described in Section
8.1. On the other hand, comparison
of the quasi-linear estimate of $n(>M)$ (described below) with estimates
from numerical simulations suggests a smaller value, Carlberg and Couchman
(1989) advocating $\delta_c=1.44$ and Efstathiou and Rees (1988) advocating
$\delta_c=1.33$ (but see also Brainerd and Villumsen (1992)
and Katz, Quinn and Gelb (1992)).

In the collapsed regions, $\delta(M,\bfx)$ is more than $\nu$
standard deviations above zero, where
\bea \nu(z,M)\eqa \delta_c/\sigma(z,M) \label{nufirst} \\
\eqa \delta_c (1+z)/\sigma_0(M) \label{nusec} \\
\eqa \delta_c\frac{1+z}{1+z\sub{nl}(M)} \label{nudef} \eea
We are working in the linear regime, corresponding to $\nu
>\delta_c>1$.

The collapsed regions occupy a volume fraction $V$ given by the Gaussian
distribution,
\be \frac{{\rm d}V}{{\rm d}\nu}=\frac1{\sqrt{2\pi}} e^{-\nu^2/2}
\label{dgauss} \ee
leading to
\bea V(\nu)
\eqa\rm{erfc}(\nu/\sqrt2)/2\\
\eqa (2\pi)\mhalf\nu\mone e^{-\nu^2/2}
(1-\nu\mtwo+O(\nu\mfour) ) \label{gauss} \eea
For $\nu=1,2,3,4$ the volume fraction is $V=.16,.023,.0013,.000031$. In
practice one is not interested in values $\nu\gsim4$, because the collapsed
regions are then too rare to be physically significant. The corresponding mass
fraction is about $(1+\delta_c)=2$ to $3$ times bigger than the volume
fraction.

To say more one needs to know the shape of the spectrum. We shall list the
relevant results given by BBKS. They involve only two moments of the spectrum,
defined by
\bea
\langle k^{2}(M)\rangle
\eqa \sigma \mtwo(M) \int^\infty_0 k^{2} \exp(-k^2 R_f^2) {{\cal P}}_\delta(k)
\frac{{\rm d}k}{k} \label{ksqm}
\\
\langle k^{4}(M)\rangle
\eqa \sigma \mtwo(M) \int^\infty_0 k^{4} \exp(-k^2 R_f^2) {{\cal P}}_\delta(k)
\frac{{\rm d}k}{k} \eea
The quantity $\langle k^2 \rangle$ is the mean
of the $\nabla^2$ operator, ie., of
the quantity $\delta\mone\nabla^2\delta$.
Similarly, $\langle k^4 \rangle$ is the mean
of $\nabla^4$.

A relevant length
scale is defined by $R_*^2=3\langle k^2 \rangle/\langle k^4 \rangle$. For any
spectral index $n>-1$, it is easy to show that in the limit
of small filtering scale $R_f$,
\be \frac{R_*}{R_f}=\rfrac6{1+n} \half \ee
For the case of CDM with $.7<n<1$, the ratio is
in the range $1$ to $3$ for the entire range of cosmologically interesting
masses.

Another relevant length scale
is $\langle k^2 \rangle\mhalf$. On large filtering scales,
such that
${{\cal P}}_\delta(k)$ is increasing fairly strongly at $k\mone\simeq R_f$,
the ratio $\langle k^2 \rangle\mhalf/R_f$ is close to 1.
As the scale is reduced it increases, but is $\lsim 10 $
for $M>10^6\msun$.

Finally, it is convenient to define the dimensionless parameter
\be \gamma(M)=\langle k^2 \rangle/\langle k^4 \rangle\half \label{gamm} \ee
It falls from about $.7$ to about $.3$ as $M$ decreases from $10^{15}\msun$ to
$10^6\msun$, for $.7<n<1$.

For sufficiently large $\nu$, each collapsed region is a sphere surrounding a
single peak of $\delta$. However, the departure from sphericity is
considerable in the cosmologically interesting regime. BBKS show that a
quantity $x\mone$, which is roughly the fractional departure from sphericity,
is well approximated by
\be
x=\gamma\nu+\theta(\gamma,\gamma \nu)
\ee
where
\be
\theta(\gamma,\gamma \nu)=\frac{3(1-\gamma^2)+(1.216-.9\gamma^4)
\exp[-\gamma/2(\gamma\nu/2)^2]}
{\left[3(1-\gamma^2)+.45+(\gamma\nu/2)^2\right]\half+\gamma\nu/2}
\ee
The asphericity is plotted in Figure 10 for $M=10^8\msun$ and $10^{14}\msun$,
for both $n=1$ and $n=.7$, and is seen to be $\gsim .3$ even at $\nu=4$
and $10^{14} \msun$. We emphasise that this is the asphericity seen in
the linearly evolved, filtered density contrast. The asphericity in
the true, unfiltered density contrast will be bigger, and will increase
during collapse.

Three useful number densities are given by BBKS. First, the density $n_\chi$
of the Euler number of the surfaces bounding the collapsed regions is
\be
\frac12 n_\chi(\nu,\langle k^2\rangle )=
\frac{(\langle k^2 \rangle/3)\threehalf}{(2\pi)^2}
(\nu^2-1) e^{-\nu^2/2} \label{nchi}
\ee
Second, the number density of upcrossing points on these surfaces is
\bea
&& n\sub{up}(\nu,\langle k^2\rangle ,\gamma)
=\frac{(\langle k^2 \rangle/3)\threehalf}{(2\pi)^2}\times\nonumber\\
&&\left[ \nu^2-1+\frac{4\sqrt3}{5\gamma^2(1-5\gamma^2/9)\half}
e^{-5\gamma^2\nu^2/18}\right] e^{-\nu^2/2} \label{nup}
\eea
An upcrossing point on a surface of constant $\delta$ is defined as one where
$\del \delta$ points along some arbitrarily chosen reference direction.

The third  number density is $n\sub{peak}$, the number density of peaks
which are more than $\nu$ standard deviations high. BBKS give
expressions for $n\sub{peak}$, but they point out also that in the
cosmologically interesting regime it is quite well approximated by
$n\sub{up}$. We shall use this approximation in what follows. It suggests that
if a collapsed region contains several peaks, they are not buried deep inside
it; rather, the boundary of the region is presumably corrugated, wrapping
itself partially around each peak.

In the limit $\nu\gamma\gg1$, $n\sub{peak}=n\sub{up}=\frac12 n_\chi$. This is
in accordance with the fact that in this regime, each surface is a sphere
surrounding a single peak. It contributes $+1$ to the number of upcrossing
points, and $+2$ to the Euler number. As $\nu$ decreases the surface becomes
deformed, but at first its contributions to $n\sub{up}$ and $n_\chi$ are not
affected. Eventually though, it may become so corrugated that it has more than
one upcrossing point, and may become a torus so that its Euler number is less
than $2$ (of course it then has more than one upcrossing point whatever its
shape). As a result, its contribution to $n_\chi/2$ becomes less than its
contribution to $n\sub{up}$. The ratio $2n\sub{peak}/n_\chi$
(approximated as
$2n\sub{up}/n_\chi$) is plotted in Figure 11.%
\footnote
{We are not interested in the
regime $\nu<1$, but for the record $n_\chi$ is negative there, indicating that
the surfaces $\delta(M,\bfx)=\nu\sigma$ definitely do not have spherical
topology; in fact they percolate leading to a sponge--like topology
(Melott 1990).}

One would like to know the number density $n\sub{coll}$ of the collapsed
regions, but in general there does not seem to be an analytic formula
for it. It satisfies the inequality $n_\chi/2<n\sub{coll}<n\sub{peak}$, and
so is equal to $n_\chi/2$ in the limit $\nu\gamma\gg1$. According to Figure
11, it is within a factor 2 of $n_\chi$ for $\nu>3$. The average number $N$ of
peaks per collapsed region satisfies $1< N < 2n\sub{peak}/n_\chi$, so
according to Figure 11 it is no bigger than 2 for $\nu\gsim 3$.

\subsection{The mass of the collapsed regions}

As the filtered density contrast does not contain structure on scales much
less than the filtering scale $R_f$, we expect the average
radius of a peak in it to
be $\gsim R_f$. It can be estimated from the number density of peaks with
arbitrary height $n \sub{peak}(-\infty)$, which is equal to $.016R_*\mthree$
where the length $R_*$ was defined at the beginning of the last subsection.
This is also the number density of troughs, so a rough estimate of the
average
radius of a peak or trough at half height is
$(2n
\sub{peak}(-\infty))\mthird /4=.8R_*$.
As we discussed earlier, $R_*/R_f\simeq1$ to 3
for all cases of interest, so the average peak size is roughly of order $R_f$.
Since there are few peaks with radius less than $R_f$, this suggests
that the probability distribution of peak sizes is
fairly narrow, most peaks having a radius around the average.

We would like to compare the average peak radius $R_f$ with the average
radius
of a collapsed region. Equivalently, we would like to compare
the filtering volume $V_f$ with the
average volume of a collapsed region.
The latter is equal to the volume
fraction occupied by the collapsed regions, \eq{gauss}, divided by their number
density $n\sub{coll}$. In general we do not know $n\sub{coll}$, but we
do know the peak number density $n\sub{peak}$, so we can calculated
the average volume
$V\sub{peak}$ \it per peak \rm of a collapsed region. It is given by
\eq{gauss}, and as a fraction of the filtering volume $V_f$ becomes
\be
V\sub{peak}/V_f
=\frac12\mbox{erfc} (\nu/\sqrt2)/
(n\sub{peak}V_f) \label{volume}
\ee
The corresponding mass fraction is roughly $M\sub{peak}
/M =(1+\delta_c)V\sub{peak}/V_f$. The
volume fraction is plotted in Figure 12, and one sees that
except for large $\nu$ it is bigger than 1.

At large $\nu$, the ratio may be calculated using
the approximation $n\sub{up}=n_\chi/2$ together with \eq{gauss},
\bea
V\sub{peak}/V_f
\simeqa (2\pi)\threehalf (\langle k^2 \rangle/3)\mthreehalf \nu\mthree/V_f \\
\eqa
\rfrac{3}{R_f^2 \langle k^2 \rangle}\threehalf \nu\mthree \label{vpea}
\eea
The fact that this ratio is small does not mean that the filtered density
contrast is seeing structure on a scale much less than the filtering
scale.%
\footnote
{We are indebted to Peter Thomas for a clarifying conversation about
this question.}
Rather, one is probably dealing with density enhancements with size is
of order $R_f$, but with mean overdensity only barely exceeding the
threshold. When one moves just a little distance away from the peak, the
mean overdensity within a sphere of radius $R_f$, represented by the
filtered density contrast, falls below the threshold.

\subsection{The number density $n(>M)$}

The main application of these results is to estimate the number density
$n(>M)$ of gravitationally bound systems with mass bigger than $M$, at a given
epoch before $z\sub{nl}(M)$. The systems
are supposed to be identifiable by looking at
the linearly evolved density contrast $\delta(M,\bfx)$. Each collapsed
region, defined as one in which $\delta(M,\bfx)>\delta_c$, is supposed to
contain one or more systems with mass bigger than $M$.

If each collapsed region is identified with a single system, then
$n(>M)=n\sub{coll}$. In general this recipe is useless for lack of an
expression for $n\sub{coll}$. A different prescription, which does lead to a
calculable expression, is to identify each peak within a collapsed region with
a different collapsed object,
\be n(>M)= n\sub{peak} \simeq n\sub{up} \label{peak} \ee
This estimate (usually without the simplifying second equality) is
widely used in the literature.
It is certainly
the same as the estimate $n(>M)=n\sub{coll}$ for large $\nu$, where we know
that there is just one peak per collapsed region. To what extent the
prescriptions are the same for lower $\nu$ is not known, because the number of
peaks per collapsed region is not known.

If at some epoch the linearly evolved density contrast does have many
peaks within a collapsed region, an interesting
situation arises. At a somewhat earlier epoch, $\delta(M,\bfx)$ was smaller,
and a separate contour $\delta(M,\bfx)=\delta_c$ was wrapped around each
peak.
In other words, each peak of the linearly evolved density contrast,
filtered on scale $M$, was inside a single collapsed region, and presumably
represented a separate gravitationally bound system. At the later epoch when
the collapsed region encompasses many peaks of the linearly evolved density
contrast, we have a bigger gravitationally bound system. If the original
systems
survive, the identification of each peak with a separate system is correct,
but it misses the larger system which contains the original systems. Of
course, missing this one system does not affect the total count much, so if
this case is typical of collapsed regions containing many peaks the estimate
$n(>M)=n\sub{peak}$ is better than the estimate $n(>M)=n\sub{coll}$. If, on
the other hand, the original systems have merged, that identification is
wrong, and the whole of the collapsed region should be identified with just
one
gravitationally bound system. If this case is typical, the estimate $n(>M)
=n\sub{coll}$ would be better, if only we had a formula for it. Which
case is the more likely? A clue is provided by the observation made earlier,
that if there are several peaks in a collapsed region they typically seem to
lie near the surface of the region, a part of the surface wrapping itself
around each peak. This picture would suggest that the estimate
$n(>M)=n\sub{peak}$ is the more reasonable, the peaks of the linearly evolved
density contrast in a typical collapsed region representing structures which
have not existed long enough to merge.

A different prescription was used by Press and Schechter (1974), to derive a
widely used alternative formula. They worked with the differential number
density,
\be \frac{{\rm d}n}{{\rm d}M}\equiv \frac{\rm d}{{\rm d}M} n(>M)
\label{dndm} \ee
At a given epoch, if the filtering mass $M$ is increased by an amount ${\rm
d}M$ then $\nu$ is increased by an amount ${\rm d}\nu$, and the volume
fraction occupied by the collapsed regions is reduced by an amount ${\rm d}V$
given by \eq{dgauss}. Press and Schechter suppose that the eliminated volume
consists of objects with mass between $M$ and $M+{\rm d}M$, corresponding to
the idealisation that filtering the density contrast on any mass scale $M$
cuts out precisely those objects with mass less than $M$ while leaving
unaffected objects with mass bigger than $M$. Ignoring the overdensity
$\simeq(1+\delta_c)$ of the collapsed regions this implies that the number
density ${\rm d}n$ of such objects is given by
\bea M\frac{{\rm d}n}{{\rm d}M}\eqa \left[M\frac{{\rm d}(R_f^2)}{{\rm d}M}
	\right] \frac{{\rm d}(\sigma^2(M))}{{\rm d}(R_f^2)} \frac{{\rm d}\nu}
	{{\rm d}(\sigma^2(M))} \frac{{\rm d}V}{{\rm d}\nu} \frac{{\rm d}n}
	{{\rm d}V} \\
\eqa \left[\frac{2R_f^2}{3} \right] \left[-\sigma^2(M)\langle k^2
\rangle\right]
	\left[-\frac{\nu}{2\sigma^2(M)} \right]
	\left[\frac1{\sqrt{2\pi}}e^{-\nu^2/2}\right]
	\left[\frac1{V_f}\right]\\
\eqa \frac{R_f^2 \langle k^2 \rangle}{3}\frac{1}{4\pi^2 R_f^3} \nu e^{-\nu^2/2}
	\label{psch} \eea

Press and Schechter multiplied this formula by a factor 2, so that when
integrated over all masses it would give the  total mass density of the
universe, rather than just the half corresponding to the regions of space
where the linearly evolved density contrast is positive. They thus arrived at
the estimate
\be n(>M)\simeq n\sub{ps}
\equiv\int_M^\infty
\frac{\langle k^2 \rangle^\prime }{6\pi^2 R_f^\prime }
\nu^\prime
e^{-\nu^{^\prime 2}/2} \frac{dM^\prime }{M^\prime } \label{nps} \ee
In this equation, $R_f^\prime =R_f(M^\prime )$, and similarly for
$\langle k^2\rangle ^\prime $ and $\nu^\prime $.
The factor 2 inserted by Press and Schechter is not justified by their
argument, because the linearly evolved density contrast has nothing to do with
reality in the non-linear regime $\sigma(M)>1$. On the other hand, the
neglected overdensity gives a factor $\simeq(1+\delta_c)=2$ to 3. Thus the
factor 2 goes in the right direction, and the Press-Schechter formula is
reasonably well founded theoretically. A somewhat different justification for
the formula has been given by Bond {\em et al} (1991b).

{}From \eq{nufirst},
\be \frac{\nu^\prime }{\nu}=\frac{\sigma(M)}{\sigma(M^\prime )} \ee
The right hand side is independent of $\sigma_8$, $\delta_c$ and $z$, so it
follows that $n\sub{ps}$, like $n\sub{peak}$, depends on these quantities
through $\nu $, which involves the combination $(\delta_c/\sigma_8)(1+z)$. In
Figure 13 is plotted the ratio of the two alternative estimates
$n(>M)=n\sub{ps}$ and $n(>M)=n\sub{peak}$, for $M=10^{10}\msun$ and for
$M=10^{15}\msun$. One sees that the estimates agree to better than a factor 2
for $\nu\lsim 2$. Presumably, this indicates that in this regime the
assumptions underlying the two estimates are compatible, in that increasing
$M$ by a small amount cuts out portions of the collapsed regions which have
mass of order $M$ and are centred on peaks with height of order $\nu(M)$.

For large $\nu$, the Press-Schechter estimate falls below $n\sub{peak}$. This
can be understood analytically, from the expression
\be \frac{{\rm d}n\sub{peak}}{{\rm d}M}
	\simeq \frac{{\rm d}n\sub{up}}{{\rm d}M}
	=\pdif{n\sub{up}}{\nu}
	\frac{{\rm d}\nu}{{\rm d}M} + \pdif{n\sub{up}}{\langle k^2 \rangle}
	\frac{{\rm d}\langle k^2 \rangle}{{\rm d}M} + \pdif{n\sub{up}}{\gamma}
	\frac{{\rm d}\gamma}{{\rm d}M}
\ee
The first term dominates for large $\nu$, leading to the ratio
\be \frac{{\rm d}n\sub{ps}/{\rm d}M}{{\rm d}n\sub{peak}/{\rm d}M}=2\rfrac{3}
	{R_f^2\langle k^2\rangle} \threehalf\nu\mthree
\label{ratio} \ee
Apart from the factor 2, this is just the filter volume divided by the average
volume of a collapsed region (\eq{vpea}).

As we saw earlier, the smallness of the size of the `collapsed regions' is an
artefact of the filtering. The conclusion is that for very rare fluctuations,
$n\sub{peak}$ is a better estimate that $n\sub{PS}$, the latter being
considerably too small (Thomas \& Couchman 1992). As we shall see
however, other sources of error are likely to be more important than the
difference between $n\sub{peak}$ and $n_{PS}$.

\subsection{Bias factors}

Another result given by BBKS is the bias factor at the present epoch, for a
given class of objects. The factor is present for any class of objects forming
when $\nu$ is significantly bigger than 1 ie., which originate as
exceptionally high peaks of the density contrast. In general the bias occurs
partly through the excess clustering of these peaks, and partly through
additional non-linear clustering after the objects have formed. An estimate
including both effects is
\be b\sub{obj}=1+\frac{\tilde\nu}{\sigma(M)}\label{bbks}\ee
where
\be
\tilde\nu=\nu-\frac{\gamma\theta(\gamma,\gamma \nu)}{1-\gamma^2}
\ee
In these expressions, everything on the right hand side is
to be evaluated at the epoch when the objects form.

\subsection{The virial velocity in a galaxy or cluster}

The quasi-linear
approach has been pushed further by some authors, to try to predict
the virial velocity $v(M)$ of structures of mass $M$. The virial velocity of a
gravitationally bound system is defined as the rms velocity of its
constituents in the centre of mass frame. According to the virial theorem it
is given by
\be v^2=\frac{GM}{R_g} \label{virial} \ee
where $M$ the mass of the system and $R_g$ is its gravitational radius,
defined by the requirement that its potential energy is $-GM^2/R_g$. The idea
is to relate $R_g$ to the comoving size $R\sub{com}$ of the object in the
early universe, defined by
\be M=(4\pi/3) R\sub{com}^3 \rho_0 \label{mass} \ee
where $\rho_0=3H_0^2/(8\pi G )$ is the present mass density. To obtain such a
relation one can use the spherical collapse model of Section \ref{NONLIN}.1
in the following manner. First, assume that the system has virialised, with no
energy loss, before the epoch $z\sub{form}$ when in the absence of pressure
forces the system   would  have contracted to a point. Next, assume that the
system collapsed from an initial configuration at rest, with a density profile
of the same shape as the profile after virialisation and without energy loss
(dissipation). It then follows from the virial theorem, \eq{virial}, that
$R_g$ is equal to one half of the initial gravitational radius. Finally, set
the initial gravitational radius equal to the initial radius of the edge of
the object, defined as the sphere containing mass $M$. Using the results
already quoted in Section \ref{NONLIN}.2  this gives
\bea R_g\mthree\eqa 8\frac{9\pi^2}{16} R\sub{com}\mthree
	(1+z\sub{max})\mthree \\
\eqa 32\frac{9\pi^2}{16} R\sub{com}\mthree (1+z\sub{form})\mthree\\
	\eqa 178 R\sub{com}\mthree  (1+z\sub{form})\mthree \eea
Then \eqs{virial}{mass} give
\be \rfrac{v}{126\km\,\sunit\mone}^2=\rfrac{M}{10^{12}\msun}\twothird
(1+z\sub{form}) \label{sthtw} \ee
This is the expression quoted by several of the authors mentioned above
(Evrard 1989; Henry \& Arnaud 1991; Evrard \& Henry 1991), and the others
presumably used a similar expression though they are less explicit. Obviously,
the assumptions leading to this relation are at best extremely crude
approximations. It seems clear, in fact, that {\it any} unique relation
between the mass, virial velocity and formation epoch of gravitationally bound
systems can only be a rough approximation.

In Figure 14, the arrows on the horizontal axes show $z\sub{form}(v)$,
calculated from  \eq{sthtw}, corresponding to estimates of the upper and
lower observational limits on observational values of $v$ for objects
of mass $M$. These estimates of $v$ are very rough. For galaxies, they
correspond roughly to those given by Blumenthal {\it et al} (1984) without any
attempt to update that analysis in the light of more recent observations. For
galaxy
clusters, the upper limit corresponds roughly to X-ray observations (Henry \&
Arnaud 1991); the observational lower limit of about $1000\km\,\sunit
\mone$ would
in that case correspond to negative $z$, a value of about $1300\km\,
\sunit\mone$
corresponding to $z=0$.

The most direct way of utilising the relation $z\sub{form}(M,v)$ is to set
$z\sub{form}$ equal to $z\sub{nl}(M)$. This gives the virial velocity $v(M)$
of systems formed when the scale $M$ goes nonlinear. Blumenthal {\it et al}
(1984) identified such systems with luminous galaxies and with galaxy
clusters, but according to the biased galaxy formation theory that
identification is wrong in the former case. Alternatively, the relation can be
combined with a theoretical estimate of the number density $dn(M,z)$ of
objects with mass between $M$ and $M+dM$, which form at epochs between \rm $z$
and $z+dz$, to calculate the number density $n(z,>v)$ of objects with virial
velocity bigger than $v$, which exist at redshift $z$. This is essentially the
approach taken by the other authors mentioned above.

\subsection{Comparison with observation}

The above estimates of number densities, sometimes combined with the \eq{bbks}
for the bias, have been compared with observation by several authors. The
statistic $n(>M)$ has been considered for galaxy clusters by Kaiser (1984),
Dalton {\it et al} (1992), Nichol {\it et al} (1992), Efstathiou, Bond \& White
(1992), Adams {\it et al} (1993) and White, Efstathiou and Frenk (1993), and
for both clusters and bright galaxies by Bardeen {\em et al} (1986), and
Bardeen, Bond and Efstathiou (1987). Statistics like $n(>v)$ have been
considered by Blumenthal {\it et al} (1984), Bardeen {\it et al} (1987), Evrard
(1989), Henry and Arnaud (1991). Evrard and Henry (1991) and Adams {\it et al}
(1993).

We argue in this section that comparisons of this kind, based on the
quasi-linear approach, are subject to rather large uncertainties
which make them of very limited use. In doing so we consider only
the statistic $n(>M)$, leaving aside the whole new range of
uncertainties which come into play when one goes to the statistic
$n(>v)$.

Typically, one will wish to compare the theoretical $n(>M)$ with the
observed quantity in order to determine the normalisation $\sigma_8$.
Because of the strong dependence on $\sigma_8^2$ this can indeed be
done very accurately {\em if we know accurately the quantities
$\delta_c$ and $M$}. The problem is that these quantities are not known
accurately, and in fact are not even well defined to high accuracy.
In Section \ref{NONLIN}.1 the collapse threshold $\delta_c$
was estimated to be $.69$ on the basis of a
spherical collapse model, but no-one would argue that this result
is not accurate, or even meaningful, to even one significant figure.
As for the filtering mass $M$, the prescription that the filtered
density contrast $\delta(M,\bfx)$
`sees' only those structures with mass bigger than $M$ is clearly
a crude approximation so one should allow considerable uncertainty
in this quantity. Yet it is clear from Figure 3b that $\sigma_0(M)$
depends quite sensitively on $M$, and even on the type of filter.
A related point is that the observed quantity $n(>M)$ depends
sensitively on the minimum mass $M$ of the objects counted, which
again is not accurately known.

The existence of these uncertainties is well known, and has of course been
pointed out before (for instance by Efstathiou and Rees (1988)), but at the
same time it often does not seem to be taken seriously. For instance, many of
the studies of galaxy clusters that we have mentioned consider the
astrophysics in considerable detail, while at the same time adopting  with
little reservation a prescription $n\sub{peak}$ or $n\sub{PS}$ with a more or
less arbitrarily chosen threshold like $\delta_c = 1.69$ or $\delta_c=1.33$. A
similar approach is taken in a recent discussion of the scale dependence of
the bias parameter (Bower {\it et al} 1993), where it is advocated that the
threshold $\delta_c$ varies by a few percent according to environment.

To understand the basic source of the uncertainty, it is enough to consider
the high peak limit, in which the prediction \eq{nchi} for $n(>M)$ depends
only on $\nu$, which is given by \eq{nusec}
\be \nu=\delta_c (1+z) \sigma_8\mone (\sigma_8/\sigma_0(M)) \ee
Let us suppose for the moment that the observational quantity $n(>M)$ is known
exactly, and estimate the uncertainty in $\Delta b_8$ due to the uncertainties
in $\delta_c$ and in the {\em theoretical} value of $M$, the latter arising
because of the identification of $M$ with the filtering mass. One has
\bea \Delta (\ln \sigma_8)\eqa\Delta(\ln \delta_c)-\Delta\left(
\ln (\sigma_0/\sigma_8)\right)\\
\eqa \Delta(\ln \delta_c)-\pdif{(\sigma_0(M)/b_8)}{M}\Delta (\ln M)
\eea
{}From \eqs{gfdis}{ksqm},
\be \pdif{(\sigma_0(M)/\sigma_8)}{M}=\frac13 R^2 \langle k^2\rangle \ee
With $n=1$, $R^2 \langle k^2\rangle=1.0$ for galaxy clusters
(mass $M=10^{15}\msun$) and .76 for galaxies (mass $M=10^{12}\msun$).

We noted in Section \ref{NONLIN}.3 that values of $\delta_c$ to be found in
the literature span a range $1.3\lsim \delta_c\lsim 1.7$, so let us take
$\delta_c=1.5\pm.2$. Taking the theoretical and observational uncertainties on
$M$ to be each a factor of 2 one finds for galaxy clusters
\be \frac{\Delta \sigma_8}{\sigma_8} =\pm.13\pm.23\pm.23 \label{unce} \ee
Combining these uncertainties in quadrature gives a total uncertainty of
35\%.

The observational uncertainty in $n(>M)$ at fixed $M$ is an additional
source of uncertainty, and so is the validity of the prescription
$n(>M)=n\sub{peak}$ as opposed to, say, $n(>M)=n\sub{PS}$
(we argued earlier that the Press-Schechter prescription is too low
in the large $\nu$ regime,
though it is used by most authors).
These uncertainties are likely
to be less important than the ones we discussed, because
of the exponential dependence of the prediction on $\nu$.

To be more explicit, we now compare theory with observation in some
detail, asking if the comparison can distinguish between the
three parameter choices $\{n,\sigma_8\}=\{1,1\}$, $\{1,.62\}$ and
$\{.7,.62\}$ (these values mark the corners of the roughly triangular region
of Figure 15 allowed by the bulk flow and cmb (COBE) data --- this region is
introduced properly in Section \ref{MDM}).
Using the prescription $n(>M)=n\sub{peak}$, and the formula \eq{bbks}
for the biases $b_g$ and $b\sub{cluster}$, we compare theory with
observation, first using
$\delta_c=1.33$
and $M=10^{15}$ (rich clusters) and $M=10^{12}$ (bright galaxies),
and then looking  at the effect of changing $\delta_c$ to 1.69 and
changing $M$ by a factor 2. We will find that the effect is so big that
the comparison cannot make the required distinction. Note that this is
without asking about the effect of changing to a different prescription
for $n(>M)$, such as the Press-Schechter one.

The prediction $n(>M)=n\sub{peak}$ is plotted against redshift in Figure 14,
for masses $M=10^{15}\msun$, $10^{12}\msun$, $10^{10}\msun$ and
$10^8\msun$, with $n\sub{peak}$ given to sufficient accuracy by
\eq{nup}. For each case,
three curves are given corresponding to the three parameter choices.
Each curve ends at the epoch $z\sub{nl}(M)$, when the linear approach
ceases to be valid. The threshold has been chosen as $\delta_c=1.33$.

\subsubsection*{Galaxy clusters}

Let us consider first the case  $M=10^{15}\msun$, which corresponds to very
large galaxy clusters.  Since this mass corresponds to the normalisation
scale $R_f\sim10h\mone\Mpc$ there is little dependence on $n$.

Since $\sigma(10^{15}\msun)=.8\sigma_8$, we assume that the filtered density
contrast on this scale is still evolving linearly at the present epoch. The
quantities of interest are the number density  $n(>M)$ and the ratio $b_c/b_g$
of the galaxy cluster bias factor to the galaxy bias factor. We will
make the identification $b_g=1/\sigma_8$.
The prediction for $n(>M)$ is $5.0\times
10^{-6}\Mpc\mthree$ if $\sigma_8=1$, and $1.0\times 10^{-6}\Mpc\mthree$ if
$\sigma_8=.63$. Following for instance BBKS and Bardeen, Bond and Efstathiou
(1987), we identify galaxy clusters in this mass range with Abell clusters of
richness class $>1$, and hence observed number density
$7.5\times10^{-7}\Mpc\mthree$ (Bahcall \& Soniera 1983). Thus, the prediction
for $n(>M)$ is about right if $\sigma_8=.6$, but too big if $\sigma_8=1$.
Coming to
$b_c/b_g$, one finds from \eq{bbks} that if $\sigma_8=1.6$,
then $b_c=4.0$ giving
$b_c/b_g=2.5$. If, on the other hand, $\sigma_8=1.0$, then $b_c=1.5$
which gives
$b_c/b_g=1.5$. Recent estimates (Dalton {\it et al} 1992; Nichol {\it et al}
1992) give $\xi_{cc}(r)=(r_0/r)^{1.9\pm.3}$ with $r_0=(13\pm5)h\mone\Mpc$.
Dividing by the galaxy correlation function $\xi_{gg}=(5h\mone\Mpc/r)^{1.8}$
gives therefore $b_c/b_g=2.4\pm.9$ as an observational estimate. Again,
$\sigma_8=.6$ is preferred.

So far so good, but what about the uncertainty? Unfortunately it is big.
First, suppose that we take $\delta_c$ to be equal to $1.69$, instead of the
$1.33$ used in the above estimates. For $\sigma_8=1$
(.62) this multiplies $n(>M)$
by a factor .56 (.16), and multiplies $b_c$ by a factor $1.4$ ($1.4$). Second,
suppose instead that we multiply $M$ by a factor $2$, on the ground that there
may be this amount of uncertainty in the observational value of the masses of
galaxy clusters (and remembering also that filtering the density contrast on
mass scale $M$ does not completely eliminate all structure with mass less than
$M$). This multiplies $n(>M)$ by a factor $.30$ ($.09$) and multiplies $b_c$
by a factor $1.6$ ($1.7$). Combining these two perfectly reasonable changes
gives $n(>M)=8\times 10^{-7}$ and $b_c/b_g=3.3$ if $\sigma_8=1$,
but $n(>M)=7\times
10^{-8}$ and $b_c/b_g=6$ if $\sigma_8=.62$. The former choice of $\sigma_8$
is now strongly preferred by both pieces of data!

\subsubsection*{Bright galaxies}

For galaxies, the linear epoch ends before the present. The number density
$n(>M)$ at the end of the linear epoch cannot be estimated reliably, and it
will also evolve with time because of merging and other non-linear phenomena.
As a result, no reliable prediction is possible for the quantity $n(>M)$ at
the present epoch.

However, in the biased theory of galaxy formation, the formation of {\it
luminous} galaxies is supposed to stop during the linear regime,
which gives a bias $b_g$
given roughly by \eq{bbks}. By demanding that
$b_g$ value reproduces the value
$b_8=1/\sigma_8$ one obtains the epoch of luminous
galaxy formation, and hence the observed number density, as a function of
$\sigma_8$. This approach has been implemented for $n=1$ by
BBKS and by Bardeen,
Bond and Efstathiou (1987). Here we extend the calculation to $n=.7$, and
comment on the uncertainty.

The stars in Figure 14b for $\sigma_8=.62$ indicate the epoch of formation of
luminous galaxies which is needed to reproduce the bias factor
$b_g=1/\sigma_8$, according to
\eq{bbks}. This epoch is $z=4.7$ (2.8) for $n=1$ (.7), and the corresponding
value of $n(>M)$ is $4.8\times 10 \mfour\Mpc\mthree$ ($7.9\times
10\mfour\Mpc\mthree$). Ignoring merging etc., this comoving number density
should be equal to the presently observed number density $n_g(>M)$ of luminous
galaxies for $M=10^{12}\msun$. An observational estimate of $n_g(>M)$ is the
Schechter parametrisation, which taking the ratio of luminosity $L$ to mass
$M$ to be independent of $M$ gives (Ellis {\it et al} 1988)
\be
M \frac{{\rm d}n_g}{{\rm d}M} =\phi_*\rfrac{M}{M_*}^{-.07} e^{-M/M_*}
\ee
In this formula
\be
\phi_*=1.56\times10\mtwo h^3 \Mpc \ee
(This provides a good estimate for the case $M=10^{12}\msun$ that we are
discussing here. For lighter galaxies the assumption of constant $M/L$ is not
very good (Ashman, Salucci \& Persig 1993).) Integrating $M {\rm d}n_g/{\rm
d}M$ over all masses gives $M_*= 1.6\Omega\sub{gal}\times 10^{13}
h\mone\msun$, where $\Omega\sub{gal}$ is the contribution to $\Omega$ of
luminous galaxies (including the dark halos) and we take $\Omega\sub{gal}=.1$.
For $M=10^{12}\msun$, this gives an observational estimate $n_g(>M)=1.75\times
10\mthree\Mpc\mthree$, which is marked by an arrow on the $y$-axis of Figure
14a.

According to this calculation, the predicted number density of luminous
galaxies is somewhat too small for $\sigma_8=.62$, and in fact one needs a
value $\sigma_8\simeq .8$ to reproduce it. But now set $M= 5\times
10^{11}\msun$ in the theoretical calculation, on the ground that the
observational value of $M$ could be a factor 2 too high. With $\sigma_8=.62$
this gives a somewhat earlier epoch of formation $z=5.9$ (3.6) for $n=1$ (.7).
The corresponding number densities are $n_g(>M)=5.8 \times 10\mfour$
($1.2\times 10\mthree$), which look much more healthy, and the actual value of
$\sigma_8$ needed for consistency is now around $.7$. Even without looking at
any other sources of uncertainty (such as galaxy merging, which BBKS
emphasise), it seems clear that it is very difficult to pin down $\sigma_8$
within the range $.6<\sigma_8<1$ from these considerations.

\subsubsection*{Small galaxies}

Finally, consider the results for $M=10^{10}\msun$ and $10^8\msun$, as shown
in Figures 14c and d. The observed number densities $n_g(>M)$ according to the
Schechter parametrisation are again shown by arrows. (though they become
increasingly uncertain as the mass is reduced). Within the biased galaxy
formation theory one expects luminous galaxies with these masses to form
before the epoch $z\sub{nl}(M)$, but the observational bias factor for them is
not known, and there is no reason why it should be equal to the result
$b\simeq1/\sigma_8$ which is observed for bright galaxies. However, the epoch
of formation would need to be a lot earlier than $z\sub{nl}(M)$ to give
agreement with the observed $n_g(>M)$. It is presumably more reasonable to
suppose that merging has taken place, reducing $n_g(>M)$ from its original
value. Again, there does not seem to be a useful constraint on the parameters
$n$ and $\sigma_8$.

\subsubsection*{The quasar density}

So far we have focussed on observations at small redshift. One can also ask
about observation at $z\gsim 1$. In particular, quasars have now been seen out
to a redshift of about 5, and are the oldest observed objects in the universe.
In order to produce the observed luminosity, some galaxies must have evolved
to contain a sufficient concentration of mass-energy. One then needs to
estimate the minimum mass required and the number density, in order to
discover if the CDM cosmogony can explain the observations.  Such a comparison
was made for $n=1$ by Efstathiou and Rees (1988). Using both the $n\sub{peak}$
and Press-Schechter prescriptions they calculated $n(>M)$ for various masses
as a function of redshift. Our results are far higher than theirs, but the
difference can be explained if they chose $b=2.5$ and defined $b$ with
respect to $J_3$ normalisation instead of $\sigma_g$ normalisation (they are
not explicit about these choices, but our calculation essentially agrees with
theirs if these changes are made in it). As in the case of large galaxy
clusters at
the present epoch, the extreme sensitivity to normalisation is caused by the
fact that $\nu$ is substantially bigger than 1.

The number density of quasars assumed by Efstathiou and Rees was $10^{-6}
h^{3}$ Mpc$^{-3}$, out to a redshift of around $4$. Since then, observations
have become more stringent, and to be consistent with present data one
requires this number density out to a redshift of $5$ (Martin Rees, private
communication). This value also assumes that the quasar lifetimes are not too
short, in which case one needs multiple generations. One also needs to know
the mass which is required to be evolving nonlinearly to harbour the quasar,
and they estimated at the time that $10^{12} \msun$ was the smallest that one
could safely get away with. However, recent work (Martin Rees, private
communication) has suggested that only $10^{10} \msun$ is required, which
corresponds to a large loosening of the constraint. As one sees from Figures
14b or 14c, the upshot of all this is that enough quasars may form at high
redshift, even if $n=.7$ and $\sigma_8=.62$.

Another cosmological requirement on high redshifts is the Gunn-Peterson
constraint on the amount of
neutral hydrogen in the inter-galactic medium. In the CDM cosmogony this
implies that some structure has formed before $z= 5$, in order to re-ionise
the hydrogen (Schneider, Schmidt \& Gunn 1989), but it is not clear how much
or of what kind. Even with $n=.7$, structure with $M\lsim 10^{10}\msun$ forms
at the epoch $1+z\simeq 9\sigma_8$ so there seems to be time
to form structure with this $n$ and with $\sigma_8\simeq .6$.

\subsubsection*{Quasi-linear approach plus numerical simulations}

Many papers calculate $n(z,>M)$ in both the quasi-linear approach also with
numerical simulations. In a limited regime of $\{z,M\}$ space, usually
corresponding to values of $\nu$ which are not too big, rough agreement is
typically found for a choice of threshold $\delta_c$ somewhere in the range
$.3$ to $.7$. Where the numerical simulations actually cover the regime of
parameter space which is required for the purpose at hand, the quasi-linear
approach should probably be regarded as redundant. On the other hand, if the
quasi-linear prediction is pushed far beyond the regime where it has been
checked, as is often the case, the uncertainties that we have discussed still
apply.

\subsubsection*{Excluding very low normalisations}

Although the quasi-linear approach is subject to rather large uncertainties,
it is capable of ruling out very low normalisations. In particular, it seems
to rule out $n\lsim .6$ with the COBE normalisation (excluding gravitational
waves) on the ground that there are too few high redshift objects to be
consistent with, for example, the abundance of quasars (Cen {\em et al} 1992;
Adams {\em et al} 1993; Haehnelt 1993). As we shall recap at the beginning of
Section \ref{MDM} such values are also excluded by the bulk flows, but this
independent exclusion from different considerations is certainly significant.

\subsection{Numerical simulations}

Let us now turn to numerical simulations. The idea here is to follow the
evolution of a gas containing a finite number of particles, which is supposed
to represent a finite comoving region of the universe. The size of the region
considered is taken to be of order 10 to 100\Mpc, depending on the
observational quantity of interest as well as on the available computing
power. Starting in the linear regime, the particles are given positions and
velocities which provide a realisation of the required Gaussian density
contrast. They are then allowed to move under the action of gravity, perhaps
with non-gravitational interactions taken into account in some way.

For the standard $n=1$ CDM model, many simulations have been reported (Davis
{\it et al} 1985; Gott {\it et al} 1986; White {\it et al} 1987; Carlberg \&
Couchman 1989; Frenk {\em et al} 1988; Frenk {\it et al} 1990; Carlberg,
Couchman \& Thomas, 1990; Bertschinger \& Gelb 1991; Couchman \& Carlberg
1992; Bahcall \& Cen 1992; Cen \& Ostriker 1992a; Cen \& Ostriker 1992b; Suto,
Cen \& Ostriker 1992; Katz, Quinn \& Gelb 1992; Katz, Hernquist \& Weinberg
1992; Gelb 1992; Gelb \& Bertschinger 1992; Gelb, Gradwohl \& Frieman 1993;
Gelb \& Bertschinger 1993; Klypin {\it et al} 1993; Katz \& White 1993; White,
Efstathiou \& Frenk 1993. A useful survey of those prior to 1992 is given by
Davis {\it et al} (1992a). Recently there have also been numerical simulations
of the CDM model which allow a tilted spectrum, $n<1$ (Vogeley {\it et al}
1992; Park {\it et al} 1992; Cen {\it et al} 1992; Cen \& Ostriker 1992c).

Numerical simulations can estimate most of the quantities listed at the
beginning of this section. The most powerful comparison seems to be with the
dispersion $\sigma_\parallel(r)$ of the line-of-sight relative velocity
between a randomly chosen pair of galaxies separated by distance $r$. It can
be measured on scales $r\sim1$ Mpc, where its values is $300 \pm 50\, {\rm
km\, s}^{-1}$ (Davis \& Peebles 1983). For the well-studied $n=1$ case, almost
all simulations agree that a low normalisation $.3<\sigma_8<.5$ is required to
fit this value. The main exception is the one reported by Couchman and
Carlberg (1992), who claim that $\sigma_8=1$ fits the data on account of a
large velocity bias. This claim is not generally accepted.

Going on to the case $n<1$, one might expect that tilting the spectrum with
$\sigma_8$ fixed would reduce $v_\parallel$ on the scale $k\mone\sim1\Mpc$.
This is certainly true in linear theory, which gives from
\eqss{159}{DISP}{corr} the result
\bea \sigma_\parallel^2(r)\equiva
\frac13\langle|{\bfv}(\bfr)-\bfv({\bf 0})|^2\rangle\\
\eqa \frac{2H_0^2}{3} \int^\infty_0
\frac{{{\cal P}}_\delta(k) }{k^2}
\left[ 1-\frac{\sin kr}{kr} \right] \frac {dk}{k} \eea
Compared with \eq{sigv}, which gives the dispersion of the peculiar velocity
itself, this expression contains the square bracket and as a result
$\sigma_\parallel(r)$ probes the spectrum mainly on scales $k\mone\sim r$. The
linear approach is not however valid on the scale $1\Mpc$  where observations
exist, and on the basis of numerical simulations Gelb {\em et al} (1993) claim
that $\sigma_\parallel(1\Mpc)$ is sensitive to scales well in excess of
$10\Mpc$. As a result, the prediction with $\sigma_8$ fixed at $0.5$ is rather
insensitive to tilt, in the range $.7<n<1$.

Of the other comparisons with observation, we mention only the number density
$n(>M)$ of galaxy clusters. We have already discussed the considerable
uncertainty in the quasi-linear estimate of this quantity. These arise
primarily from the uncertainty in the threshold $\delta_c$, the uncertainty in
the theoretical mass $M$ (reflecting the uncertainty in the prescription of
identifying the filtering mass with the observed mass), and the uncertainty of
the observed mass. Numerical simulation can in principle eliminate the first
two uncertainties, but the third remains. The estimate that we made of  it in
\eq{unce} should remain roughly valid for the numerical simulations, because
they reproduce roughly the quasi-linear results for a suitable choice of
$\delta_c$.

The most recent comparison with observation of the number density of galaxy
clusters is due to  White, Efstathiou and Frenk (1993). They compare their
simulations with two different observational estimates, $M=1.8\times
10^{14}h\mone \msun$ and $M=1.4\times10^{14}h\mone\msun$ of the value of $M$
for which $n(>M)=4\times8\times 10^{-6} h^3 \Mpc\mthree$. They deduce
normalisations respectively $\sigma_8=.62$ and $.54$. They emphasise, however,
that these estimates of $M$ are probably too high, by a factor of perhaps 2 to
3, which would lower $\sigma_8$. Using the estimate \eq{unce} a reduction by a
factor 2 from the lower value gives $\sigma_8=.41$, and a reduction by a
factor 3 gives $\sigma_8=.32$. The conclusion is therefore that a comparison
of simulations with the observed number  density of galaxy clusters gives the
result $.3\lsim\sigma_8\lsim .6$. This is about the same range as the one
coming from simulations of the pairwise velocity dispersion, except that the
upper bound on $\sigma_8$ is somewhat weaker.

\section{Alternatives to pure cold dark matter}
\label{MDM}
\setcounter{equation}{0}
\renewcommand\theequation{\thesection.\arabic{equation}}

In this final section, we assess the extent to which observation requires a
modification of the CDM model. We then look briefly at the two modifications
which at the present time have been investigated in some detail. They both
invoke the assumption that the cold dark matter density is significantly less
than the critical value. In the {\em cosmological constant} model, the
difference is made up by a cosmological constant, whereas in the {\em mixed
dark matter} (MDM) model it is made up by hot dark matter.

\subsection{Comparison of the CDM model with observation}

Figure 15 shows the main constraints which have been assembled in the last
three sections. Assuming cold dark matter with $\Omega_m=1$, it gives some
observational constraints in the $n-\sigma_8$ plane, where $n$ defines the
shape of the spectrum of the primeval density contrast and
$\sigma_8\equiv1/b_8$ defines its normalisation. The standard CDM model has
spectral  index $n=1$, but we have allowed the possibility of `tilt' away from
this value.

The significance to be attached to the various results is probably best left
to the reader, as the observational situation remains fluid. The lines plotted
omit the mean value; instead they indicate the allowed spread about that
unplotted mean. The constraints plotted are
\begin{itemize}
\item The limits from the COBE experiment. The COBE prediction depends on the
gravitational wave amplitude produced, and we have plotted the constraints for
the case where it is negligible (dashed line) as in the case of `natural'
inflation ($n<1$) and two-scale inflation ($n>1$), and also where it is given
by power-law or extended inflation (solid line). The lower line corresponds to
the observational upper limit from COBE. Prejudice from the South Pole null
result (Gaier {\it et al} 1992) and from the MIT-balloon experiment suggests
there is little room for the true result to be much above the mean. This can
be seen as especially troublesome in the power-law case if one wants to fit
the APM data.
\item The limits from the QDOT survey (dot--dashed), obtained in the manner
discussed in Section \ref{CLUSTER}.2.
\item The limits on $n$ (independent of $\sigma_8$) obtained from requiring a
fit to the APM data (dotted). We have taken our constraint to be $0.3 < n <
0.6$ (see Figure 7), though we note that Efstathiou, Bond and White (1992)
allow a range for their $\Gamma$ which is roughly equivalent to letting $n$ be
as large as 0.67, which is a very conservative constraint. These lines should
be taken very seriously as limits outwith which the theory appears to fail to
fit the APM data.
\item The limits from pairwise velocities (dot--dot--dot--dashed). According
to Gelb, Gradwohl and Frieman (1993) these are roughly independent of $n$ for
reasonable values, and we have plotted as such, though one should treat the
extrapolation much below $n=0.7$ with skepticism. Note the inherent
contradiction with the QDOT lines in the CDM cosmogony. Again, it would appear
that one really has to lie between these lines in order to satisfy the present
observational data.
\end{itemize}

The simplest conclusion that one could draw is that standard CDM cannot fit
all these data. Perhaps one might hope to stretch the QDOT and APM lines
towards $n=0.7$, $\sigma_8$= 0.5 if the gravitational waves are negligible.
However, if one takes the QDOT data seriously at say the $2$-sigma level this
possibility appears to be excluded even for natural inflation. In any event,
the situation appears hopeless for power-law inflation due to the large
gravitational wave contribution.

\subsection{The cosmological constant model}

Now let us consider the effect of making the present matter density $\Omega_m$
less than the critical value 1, making up the difference with a cosmological
constant contribution $\Omega_\Lambda$ (Peebles 1984b; Kofman \& Starobinsky
1985; Turner 1991b; Gorski {\em et al} 1992; Lilje 1992; Wright {\em et al}
1992; Efstathiou {\em et al} 1992; Kofman, Gnedin \& Bahcall 1992; White,
Efstathiou \& Frenk 1993). Following most of these authors we set the spectral
index $n$ equal to 1.

The cosmological constant may be thought of as a contribution with
homogeneous, time-independent energy density and pressure related by
$\rho_\Lambda+p_\Lambda=0$. Since $\rho_m\propto a\mthree\propto (1+z)^3$,
the matter density is equal to $\rho_\Lambda$ at the epoch
\be (1+z_\Lambda)= (\Omega_\Lambda/\Omega_m)\third \ee
Until just before this epoch the effect of the cosmological constant is
negligible, being $\lsim 10\%$ of the matter density before the epoch
$(1+z)=2(1+z_\Lambda)$. The theory developed in Section \ref{DENSPER} is then
an adequate approximation. The spectrum of the matter density is specified by
the transfer function, but its shape is affected because of the fact that the
epoch of matter-radiation equality is earlier, $1+z\sub{eq}$ being increased
by a factor $1/\Omega_m$. As long as Silk damping is neglected the shape of
the transfer function is {\em unchanged} when written as a function of
$k/k\sub{eq}$, or equivalently (assuming $z\sub{eq} \gg z_\Lambda)$ as a
function of $k/\Omega_m$, as exhibited in \eq{tran}.

Around the epoch $z_\Lambda$ the effect of the cosmological constant starts to
become important. The growth of the density contrast slows down, and ceases
altogether well after $z_\Lambda$. As a result the normalisation of the
transfer function begins to fall, but {\em its shape is retained}. According
to the discussion in Section \ref{CLUSTER}.1, this means that the slope of the
galaxy correlation function is fitted with a value of $\Omega_m$ in the range
0.3 to 0.6.

To calculate the effect of the  cosmological constant on other observable
quantities one has to follow the evolution of the transfer function to the
present, and calculate the associated peculiar velocity and the Sachs-Wolfe
effect (Kofman \& Starobinsky 1985). The authors cited earlier find find at
least marginal consistency with the data, for $\Omega_m \simeq .4$,
 corresponding to $\Omega_\Lambda \simeq .6$.
The most serious potential problem is with the bulk flows, for which the
prediction seems to be too low.

The cosmological constant model with $\Omega_m\sim\Omega_\Lambda$ may be
viewed as unnatural, in that we are living at the epoch when the cosmological
constant is starting to be significant. It does not seem possible to offer
even an anthropic justification for this fact, since the cosmological constant
is negligible until $z\sim1$, by which time galaxy and star formation are well
under way. The most effective way to constrain the scenario appears to be via
the abundance of lensed quasars (Fukugita {\it et al} 1992), which measures
the volume at high redshift, expected to be much larger in the cosmological
constant model than in the standard case. New constraints of this type should
shortly be available from the Hubble space telescope snapshot survey.

One could, of course, consider the possibility that $\Omega_m$ is less than 1
without making up the difference at all,
$\Omega<1$. In this `open' model (so called
because the spatial sections are open, as opposed to the compact
case $\Omega>1$) the growth of the density contrast is reduced more than in
the cosmological constant case, because one has to go to higher redshift
before the matter dominated $\Omega$ achieves its early time value $\Omega
_m\simeq1$. As a result the model predicts a bigger cmb anisotropy for a given
normalisation $\sigma_8$, which requires higher values of $\Omega_m$
(Blumenthal, Dekel \& Primack 1987).

This is a good place to mention another possibility, which is to reduce the
Hubble parameter. The effect of such a reduction is again to alter the scale
$k\sub{max}$, so that the transfer function remains the same when written in
terms of the variable $k/h^2$ as exhibited in \eq{tran}. The galaxy
correlation function, which is known as a function of $h\mone r$ is therefore
fitted with a value $h\simeq .2$. The problem, of course, is that such a value
is ruled out by direct observation, the lowest reported value being around
$.4$.

\subsection{The mixed dark matter (MDM) model}

Many authors, especially since the advent of the COBE data, have considered
the possibility that the dark matter has a hot component (Shafi \& Stecker
1984; Fang, Li \& Xiang 1984; Valdarnini \& Bonometto 1985; Achilli,
Occhionero \& Scaramella 1985; Ikeuchi, Norman \& Zhan 1988; Schaefer, Shafi
\& Stecker 1989; Holtzman 1989; Schaefer 1991; Wright {\em et al} 1992; van
Dalen \& Schaefer 1992; Schaefer \& Shafi 1992; Taylor \& Rowan-Robinson 1992;
Davis, Summers \& Schlegel 1992; Klypin {\it et al} 1992; Shaefer \& Shafi
1993; Holtzman \& Primack 1993). We discuss this possibility now, again
following the majority of authors by setting the spectral index $n$ equal to
1.

Loosely speaking, the term hot dark matter denotes dark matter whose primeval
perturbation free streams away on scales up 100 Mpc or so, as opposed to warm
dark matter where the maximum scale is appreciably lower and cold dark matter
where it is too low to be cosmologically significant. In practice the term is
more specific, denoting matter which consists of a massive stable particle
species which falls out of thermal equilibrium while still relativistic. The
only known candidates are the three neutrino species $\nu_e$, $\nu_\mu$ and
$\nu_\tau$.

We briefly summarise some of the main features of such `neutrino' hot dark
matter, following Davis, Summers \& Spergel (1992). Denoting its mass by
$m_\nu$, the present mass density of the hot dark matter is
\be \Omega_\nu=.43 \frac{m_\nu}{10\eV} \label{omnu}
\ee
It becomes non-relativistic at the epoch
\be 1+z\sub{nr}=(4.2\times10^4)\Omega_\nu \ee
For $\Omega_\nu\gsim .1$, this is sufficiently early that the epoch
$z\sub{eq}=.62\times10^4$ of matter-radiation equality is the same as for pure
CDM. In that case the scale entering the horizon at $z\sub{nr}$ is given by
\be k\mone\sub{nr}=\rfrac{1+z\sub{nr}}{1+z\sub{eq}}\half
k\sub{eq}\mone = 46 \Omega_\nu\half \Mpc \ee
On scales smaller than this the perturbation of the HDM free-streams away upon
horizon entry, causing the growth of the CDM perturbation to slow down.

This free-streaming makes pure HDM completely unviable. However, if CDM is
also present, then on becoming nonrelativistic the HDM can fall into the
potential wells already created by the CDM so that its density contrast grows
to match that of the CDM. At any epoch, this can occur on scales below an
effective Jeans
length
\be k_J\mone=.11(1+z)\half\Omega_\nu \Mpc \ee
By the present time, the CDM  and HDM have a common density contrast on scales
$k\mone\gsim .1\Mpc$, which are the only ones of interest.

Because the HDM continues to have significant random motion even after matter
domination, the transfer function for MDM continues to evolve right up to the
present. At a given epoch it has less power on short scales than the pure CDM
transfer function, in the regime $1$ to $100\Mpc$. As a result, it can
generate the `excess power' on large scales relative to short scales discussed
in Section \ref{CLUSTER}.1, if $\Omega_\nu\simeq.3$. To demonstrate this we
have plotted in Figure 16 $\sigma_0(M)$ normalised to the pure CDM value,
using the transfer function given by Klypin {\em et al} for $\Omega_\nu =
0.3$. (For illustration, we also show two tilted models, one with and one
without gravitational waves.) With the COBE normalisation, $\sigma_8=1/1.7$.
[Klypin {\it et al} quote $\sigma_8$ as 1/1.5, but this appears to be because
they have normalised to a high value of the rms quadrupole rather than
directly to the $10^0$ result as we do.] Thus one has less need of galaxy bias
than in the pure CDM model.

This same value of $\Omega_\nu$ is claimed by the above authors to fit various
other observations. One would particularly like to know the prediction for the
pairwise velocity $\sigma_\parallel(r)$. Numerical simulations with MDM are
more complicated than for pure CDM, because one has to take into account the
random motion of the neutrinos, and so far one set of simulations has been
done with sufficient resolution to look at this quantity (Klypin {\em et al}
1992). A fit is claimed, but the same simulations run for pure CDM give a
lower pairwise velocity than most other authors. Specifically, at $r=1\Mpc$
and with the normalisation $\sigma_8=1/1.5$, the simulations give
$\sigma_\parallel\simeq450 \km\sec\mone$ whereas those of Gelb, Gradwohl and
Frieman (1993) give $\sigma_\parallel\sim 600$ to $700\Mpc$ (interpolating
their results for $\sigma_8=.5$ and .75). According to the simulations of
Klypin {\em et al}, going from pure CDM to MDM with $\sigma_8$ fixed at
$1/1.5$ only reduces the pairwise velocity by around $100\km\sec\mone$. [Note
though our comments above that our COBE normalisation gives a lower $\sigma_8$
for MDM than does theirs.]

In summary, while it is clear that MDM gives a much lower pairwise velocity
than pure CDM, when normalised to COBE, it is not yet clear that the reduction
is sufficient to fit the observed pairwise velocity.

In Figure 17 the MDM and CDM predictions for $\sigma_0 (M)$ are plotted
directly. One sees that the bottom-up structure formation will proceed more or
less as with pure CDM, as discussed in Section 8.2. The reduction of small
scale power means, though, that there is a lack of high redshift objects, and
in fact it is clear that a value $\Omega_\nu$ significantly bigger than $.3$
is fatal (Klypin {\em et al} 1992; Haehnelt 1993). Whether
$\Omega_\nu\simeq.3$ is viable remains to be seen.

\subsubsection*{Fine tuning in the MDM model}

The fine tuning required in the MDM model is of a different kind from that
required in the cosmological constant model, in that it involves parameters
which one might hope eventually to calculate from a fundamental theory.
One of these is clearly the neutrino mass, which has to be around $7\eV
$ to make $\Omega_\nu=.3$. In addition, there are whatever parameters
determine the cold dark matter density $\Omega_c\sim .7$.

As a purely illustrative example, suppose that the cold dark matter consists
of axions, and take as an estimate of its density the commonly quoted result
(Kolb \& Turner 1990)
\be \Omega_c=\frac{f_a}{10^{12}\GeV} \ee
where $f_a$ is the axion decay constant. Then $\Omega_\nu/\Omega_c$ is of
order 1, provided that $f_a\sim 10^{12} \GeV$ and that $m_\nu\sim 10\eV$. It
is natural to suppose that the massive neutrino is the $\tau$ neutrino, in
which case one suggestion for its mass is the see-saw formula (Ellis, Lopez \&
Nanopoulos 1992) \be m_{\nu}=m\sub{top}^2/M \ee where $M$ is some very high
mass scale. With $m\sub{top}\sim 100\GeV$ as suggested by collider results,
one obtains $m_\nu\sim 10eV$ if $M\sim 10^{12} \GeV$. With these estimates,
the question of whether MDM is `natural' boils down to the question of whether
the existence of mass scales $m_\tau\sim 100\GeV$ and $f_a\sim M\sim
10^{12}\GeV$ is `natural'. Whether natural or not, such scales have certainly
been invoked by particle theorists and experimenters, for reasons that have
nothing to do with cosmology or dark matter (Ellis, Lopez \& Nanopoulos 1992).

Leaving aside specific formulas for $\Omega_\nu$ and $\Omega_c$, which at the
present time cannot to be taken seriously, the point we are making is that one
hopes eventually to relate the ratio of these quantities to masses and
couplings which appear in a Lagrangian describing the fundamental theory of
nature. Then the problem of `explaining' this ratio will be the same as the
problem of `explaining' these masses and couplings, and thus belongs firmly in
the realm of particle theory or, some would hope, of superstring theory.

Finally, we comment on the prospect of verifying the MDM model by direct
measurement of the neutrino properties. The model requires a neutrino,
presumably $\nu_\tau$, and a mass of order $10\eV$ and a lifetime $\gsim
10^{10}$ years. The lifetime is obviously inaccessable to any expected
astronomical observations. What about the mass? It is far too small to measure
directly through energy and momentum conservation, the present upper limit
from such techniques being about $35\MeV$. However, if they have mass the
neutrinos are expected to mix, and unless the mixing angle of $\nu_\tau$ with
$\nu_\mu$ is extremely small it should be observable in the forseeable future.
At present the limit on the mixing angle if $\nu_\tau$ has the required mass
is about $.03$, and this can be improved by a factor of 3 or so in the
proposed CHORUS and NOMAD experiments at CERN and P803 (Ellis, Lopez \&
Nanopoulos 1992). Thus one has a real prospect of verifying the model, or
ruling it out, in the forseeable future.

\subsection{Topological defect models}

Finally, let us briefly summarise the present status of the chief rivals to
the inflationary model for providing initial density perturbations --- models
based on topological defects where inhomogeneities are created in an initially
homogeneous background via the Kibble mechanism (Kibble 1976).
Possible types of defect are gauge strings (Albrecht \&
Stebbins 1992; Perivolaropoulos 1993;
Perivolaropoulos and Vachaspati 1993), global strings or textures (Pen,
Spergel \& Turok 1992), global monopoles (Bennett \& Rhie 1992; Pen, Spergel
\& Turok 1992) and domain walls forming at a very late phase transition
(Xiao-chun \& Schramm 1992; Jaffe, Stebbins \& Frieman 1992).

In these models, the seed (defect) field evolves not under gravity but instead
in accord with the scalar field dynamics appropriate to the defect field in
question. At lowest order, one can assume in fact that these defects evolve in
a Friedmann background driven by the matter content of the universe. The
defects act on the matter to provide seed inhomogeneities, but at lowest order
these inhomogeneities do not act back on the defect evolution (Veeraraghavan
\& Stebbins 1990). Once gravity is not the only force acting, the evolution of
inhomogeneities is vastly complicated, and because of this topological defect
theories have considerable ground to catch up relative to the inflation-based
models in terms of theoretical development. These complications are outside
the scope of the present paper.

At the present time, textures seem to be more or less ruled out, because the
COBE data require a normalisation $\sigma_8 \simeq .25$ (Pen, Spergel \& Turok
1992), too low to explain other types of data, given that the perturbation
spectrum induced by textures is very similar in shape to that of standard CDM.
The other cases have not been ruled out, but neither have they been confronted
with data to anything like the same extent as the CDM model. The most
promising case is that of gauge strings with hot dark matter (Albrecht \&
Stebbins 1992b). As already discussed, most theories with hot dark matter are
in deep trouble as the dark matter free-streaming removes too much small-scale
structure for them to be viable at the COBE normalisation. Cosmic strings
provide a possible exception, as small-scale structure on the strings
regenerates small-scale inhomogeneities after (or during) free-streaming. On
the other hand, this process appears to generate rather too much small-scale
power in a model with strings and cold dark matter (Albrecht \& Stebbins
1992a).

A possible problem for gauge strings is the fact that the temperature of the
universe cannot be appreciably higher than $10^{16}\GeV$ after inflation, as
we discussed after \eq{vqua}. This bound implies that cosmic strings with
enough energy to form structure probably cannot form by the usual Kibble
mechanism of a thermal phase transition. (Pollock 1987). To see this, one has
to compare two results. First, in order to form structure, gauge strings would
need to have an energy per unit length of at least $2\times(10^{16}\GeV)^2$
(Albrecht \& Stebbins 1992; Perivolaropoulos 1993). Second, the energy per
unit length of a string formed in a thermal phase transition at temperature
$T$ is typically $\simeq T^2$. Combined, these results require a temperature
after inflation $T\gsim 2\times 10^{16}\GeV$, or an energy density
$\rho\quarter>(\pi^2 g_*/ 15)\quarter \times 10^{16}\GeV$ where $g_*\gsim 100$
is the effective number of particle species. Thus, structure forming gauge
strings probably cannot be created after inflation.

Forming gauge strings, or anything else, before the observable universe leaves
the horizon during inflation is useless because inflation dilutes their
density exponentially. There remains the possibility of forming the strings
after horizon exit, but before the end of inflation. This can be done, if the
string forming field has a suitable coupling to the inflaton field or to
gravity (Shafi \& Vilenkin 1984; Vishniac, Olive \& Seckel 1987; Lyth 1987;
Yokoyama 1988; Shafi 1988; Yokoyama 1989; Lyth 1990; Hodges \& Primack 1990;
Basu, Guth \& Vilenkin 1991; Nagasawa \& Yokoyama 1992; Shafi 1993; see also
the two-scale model of Section \ref{INFL}.4).

The situation is different for global strings because they are surrounded by
a Goldstone field. As a result, their effective energy per unit length grows
like $\ln H\mone$ after formation, and can increase by a factor of order 100
by the time of structure formation, allowing them to form in a thermal
phase transition after inflation. A structure formation scenario with global
strings has yet to be properly investigated, though some tentative, and rather
unfavourable, conclusions have been drawn by Pen, Spergel and Turok (1992).

\section{Discussion and Conclusions}
\label{CON}
\setcounter{equation}{0}
\renewcommand\theequation{\thesection.\arabic{equation}}

In this paper we have provided a detailed review of the machinery which allows
one to make calculations in the Cold Dark Matter model and its variants,
including considerable material which is applicable in more general
surroundings. While it is certainly possible to have `designer' models of
inflation which are capable of adapting to almost any conceivable
observational setback, the typical outcomes of presently available models are
fairly simple. Typically, inflation provides a spectrum of adiabatic gaussian
perturbations which can be well described by a power-law spectrum, tilted from
the Harrison--Zel'dovich spectrum, normally tilted so as to provide extra
large scale power. The magnitude of the tilt may be modest or pronounced.
Long wavelength gravitational waves are also produced, in most models with
amplitude increasing with the degree of tilt in the density perturbation
spectrum.

The details of structure formation are very sensitive to the matter content of
the universe. It appears that if cold dark matter is the main constituent of
the universe, present observations require that the initial perturbations be
adiabatic --- isocurvature perturbations generate excessively large cmb
anisotropies for the same final density perturbation. Adiabatic perturbations
are exactly what inflation provides. In CDM models, the only remaining
alternative would appear to be texture seeded models, which have been placed
in jeopardy by a combination of microwave anisotropy and velocity data, though
the death blow apparently remains to be struck (Pen, Spergel \& Turok 1992).

To end this review, we summarise the present observational situation with
regard to realistic inflation--seeded CDM models. Our final constraints on
models possessing only CDM are those illustrated in Figure 15. It is clear
that nothing works on all data; adding tilt does not appear to offer any
immediate advantages over standard CDM. The most striking problem appears to
be the reconciliation of the pairwise velocity data with the bulk flow data
(though the former could also be substituted by the galaxy cluster abundance
(White, Efstathiou \& Frenk 1993)). The excess clustering of galaxies on large
scales as reflected in the slope of the galaxy correlation function is also
clearly problematical if one takes the bulk flows seriously, but as the
hardest piece of data to interpret ({\it ie} the easiest to dismiss as the
consequence of complex astrophysics (see {\it eg} Bower {\it et al} 1993) )
one could argue that it be ignored.

With nowhere providing clear agreement, it is not too clear how one should
assess the constraints on $n$. If one ignores completely the clustering and
pairwise data, but decides that the bulk flows should be believed at some
level not much more than two sigma (and assuming at least that the COBE two
sigma upper limit cannot be violated), then for natural inflation one gets
something like $n > 0.70$. For power-law and extended inflation, the large
gravitational wave contribution to COBE pushes the limit up to around $0.85$.
(Liddle \& Lyth 1992, 1993). This clearly excludes the possibility that these
models can fit the clustering data, and indeed we have already remarked
(Liddle \& Lyth 1992) that this limit rules out simple forms of extended
inflation, as they give a big bubble constraint (Liddle \& Wands 1991) $n <
0.75$. (Changing from a cold dark matter cosmogony to a different choice also
does not seem to salvage the situation (Liddle \& Lyth 1992).)

With a lot of work, one could aim for agreement around $n \simeq 0.6$ -- $0.7$
and $\sigma_8 \simeq 0.5$ -- $0.6$ in the natural inflation model. Note though
that this requires pushing all data to extremes, and in particular relies on
the true $10^0$ variance on the mwb being well above the COBE mean, which is
strongly disfavoured by other microwave results. Such a model does however
seem to deal adequately with the epoch of formation of structure (contrary to
our more pessimistic assessment in Liddle, Lyth and Sutherland (1992)).

To conclude, we have found strong constraints on the slope of the primeval
spectrum when generated by various inflationary models. Most extended
inflation models appear to be ruled out completely. Power-law inflation is
perhaps viable, but only for values of $n \gsim 0.85$, too high to allow an
explanation of the clustering data. Natural inflation (and related) models are
the most promising candidates for generating useful power-law spectra ---
provided the true level of fluctuations is close to the top of the COBE range
they seem marginally able to explain the excess large scale clustering as now
seen in many optical surveys. In most aspects, such a model does at least as
well as an unbaised standard CDM model, with the advantage of a more plausible
galaxy cluster abundance as well as helping with galaxy clustering statistics.

We completed our survey by examining variants on the CDM model which may be
better suited to explaining the observational data. The standard technique is
to utilise additional matter (be it a component of hot dark matter or of a
cosmological constant) to remove short-scale power from the CDM spectrum. Hot
dark matter does this by free-streaming, a cosmological constant by delaying
matter-radiation equality. Because this power can be removed over a much
shorter range of scales than with tilt, it enables an explanation of the
observed deficit of short-scale power relative to intermediate scale power in
the spectrum.

We have not attempted a detailed investigation of these models; however, this
can be pursued using the techniques we have aimed to illustrate in this
report. We note though that provided modifications to the spectrum remove
short-scale power, then the limits on the deviation of the spectral slope from
unity will tighten.

The main moral of this paper is that if one is to take inflation seriously as
the originator of the adiabatic density perturbation, then one must deal with
the range of possible predictions that inflationary models may make. With
options both of tilt and gravitational waves, the uncertainty in the
inflationary prediction (which can easily halve the short-scale power, for
example) appears greater than the uncertainty from any other source. In
general, the predictions for tilt and gravitational wave contribution to COBE
from inflation are independent, corresponding to choices for the $\epsilon$
and $\eta$ slow-roll parameters, thus feeding two new parameters into large
scale structure studies.

Indeed, given two new parameters, it is in some ways remarkable that this does
not appear to provide enough freedom to fix up the CDM model, as Figure 15
illustrates. MDM adds yet another new parameter, roughly speaking an ability
to remove short-scale power from the spectrum while leaving large scales
untouched, and may be necessary should all present observations stand up. It
appears likely that MDM will however need an initial spectrum close to $n=1$
with no gravitational waves if it is to succeed. Should MDM be vindicated,
this appears to raise the possibility that a wide range of inflation models
not providing these conditions with sufficient accuracy may be excluded.

\section*{Appendix A}
\setcounter{equation}{0}
\renewcommand\theequation{A.\arabic{equation}}

In the text we found it convenient to  regard a choice of the time
coordinate $t$ as a slicing of spacetime into hypersurfaces labelled by
$t$. Each such slicing was referred to as a gauge, and we dealt
explicitly with two choices; the comoving gauge where the hypersurfaces
are orthogonal to the comoving worldlines, and the synchronous gauge
where they are orthogonal to geodesics which start out comoving
in the limit of very early times. Here we derive some results relating
different gauges, the main object being to prove a couple of results which
were used in the text.

\subsubsection*{Proper time versus coordinate time}

Consider a slicing of spacetime into hypersurfaces labelled by a
parameter $t$. The gradient $\pa_\mu t$ is some multiple of the unit
normal vector $n_\mu$,
\be fn_\mu=\pa_\mu t \label{aone} \ee
It follows that $\pa_\nu(f u_\mu)$ is antisymmetric.
\be [(\pa_\nu f) u_\mu
-(\pa_\mu f) u_\nu]
 + [f \pa_\nu u_\mu-
f \pa_\mu u_\nu] =0 \label{atwo} \ee

First take the hypersurfaces to be orthogonal to the comoving
worldlines, so that $n^\mu$ is the fluid 4-velocity $u^\mu=
dx^\mu/d\tau$, where $\tau$ is proper time along a comoving worldline.
Taking the inner product of both sides of \eq{aone} with $u^\mu$ gives
$f=dt/d\tau$. Then, in a basis such that $u^\mu=(1,0,0,0)$, \eq{atwo} becomes
\bea \omega_{ij}=0 \\
\pa_i{f}-  f \frac{\pa _i p}{\rho + p} \eqa  0
\eea
The first equation is the necessary and sufficient condition that comoving
hypersurfaces exist. The second determines $f$. To first order in the
perturbations,
\be f=\mbox{constant} \left[ 1+ \frac{\delta p}{\rho+p}
\right] \ee
Setting the constant equal to 1 corresponds to choosing $t$
to be the average of $\tau$ over each hypersurface.

\subsection*{The relation $\delta {{\cal R}} =-H \delta t$}

The immediate purpose of the following discussion is to prove the relation
${{\cal R}}= (H/\dot\phi)\delta \phi$, which was used
in deriving \eq{512}. It was first derived by Sasaki (1986) using the metric
perturbation formalism of Bardeen (1980), but it is easy also to derive it
using the more  geometrical viewpoint adopted here.

The quantity on the left hand side of the relation is related to the curvature
of comoving hypersurfaces by \eq{122}. The inflaton field perturbation
vanishes on such hypersurfaces, because its spatial gradient corresponds to
momentum flow and in the text we said that $\delta\phi$ on the right hand side
of the equation is to be defined using any gauge (choice of hypersurfaces)
which is non-singular in the slow-roll limit. Before justifying this statement
in general, we show that it holds in the gauge where the hypersurfaces have
zero intrinsic curvature scalar. This will turn out to be true irrespective of
the slow-roll limit, as long as $\phi$ dominates the momentum density, and
will allow us in the next Appendix to extend the calculation to that case.

The relation we want is in fact a special case of a more general relation
$\delta {{\cal R}} =-H \delta t$, giving the change in the intrinsic curvature
of a spacelike hypersurface when it is displaced by a proper time interval
$\delta t$. {\em In this more general relation $H$ and ${{\cal R}}$ do {\em
not} refer to a hypersurface orthogonal to comoving worldlines. Rather, they
refer to an arbitrary hypersurface.} Thus, for each hypersurface we define $H$
and ${{\cal R}}$ as in Section 2, \eqs{14}{122}, except that $u^\mu$ is taken
to denote the unit normal to the hypersurface rather than the 4-velocity of
the fluid. Using a locally inertial coordinate system such that
$u^\mu=(1,v^i)$ with $v^i$ of first order, $3H=\pa_i v^i$, and from \eq{122}
$R^{(3)}=-4\nabla^2{{\cal R}}$ where $\nabla^2=\pa_i \pa^i $ and $R^{(3)}$ is
the curvature scalar of the hypersurface.

The proof of the required relation is very simple, if we take as a starting
point the following relation (Hawking \& Ellis 1973)
\be \frac13 G^{00}=H^2-\frac16 R^{(3)} \ee
In this relation, $G^{00}=R^{00}-\frac12 g^{00} R$ is the component of the
Einstein tensor along the unit normal to the hypersurface,
where $R$ is the curvature scalar of spacetime, and $R^{(3)}$ is
the curvature scalar of the hypersurface. The quantity $H$ is
essentially the extrinsic curvature scalar of the hypersurface, so the
relation we are using connects the intrinsic and extrinsic curvature
scalars of a hypersurface, given the curvature of the spacetime
in which it is embedded.
The relation is
valid for any choice of hypersurfaces, to first order in the
perturbations.

For a comoving hypersurface,
Einstein's field equation says that the left hand side is equal to $8\pi G
\rho/3$, where $\rho$ is the energy density in a comoving frame.
It is therefore not necessary  that the unit normal
$n^\mu$ is orthogonal to the hypersurface, it need only be of the form
$n^\mu=(1,v^i)$ where $v^i$ is of first order. The energy density
$\rho$ and therefore $G^{00}$ is the same for any such choice, to first
order.

Now consider two different choices of hypersurface, and consider the one
from each choice which goes through a given point in spacetime.
The left hand side is the same for both choices
so the differences $\delta H$ and $\delta R^{(3)}$ are
related by
\be 2H\delta H=\frac16 \delta R^{(3)} \label{afirst} \ee
We will now calculate $\delta H$, and hence deduce the required change
$\delta {{\cal R}}$.

To calculate the change in $H$, start with
\eq{aone}, which is valid for any gauge. In the limit where the
perturbations vanish all gauges have to become equivalent, so
$f=1+x$ where $x$ is of first order in the perturbations.
Now consider
two different hypersurface choices,
\bea (1+x) n_\mu\eqa \pa_\mu t \\
(1+x^\prime ) n_\mu^\prime \eqa \pa_\mu t^\prime \eea
In any basis such that $n^\mu=(1,0,0,0)$ up to first order corrections,
the relative velocity
$v_i\equiv n_\mu^\prime -n_\mu$ of the normals to the hypersurfaces  is
given in terms of the time displacement $\delta t\equiv t^\prime -t$
between them by
\be v_i=\pa_i \delta t \label{96} \ee
The required change in $H$ is therefore given by
$3\delta H=\pa_i v^i=\nabla^2 \delta t$

{}From \eq{afirst} it follows that
\be H\nabla^2 \delta t=-\nabla ^2 {{\cal R}} \ee
As we are working with Fourier series (periodic boundary conditions),
the unique solution of this equation is the advertised relation.

Now consider inflation. The field perturbation $\delta\phi$ vanishes on
comoving hypersurfaces, so it follows that their curvature perturbation is
given as advertised by
\be {{\cal R}}=(H/\dot\phi)\delta\phi \label{asec} \ee
where $\delta\phi$ is the field perturbation defined on those hypersurfaces
where the curvature perturbation vanishes.

In the slow-roll limit the comoving hypersurfaces become infinitely distorted,
as witnessed by the divergence of their displacement $\delta
t=\delta\phi/\dot\phi$ from the undistorted zero curvature hypersurface, as
well as the divergence of their curvature ${{\cal R}}$. It follows that the
displacement between gauges remaining finite in the slow-roll limit becomes
negligible compared with $\delta t$, so that \eq{asec} becomes valid in all of
them.

\section*{Appendix B}
\setcounter{equation}{0}
\renewcommand\theequation{B.\arabic{equation}}

We first explain why the effect of the metric perturbation on the field
equation \eq{510} for $\delta\phi$ is of the same order as other departures
from slow-roll (Lyth \& Stewart 1990a). Including both metric and field
perturbations, the field equation is  $\Box\phi+V^\prime =0$, where
\be \Box=(-g)\mhalf\partial_a(-g)\half g^{ab}\partial_b \ee
To first order, the metric perturbation adds to the left hand side of \eq{510}
a term $(\delta \Box) \bar \phi$. The unperturbed field $\bar\phi$ is position
independent so that the spatial derivatives in $\delta \Box $  have no effect
on it. If it were time-independent as well the metric perturbation would have
no effect at all. Thus the effect of the metric perturbation vanishes in the
limit where the slow-roll approximation becomes exact, for any coordinate
choice which remains non-singular in that limit. (This argument also shows
explicitly what one usually takes for granted, that there is no back-reaction
for a massless (non-inflaton) scalar field; there too, the background field
has no time dependence.)

The exact field equation for $\delta\phi$, including the effect of the metric
perturbation and making no slow-roll assumption, has been derived by Mukhanov
(1985) in a coordinate system with flat spatial hypersurfaces. In terms of
$u=a\delta\phi$ the equation is
\be u\sk^\prime +(k^2-z^\prime /z) u\sk = 0 \ee
where $z=a\dot\phi/H$ and the prime denotes differentiation with respect to
conformal time $\eta$, $d\eta=dt/a$. Mukhanov derived this equation using the
metric perturbation approach, but it follows straightforwardly in the fluid
flow approach that we are using, from \eqsss{532}{118}{121}{122}. (In deriving
it one has to note that the last two equations are valid even when ${{\cal
R}}$ is not constant, and that  on comoving hypersurfaces $\delta \rho=\delta
P $, because $V(\phi)$ is unperturbed.)

One easily checks that the equation reduces to \eq{510} in the slow-roll
approximation, in accordance with the general argument above. For the general
case, all of the formalism in Section \ref{INFL} goes through, with $a w_k$ the
solution of the above equation satisfying \eq{519} well before horizon entry.
It has been used to calculate the exact spectrum for power-law inflation (Lyth
\& Stewart 1992a), and to estimate the error in the slow-roll approximation
for an arbitrary inflaton potential, in the slow-roll regime (Stewart \&
Lyth 1993).

\section*{Acknowledgements}

We would like to express special thanks to Lauro Moscardini, David Salopek,
Ewan Stewart and Will Sutherland for many discussions relating to this work.
We would also like to thank Peter Coles, Ed Copeland, Rod Davies, George
Efstathiou, George Ellis, Carlos Frenk, Richard Frewin, Josh Frieman, Martin
Hendry, Jon Holtzman, Lev Kofman, Rocky Kolb, Tony Lasenby, Jim Lidsey, Andrei
Linde, Joel Primack, Martin Rees, Varun Sahni, Misao Sasaki, David Spergel,
Paul Steinhardt, Andy Taylor, Peter Thomas and Simon White for helpful
discussions and comments at various stages. ARL is supported by the SERC, and
acknowledges the use of the STARLINK computer system at the University of
Sussex.

\section*{References}
\frenchspacing
\begin{description}
\item Abbott, L. F. and Wise, M. B. 1984a, {\it Astrophys J Lett} {\bf 282},
	L47.
\item Abbott, L. F. and Wise, M. B. 1984b, {\it Nucl Phys} {\bf B244}, 541.
\item Abbott, L. F. and Wise, M. B. 1984c, {\it Phys Lett} {\bf B135}, 279.
\item Achilli, S., Occhionero, F. \& Scaramella, R. 1985, {\it Astrophys J}
	{\bf 299}, 577.
\item Adams, F. C. and Freese, K. 1991, {\it Phys Rev D}{\bf 43}, 353.
\item Adams, F. C., Bond, J. R., Freese, K., Frieman, J. A. and Olinto, A.
	V. 1993, {\it Phys Rev D}{\bf 47}, 426.
\item Albrecht, A. and Stebbins, A. 1992a, {\it Phys Rev Lett} {\bf 68}, 2121.
\item Albrecht, A. and Stebbins, A. 1992b, {\it Phys Rev Lett} {\bf 69}, 2615.
\item Albrecht, A., Ferreira, P., Joyce, M. and Prokopec, T. 1993,
	``Inflation and Squeezed Quantum States'', IC London preprint.
\item Ashman, K. M., Salucci, P. and Persig, M. 1993,
	{\it Mon Not Roy astr Soc}, in press.
\item Babul, A. and White, S. D. M. 1991, {\it Mon Not Roy astr Soc} {\bf
	253}, 31p.
\item Bahcall, N. and Soniera, R. 1983, {\it Astrophys J} {\bf 270}, 20.
\item Bardeen, J. M. 1980, {\it Phys Rev D}{\bf 22}, 1882.
\item Bardeen, J. M. 1986, in {\sl Proc. Inner Space/Outer Space Conf.},
	eds Kolb, E. W. {\it et al} (Univ. of Chicago Press).
\item Bardeen, J. M., Bond, J. R., Kaiser, N. and Szalay, A. S. 1986, {\it
	Astrophys J} {\bf 304}, 15 [BBKS].
\item Bardeen, J. M., Bond J. R. and Efstathiou, G. 1987, {\it Astrophys
	 J} {\bf 321}, 28.
\item Bardeen, J. M., Steinhardt, P. S. and Turner, M. S. 1983,
	{\it Phys Rev D}{\bf 28}, 679.
\item Barrow, J. D. and Ottewill, A. C. 1983, {\it J Phys} {\bf A16}, 2757.
\item Barrow, J. D. 1987, {\it Phys Lett} {\bf B187}, 12.
\item Barrow, J. D. 1990, {\it Phys Lett} {\bf B235}, 40.
\item Barrow, J. D. and Saich, P. 1990, {\it Phys Lett} {\bf B249}, 406.
\item Barrow, J. D. and Liddle, A. R. 1993, ``Perturbation Spectra from
	Intermediate Inflation'', Sussex preprint SUSSEX-AST 93/2-1.
\item Basu, R., Guth, A. H. and Vilenkin, A. 1991, {\it Phys Rev D}{\bf 44},
	340.
\item Bennett, D. P. and S. H. Rhie, ``COBE's Constraints on the Global
	Monopole and Texture Theories of Structure Formation'', Berkeley
	preprint 1992.
\item Bennett, D. P., Stebbins, A. and Bouchet, F. R. 1992, {\it Astrophys J
	Lett} {\bf 399}, L5.
\item Bertschinger, E. 1985, {\it Astrophys J Supp} {\bf 58}, 39.
\item Bertschinger, E. and Dekel, A. 1989, {\it Astrophys J Lett} {\bf 336},
	L5.
\item Bertschinger, E., Dekel, A., Faber, S. M., Dressler, A. and
	Burstein, D. 1990, {\it Astrophys J} {\bf 364}, 370.
\item Bertschinger, E. and Gelb, J. M. 1991, {\it Computers Phys} {\bf 5}, 164.
\item Birrell, N. D. and Davies, P. C. W. 1982, {\sl Quantum Field Theory in
	Curved Space-time}, Cambridge University Press.
\item Blumenthal, G. R., Faber, S. M., Primack, J. R. and Rees, M. J. 1984,
	{\it Nature} {\bf 311}, 517.
\item Blumenthal, G. R. and Primack, J. R. 1984, in {\sl Fourth Workshop on
	Grand Unification}, eds Weldon, H. A. {\it et al} (Birkhauser).
\item Blumenthal, G. R., Dekel, A. \& Primack, J. R. 1987,
	{\it Astrophys J} {\bf 326}, 539.
\item Bond, J. R. and Efstathiou, G. 1984, {\it Astrophys J} {\bf 285}, L45.
\item Bond, J. R. and Efstathiou, G. 1987, {\it Mon Not Roy astr Soc} {\bf
	226}, 655.
\item Bond, J. R. 1989, in {\sl Highlights in Astronomy} vol. 9
	Proceedings of the IAU Joint discussion, Ed. J. Bergeron
	(Buenos Aires Gen Ass).
\item Bond, J. R. and Efstathiou, G. 1991, {\it Phys Lett} {\bf B265}, 245.
\item Bond, J. R., Efstathiou, G., Lubin, P. M. and Meinhold, P. R.
	1991a, {\it Phys Rev Lett} {\bf 66}, 2179.
\item Bond, J. R., Cole, S., Efstathiou, G. \& Kaiser, N. 1991b,
	{\em Astrophys J} {\bf 379}, 440.
\item Copeland, E. J., Kolb, E. W., Liddle, A. R and Lidsey, J. L. 1993,
	``Reconstructing the Inflaton Potential --- In Principle and in
	Practice'', Sussex/Fermilab preprint SUSSEX-AST 93/3-1;
	FNAL-PUB-93/029-A.
\item Crittenden, R., Bond, J. R., Davis, R. L., Efstathiou G. and
	Steinhardt, P. J. 1993 ``The Imprint of Gravitational Waves on the
	CMB'' Pennsylvania preprint.
\item Bower, R. G., Coles, P., Frenk, C. S., White, S. D. M. 1993, {\it
	Astrophys J} {\bf 405}, 403.
\item Brainerd, T. G. and Villumsen, J. V. 1992, {\it Astrophys J} {\bf 394},
	409.
\item Bruni, M., Dunsby, P. K. S., and Ellis, G. F. R. 1992, {\it Astrophys
	J} {\bf 395}, 34.
\item Cardoso, G. L. and Ovrut, B. A. 1993, ``Natural Supergravity
	Inflation'' CERN preprint.
\item Carlberg, R. G. and Couchman, H. M. P. 1989, {\it Astrophys J} {\bf
	340}, 47.
\item Carlberg, R. G., Couchman, H. M. P. and Thomas, P. A. 1990, {\it
	Astrophys J} {\bf  352}, L29.
\item Cen, R. and Ostriker, J. P. 1992a, {\it Astrophys J} {\bf 393}, 22.
\item Cen, R. and Ostriker, J. P. 1992b, ``CDM cosmology with
	hydrodynamics and galaxy formation: the evolution of the IGM and
	background radiation fields'' Princeton University preprint.
\item Cen, R. and Ostriker, J. P. 1992c, ``A hydrodynamical treatment of
	the tilted cold dark matter cosmological scenario'' Princeton
	University preprint.
\item Cen, R., Gnedin, N. Y., Kofman, L. A. and Ostriker, J. P. 1992, {\it
	Astrophys J Lett} {\bf 399}, L11.
\item Couchman, H. M. P. and Carlberg, R. G. 1992, {\it Astrophys J} {\bf
	389}, 453.
\item Dalton, G. B., Efstathiou, G., Maddox, S. J. and Sutherland, W. J.
	1992, {\it Astrophys J Lett} {\bf 390}, L1.
\item Davis, M. and Peebles, P. J. E. 1983, {\it Astrophys J} {\bf 267}, 465.
\item Davis, M., Efstathiou, G., Frenk, C. S. and White, S. D. M. 1985, {\it
	Astrophys J} {\bf 292}, 371.
\item Davis, M., Efstathiou, G., Frenk, C. S. \& White, S. D. M. 1992a, {\it
	Nature} {\bf 356}, 489.
\item Davis, M., Summers, F. J. and Schlegel, D. 1992, {\it Nature} {\bf 359},
	393.
\item Davis, R. L., Hodges, H. M., Smoot, G. F., Steinhardt, P. J. and
	Turner, M. S. 1992b, {\it Phys Rev Lett} {\bf 69}, 1856.
\item Dekel, A., Bertschinger, E. and Faber, S. M. 1990, {\it Astrophys J}
	{\bf 364}, 349.
\item Dekel, A. 1991, in ``Observational Tests of Cosmological
	Inflation'', eds Shanks, T. {\it et al}, Kluwer academic,
        and in Annals of the New York Academy of Sciences, Volume 647.
\item Dekel, A., Bertschinger, E., Yahil, A., Strauss, M. A., Davis, M.
	and Huchra, J. P. 1992, ``IRAS galaxies verses POTENT mass: Density
	Fields, Biasing and $\Omega$'', Princeton preprint IASSNS-AST 92/55.
\item Dodelson, S. and Jubas, J. M. 1992, ``Microwave Anisotropies in the
	Light of COBE'', Fermilab preprint FERMILAB-PUB-92/366-A.
\item Duruisseau, J. P. and Kerner, R. 1986, {\it Class Quant Grav}
	{\bf 3}, 817.
\item Efstathiou, G. and Bond, J. R. 1986, {\it Mon Not Roy astr Soc}
	{\bf 218}, 103.
\item Efstathiou, G. and Rees, M. J. 1988, {\it Mon Not Roy astr Soc} {\bf
	230}, 5p.
\item Efstathiou, G. 1990, in ``The Physics of the Early Universe'', eds
	Heavens, A., Peacock, J. and Davies, A., SUSSP publications.
\item Efstathiou, G., Sutherland, W. J. and Maddox, S. J. 1990, {\it Nature}
	{\bf 348}, 705.
\item Efstathiou, G. 1991, in ``Observational Tests of Cosmological
	Inflation'', eds Shanks, T. {\it et al}, Kluwer academic.
\item Efstathiou, G., Bond, J. R. and White, S. D. M. 1992, {\it Mon Not Roy
	astr Soc} {\bf 258}, 1p.
\item Ehlers, J. 1961, {\it Abh Mainz Akad Eiss Lit (Math Nat Kl)} {\bf 11}, 1.
\item Ellis, G. F. R., Lyth, D. H. and Mijic, M. B. 1991, {\it Phys Lett}
	{\bf B271}, 52.
\item Ellis, R. S. and Peterson, B. A. 1988, {\it Mon
	Not Roy astr Soc } {\bf 232}, 431.
\item Evrard, A. 1989, {\it Astrophys J } {\bf 341}, L71.
\item Evrard, A. and Henry, J. P. 1991, {\it Astrophys J} {\bf 383}, 95.
\item Fabbri, R., Lucchin, F. and Matarrese, S. 1986, {\it Phys Lett} {\bf
	B166}, 49.
\item Fabbri, R. and Pollock, M. D. 1983, {\it Phys Lett} {\bf B125}, 445.
\item Fang, L. Z., Li, S. X. \& Xiang, S. P. 1984, {\it Astron Astrophys}
	{\bf 140}, 77.
\item Fisher, K. B., Davis, M., Strauss, M. A., Yahil, A. and Huchra, J. P.
	1993, {\it Astrophys J} {\bf 402}, 42.
\item Fong, R., Hale-Sutton, D. and Shanks, T. 1992, {\it Mon Not Roy astr
	Soc} {\bf 257} 650.
\item Freese, K., Frieman, J. A. and Olinto, A. V. 1990, {\it Phys Rev Lett}
	{\bf 65}, 3233.
\item Frenk, C. S., White, S. D. M., Davis, M. and Efstathiou, G. 1988, {\it
	Astrophys J} {\bf 327}, 507.
\item Frenk, C. S., White, S. D. M., Efstathiou, G. and Davis, M. 1990, {\it
	Astrophys J} {\bf 351}, 10.
\item Fukugita, M., Futamase, T., Kasai, M. and Turner, E. L. 1992, {\it
	Astrophys J} {\bf 393}, 3.
\item Gaier, T., Schuster, J., Gunderson, J., Koch, T., Seiffert, M.,
	Meinhold, P. and Lubin, P. 1992, {\it Astrophys J Lett} {\bf 398}, L1.
\item Gelb, J. M. 1992, Fermilab preprint 92/181-A, to appear in the
	proceedings of the `Groups of Galaxies' workshop held in June 1992 at
	the Space Telescope Science Institute, Baltimore, Maryland, U.S.A..
\item Gelb, J. M., Gradwohl, B.-A. and Frieman, J. A. 1993, {\it Astrophys J
	Lett} {\bf 403}, L5.
\item Gelb, J. M. and Bertschinger, E. 1993, ``Cold Dark Matter II: Spatial
	and Velocity Statistics'', Fermilab preprint FERMILAB-PUB-92/74-A.
\item Gorski, K., Silk, J. and Vittorio, N. 1992, {\it Phys Rev Lett},
	{\bf 68}, 733.
\item Gott, J. R., Melott, A. L. \& Dickinson, M. 1986, {\it Astrophys J}
	{\bf 306}, 341.
\item Grischuk, L. 1993, Washington University (St Louis) preprint,
	to be published in {\sl The Origin of Structure in the Universe},
	ed. P Nardone.
\item Groth, E. J., Juszkiewicz, R. \& Ostriker, J. P. 1989, {\it Astrophys J}
	{\bf 346}, 558.
\item Gunn, J. E. and Gott, J. R. 1972, {\it Astrophys J} {\bf 176}, 1.
\item Guth, A. H. 1981, {\it Phys Rev D}{\bf 23}, 347.
\item Guth, A. H. and S.-Y. Pi 1982, {\it Phys Rev Lett} {\bf 49}, 1110.
\item Guth, A. H. and  Pi, S.-Y. 1985, {\it Phys Rev D}{\bf 32}, 1899.
\item Halliwell, J. J. 1989, {\it Phys Rev D}{\bf 39}, 2912.
\item Harrison, R. 1970, {\it Phys Rev D}{\bf 1}, 2726.
\item Hawking, S. W. 1966, {\it Astrophys J}, {\bf 145}, 544.
\item Hawking, S. W. 1982, {\it Phys Lett} {\bf B115} 339.
\item Hawking, S. W. and Ellis, G. F. R. 1973, \sl The Large-Scale Structure
	of Space-Time \rm (Cambridge University Press).
\item Henry, J. P. and Arnaud, K. A. 1991, {\it Astrophys J} {\bf 372},
	410.
\item Hodges, H. M., Blumenthal, G. R., Kofman, L. and Primack, J. 1989,
	{\it Nucl Phys} {\bf B244}, 541.
\item Hodges, H. M. and Blumenthal, G. R. 1990, {\it Phys Rev D}{\bf 42}, 3329.
\item Hodges, H. M. and Primack, J. 1990, {\it Phys Rev D}{\bf 43}, 3155.
\item Holtzman, J. 1989, {\it Astrophys J Supp} {\bf 71}, 1.
\item Holtzman, J. and Primack, J. 1993, {\it Astrophys J} {\bf 405}, 428.
\item Hogan, C. J., Kaiser, N. and Rees, M. J. 1982, {\it Phil Trans R Soc
	Lond} {\bf A307}, 97.
\item Hogan, C. J. and Rees, M. J. 1988, {\it Phys Lett B} {\bf 205}, 228.
\item Ikeuchi, S., Norman, C. \& Zhan, Y. 1988, {\it Astrophys J} {\bf 324},
	33.
\item Jones, B. J. T. and Wyse, R. F. G. 1985, {\it Astron Astrophys} {\bf
	149}, 144.
\item Kaiser, N. 1984, {\it Astrophys J Lett} {\bf 284}, L9.
\item Kaiser, N., Efstathiou, G., Ellis, R., Frenk, C., Lawrence, A.,
	Rowan-Robinson, M. and Saunders, W. 1991, {\it Mon Not Roy astr Soc}
	{\bf 252}, 1.
\item Kashlinsky, A. and Jones, B. J. T. 1991, {\it Nature} {\bf 349}, 753.
\item Kashlinsky, A. 1992, {\it Astrophys J Lett} {\bf 387}, L5.
\item Katz, N., Quinn, T. and Gelb, J. M. 1992, ``Galaxies form at
	peaks--Not!'', Fermilab preprint.
\item Katz, N., Hernquist, L. and Weinberg, D. H. 1992, {\it Astrophys J}
	{\bf 399}, L109.
\item Katz, N. and White, S. D. M. 1993, {\it Astrophys J}, in press.
\item Kibble, T. W. B 1976, {\it J Phys} {\bf A9}, 1387.
\item Klypin, A., Holtzman, J., Primack, J. R. and Reg\"{o}s, E. 1992,
	``Structure Formation with Cold + Hot Dark Matter'', Santa Cruz
	preprint SCIPP 92/52.
\item Kodama, H. and Sasaki, M. 1984, {\it Prog Theor Phys Supp} {\bf 78}, 1.
\item Kodama, H. and Sasaki, M. 1987, {\it Int J Mod Phys} {\bf A2}, 491.
\item Kofman, L. and Starobinsky, A. A. 1985, {\it Sov Astron Lett},
	{\bf 11}, 271.
\item Kofman, L. and Linde, A. 1987, {\it Nucl Phys} {\bf B282}, 555.
\item Kofman, L., Mukhanov, V. F. and Pogosyan, D. Yu. 1987, {\it Sov Phys
	JETP} {\bf 66}, 441.
\item Kofman, L. and Pogosyan, D. 1987, {\it Nucl Phys} {\bf 282}, 555.
\item Kofman, L., Gnedin, N. and Bahcall, N. A. 1992, ``Cosmological
	Constant, COBE  CMB Anisotropy and Large Scale Clustering'', CITA
	Preprint 92/94.
\item Kofman, L., Linde, A. D. \& Starobinsky, A. A. 1993,
	``Reheating of the Universe in Inflationary Cosmology'', Stanford
	preprint.
\item Kolb, E. W. and Turner, M. S. 1990, \sl The Early Universe
	\rm (Addison-Wesley).
\item Kolb, E. W., Salopek, D. S. and Turner, M. S. 1990, {\it Phys Rev D}{\bf
	42}, 3925.
\item Kolb, E. W. 1991, {\it Physica Scripta} {\bf T36}, 199.
\item Krauss, L. M. and White, M. 1992, {\it Phys Rev Lett} {\bf 69}, 869.
\item La, D. and Steinhardt, P. J. 1989, {\it Phys Rev Lett} {\bf 62}, 376.
\item Lahav, O., Edge, A., Fabian, A. C. and Putney, A. 1989, {\it Mon Not Roy
	astr Soc} {\bf 238}, 881.
\item Liddle, A. R. 1989, {\it Phys Lett} {\bf B220}, 502.
\item Liddle, A. R. and Wands, D. 1991, {\it Mon Not Roy astr Soc} {\bf 253},
	637.
\item Liddle, A. R., Lyth, D. H. and Sutherland, W. J. 1992, {\it Phys Lett}
	{\bf B279}, 244.
\item Liddle, A. R. and Lyth, D. H. 1992, {\it Phys Lett} {\bf B291}, 391.
\item Liddle, A. R. and Lyth, D. H. 1993, to appear, Proceedings of the 16th
	Texas Symposium.
\item Lidsey, J. E. and Coles, P. 1992, {\it Mon Not Roy astr Soc} {\bf 258},
	57p.
\item Lifshitz, E. M. 1946, {\it J Phys (Moscow)} {\bf 10}, 116.
\item Lilje, P. B. 1992, {\it Astrophys J Lett} {\bf 386}, L33.
\item Linde, A. D. 1982, {\it Phys Lett} {\bf B116} 335.
\item Linde, A. D. 1983, {\it Phys Lett} {\bf B129}, 177.
\item Linde, A. D. 1987, in ``300 Years of Gravitation'', eds Hawking, S. W.
and
	Israel, W. I., Cambridge University Press.
\item Linde, A. D. 1990, \sl Particle Physics and Cosmology \rm
	(Gordon and Breach).
\item Linde, A. D. 1991a, {\it Phys Lett} {\bf B249}, 18
\item Linde, A. D. 1991b, {\it Phys Lett} {\bf B259}, 38.
\item Linde, A. D. 1992, {\it Phys lett} {\bf B284}, 215.
\item Linde, A. D. and Lyth, D. H. 1990, {\it Phys Lett} {\bf B246}, 353.
\item Linder, E. V. 1988, {\it Astrophys J} {\bf 326}, 517.
\item Loveday, J., Efstathiou, G., Peterson, B. A. \& Maddox, S. J.
	1993, {\it Astrophys J} {\bf 400}, L43.
\item Lucchin, F. and Matarrese, S. 1985, {\it Phys Rev D}{\bf 32}, 1316.
\item Lucchin, F., Matarrese, S. and Mollerach, S. 1992, {\it Astrophys J
	Lett} {\bf 401}, 49.
\item Lyth, D. H. 1984, {\it Phys Lett} {\bf B147}, 403; {\bf B150}, 465(E).
\item Lyth, D. H. 1985, {\it Phys Rev D}{\bf 31} 1792.
\item Lyth, D. H. 1987, {\it Phys Lett} {\bf B196}, 126.
\item Lyth, D. H. 1990, {\it Phys Lett} {\bf B236}, 408.
\item Lyth, D. H. 1992, {\it Phys Rev D}{\bf 45}, 3394.
\item Lyth, D. H. 1993a, ``The effect of saxinos on axion cosmology'',
	Lancaster preprint.
\item Lyth, D. H. 1993b, `Peculiar velocity and the Sachs-Wolfe effect',
	in preparation.
\item Lyth, D. H. and Mukherjee, M. 1988, {\it Phys Rev D}{\bf 38}, 485.
\item Lyth, D. H. and Stewart, E. D. 1990a, {\it Phys Lett} {\bf B252}, 336.
\item Lyth, D. H. and Stewart, E. D. 1990b, {\it Astrophys J} {\bf 361}, 343.
\item Lyth, D. H. and Stewart, E. D. 1992a, {\it Phys Lett} {\bf B274}, 168.
\item Lyth, D. H. and Stewart, E. D. 1992b, {\it Phys Rev D}{\bf 46}, 532.
\item Maddox, S. J., Efstathiou, G., Sutherland, W. J. and Loveday, J. 1990,
	{\it Mon Not Roy astr Soc} {\bf 242}, 43p.
\item Maddox, S. J., Sutherland, W. J., Efstathiou, G., Loveday, J.
	and Peterson, B. A. 1991, {\it Mon Not Roy astr Soc} {\bf 247}, 1p.
\item Makino, N. and Suto, Y. {\em Astrophys J} {\bf 405}, 1.
\item Mather, J. C. {\it et al} 1990, {\it Astrophys J Lett} {\bf 354}, L37.
\item Mathewson, D. S., Ford, V. L. and Buchhorn M. 1992, {\it Astrophys J
	Lett} {\bf 389}, L5.
\item Melott, A. C. 1990, {\it Phys Rep} {\bf 193}, 1.
\item Mijic, M B., Morris, M. S. and Suen, W.-M. 1986, {\it Phys
	Rev D}{\bf 34}, 2934.
\item Mukhanov, V. F. 1985, {\it JETP Lett} {\bf 41}, 493.
\item Mukhanov, V. F. 1989, {\it Phys Lett} {\bf B218}, 17.
\item Mukhanov, V. F., Feldman, H. A. and Brandenberger, R. H. 1992, {\it
	Phys Rep} {\bf 215}, 203.
\item Nagasawa, M. and Yokoyama, J. 1992, {\it Nucl Phys} {\bf B370}, 472.
\item Nichol, R. C., Collins, C. A., Guzzo, L. and Lumsden, S. L. 1992,
	{\it Mon Not Roy astr Soc} {\bf 255}, 21p.
\item Nilles, H. P. 1984, {\it Phys Rep} {\bf 110}, 1.
\item Olive, K. 1990, {\it Phys Rep} {\bf 190}, 307.
\item Olson, D. W. 1976, {\it Phys Rev D}{\bf 14}, 32.
\item Ostriker, J. P. 1993, {\it Ann Rev Astron Astrophys}, in press.
\item Ostriker, J. P. and Suto, Y. 1990, {\it Astrophys J} {\bf 348}, 378.
\item Padmanabhan, T. 1989, {\it Phys Rev D}{\bf 39}, 2924.
\item Park, C., Gott, J. R. and da Costa, L. N. 1992, {\it Astrophys J Lett}
	{\bf 392}, L51.
\item Parker, L. 1969, {\it Phys Rev D}{\bf 183}, 1057.
\item Peacock, J. A. and Heavens, A. F. 1990, {\it Mon Not Roy astr Soc} {\bf
	243}, 133.
\item Peacock, J. A. 1991, {\it Mon Not Roy astr Soc} {\bf 253}, 1p.
\item Peacock, J. A. and Nicholson, D. 1991, {\it Mon Not Roy astr Soc}
	{\bf 253}, 307.
\item Peebles, P. J. E. 1970, {\it Astrophys J} {\bf 75}, 13.
\item Peebles, P. J. E. 1980, \sl The Large Scale Structure of the Universe
	\rm (Princeton University Press).
\item Peebles, P. J. E. 1982a, {\it Astrophys J} {\bf 258}, 415.
\item Peebles, P. J. E. 1982b, {\it Astrophys J Lett} {\bf 263}, L1.
\item Peebles, P. J. E. 1984a, {\it Astrophys J} {\bf 277}, 470.
\item Peebles, P. J. E. 1984b, {\it Astrophys J} {\bf 284}, 439.
\item Pen, U.-L., Spergel, D. N. and Turok, N. 1992, ``Cosmic Structure
	Formation and Microwave Anisotropies from Global Field Ordering'',
	Imperial College pre\-print Imper\-ial/TP/92-93/04.
\item Perivolaropoulos , L. 1993, {\it Phys Lett} {\bf B298}, 299.
\item Perivolaropoulos , L. and Vachaspati, T. 1993, ``Peculiar Velocities
	and Microwave Background Anisotropies from Cosmic Strings'',
	Harvard-Smithsonian preprint.
\item Pollock, M. D. 1987, {\it Phys Lett} {\bf B185}, 34.
\item Press, W. H. and Schechter, P. 1974, {\it Astrophys J} {\bf 187},
	452.
\item Primack, J. R. and Blumenthal, G. R. 1984, in {\sl Clusters and Groups
	of Galaxies}, eds Mardirossian, F. {\it et al} (Reidel, Dordrecht).
\item Raychaudhuri, A. K. 1955, {\it Phys Rev} {\bf 98}, 1123.
\item Raychaudhuri, A. K. 1979, {\sl Theoretical Cosmology}
	(Clarendon, Oxford).
\item Sachs, R. K. and Wolfe, A. M. 1967, {\it Astrophys J} {\bf 147}, 73.
\item Sahni, V. 1990, {\it Phys Rev D}{\bf 42}, 453.
\item Sakagami, M. 1988, {\it Prog Theor Phys} {\bf 79}, 442.
\item Salopek, D. S., Bond, J. R. and Bardeen, J. M. 1989, {\it Phys Rev
	D}{\bf 40}, 1753.
\item Salopek, D. S. 1992a, {\it Phys Rev Lett} {\bf 69}, 3602.
\item Salopek, D. S. 1992b, in {\sl Proc of the Int School of Astrophys,
	Erice} ed. N Sanchez (World Scientific Publications).
\item Sasaki, M. 1986, {\it Prog Theor Phys} {\bf 76}, 1036.
\item Saunders, W. {\it et al} 1991, {\it Nature} {\bf 349}, 32.
\item Saunders, W, Rowan-Robinson, M and Lawrence, A. 1992, {\it Mon Not Roy
	astr Soc} {\bf 258}, 134.
\item Scaramella, R. and Vittorio, N. 1988, {\it Astrophys J Lett} {\bf 331},
	L53.
\item Scaramella, R. and Vittorio, N. 1990, {\it Astrophys J} {\bf 353}, 372.
\item Scaramella, R. 1992, {\it Astrophys J Lett} {\bf 390}, L57.
\item Schaefer, R. K., Shafi, Q. and Stecker, F. 1989, {\it Astrophys J}
	{\bf 347}, 575.
\item Schaefer, R. K. 1991, {\it Int J Mod Phys} {\bf A6}, 2075.
\item Schaefer, R. K. and Shafi, Q. 1992, {\it Nature} {\bf 359}, 199.
\item Schaefer, R. K. and Shafi, Q. 1993, {\it Phys Rev D}{\bf 47}, 1333.
\item Schneider, D. P., Schmidt, M. and Gunn, J. E. 1989,
	{\it Astron J} {\bf 98}, 1507.
\item Shafi, Q. 1988, {\it Comments Nucl. Part. Phys} {\bf 18}, 103.
\item Shafi, Q. 1993, ``A Predictive Inflationary Scenario without
	the Gauge Singlet'', Bartol preprint.
\item Shafi, Q. and Stecker, F. W. 1984, {\it Phys Rev Lett} {\bf 53}, 1292.
\item Shafi, Q. and Vilenkin, A. 1984, {\it Phys Rev D}{\bf 29}, 1870.
\item Shandarin S. F. and Zel'dovich, Ya. B. 1989, {\it Rev Mod Phys}
	{\bf 61}, 185.
\item Smoot, G. F. {\it et al} 1992, {\it Astrophys J Lett} {\bf 396}, L1.
\item Souradeep, T. and Sahni, V. 1992, {\it Mod Phys Lett} {\bf A7}, 3541.
\item Starobinsky, A. A. 1979, {\it JETP Lett} {\bf 30} 682.
\item Starobinsky, A. A. 1980, {\it Phys Lett} {\bf B91}, 99.
\item Starobinsky, A. A. 1982, {\it Phys Lett} {\bf B117}, 175.
\item Starobinsky, A. A.. 1986, in \sl Lecture Notes in
	Physics, \rm Vol. 242, eds. de Vega, H. J. and Sanchez, N.,
	(Springer, Berlin).
\item Starobinsky, A. A. and Sahni, V. 1984, in {\sl Modern Theoretical and
	Experimental Problems of General Relativity} (Moscow)
	(in Russian), quoted by Shandarin and Zel'dovich (1989).
\item Starobinsky, A. A. and Schmidt, H.-J. 1987, {\it Class Quant Grav}
	{\bf 4}, 695.
\item Stewart, E. D. and Lyth, D. H. 1993, {\it Phys Lett} {\bf B302}, 171.
\item Suen, W.-M. and Anderson, P. R. 1987, {\it Phys Rev D}{\bf 35}, 2940.
\item Sutherland, W. 1988, {\it Mon Not Roy astr Soc} {\bf 257}, 650.
\item Suto, Y. and Fujita, M. 1990, {\it Astrophys J} {\bf 360}, 7.
\item Suto, Y., Gouda, N. and Sugiyama, N. 1990, {\it Astrophys J Supp}
	{\bf 74}, 665.
\item Suto, Y., Cen, R. and Ostriker, J. P. 1992 {\it Astrophys	J} {\bf
	395}, 1.
\item Taylor, A. N. and Rowan-Robinson, M. 1992, {\it Nature} {\bf 359}, 396.
\item Thomas, P. A.  and Couchman, H. M. P. 1992, {\it Mon Not Roy astr Soc}
	{\bf 257}, 11.
\item Tormen, G., Lucchin, F. and Matarrese, S. 1992, {\it Astrophys J}
	{\bf 386}, 1.
\item Tormen, G. Moscardini, L., Lucchin, F. and Matarrese, S. 1992,
	``The Galaxy Velocity Field and CDM Models'', Padova preprint.
\item Turner, M. S. 1991a, {\it Phys Rev D}{\bf 44}, 12.
\item Turner, M. S. 1991b, {\it Physica Scripta}, {\bf 36}, 167.
\item Turner, M. S. 1993, ``On the Production of Scalar and Tensor
	Perturbations in Inflationary Models'', Fermilab preprint.
\item Turok, N. and Spergel, D. 1990, {\it Phys Rev Lett} {\bf 64}, 2736.
\item van Dalen, T. \& Schaefer, R. K. 1992, {\it Astrophys J} {\bf 398}, 33.
\item Valdarnini, R. and Bonometto, S. A. 1985, {\it Astron Astrophys}
	{\bf 146}, 235.
\item Veeraraghavan, S. and Stebbins, A. 1990, {\it Astrophys J} {\bf 365},
	37.
\item Vilenkin, A. 1981, {\it Phys Rev Lett} {\bf 46}, 1169; 1496[E].
\item Vilenkin, A. and Ford, L. H. 1982, {\it Phys Rev D}{\bf 25}, 1231.
\item Vishniac, E. T., Olive K. A. and Seckel, D. 1987, {\it Nucl Phys}
	{\bf B289}, 717.
\item Vittorio, N., Matarrese, S. and Lucchin, F. 1988, {\it Astrophys J}
	{\bf 328}, 69.
\item Vittorio, N., Meinhold, P., Muciacca, P. F., Lubin, P. and Silk, J.
	 1991, {\it Astrophys J Lett} {\bf 372}, L1.
\item Vogeley, M. S., Park, C., Geller, M. J. and Huchra, J. P. 1992, {\it
	Astrophys J Lett} {\bf 391}, L5.
\item Walker, T., Steigman, G., Schramm, D. N., Olive, K. A. and Kang, H.-S.
	1991, {\it Astrophys J} {\bf 376}, 51.
\item Watson, A. A. {\it et al} 1992, {\it Nature} {\bf 357}, 660.
\item Weinberg, S. 1972, {\sl Gravitation and Cosmology} (Wiley, New York).
\item Weinberg, S. 1993, {\sl Dreams of a Final Theory} (Pantheon/
	Hutchinson).
\item White, M. 1992, {\it Phys Rev D} {\bf 46}, 4198.
\item White, S. D. M., Frenk, C. S. and Davis, M. 1983, {\it Astrophys J Lett}
	{\bf 274}, L1; {\it Astrophys J} {\bf 287}, 1.
\item White, S. D. M., Frenk, C. S., Davis, M. and Efstathiou, G. 1987, {\it
	Astrophys J} {\bf 313}, 505.
\item White, S. D. M., Efstathiou, G. and Frenk, C. S. 1993, ``The
	Amplitude of Mass Fluctuations in the Universe'', Durham preprint.
\item Wright, E. L. {\it et al} 1992, {\it Astrophys J Lett} {\bf 396}, L13.
\item Yokoyama, J. 1988, {\it Phys Lett} {\bf B212}, 273.
\item Yokoyama, J. 1989, {\it Phys Rev Lett} {\bf 63}, 712.
\item Zel'dovich, Ya. B. 1970, {\it Astron Astrophys} {\bf 5}, 84.
\item Zel'dovich, Ya. B. 1980, {\it Mon Not Roy astr Soc} {\bf 192}, 663.
\end{description}
\nonfrenchspacing

\newpage

\begin{table}
\caption{Current models of structure formation}
\begin{tabular}{|c|c|}
\hline \hline
 non-baryonic $\Omega$ & Structure origin \\
\hline
cold dark matter & adiabatic density perturbation \\
30\% hot dark matter, 60\%
cold dark matter & adiabatic density perturbation\\
60\% cosmological constant, 40\% cdm & adiabatic density perturbation\\
hot dark matter & cosmic strings \\
\hline \hline
\end{tabular}
\end{table}

\begin{table}
\caption{Models of inflation}
\begin{tabular}{|c|c|c|c|}
\hline \hline
Model & $n$ & $R$ & $V_1\quarter/1\GeV$ \\
 & $=1+2\eta-6\epsilon$ & $=12\epsilon$ & $=6\epsilon\quarter\times 10^{16}$ \\
\hline
Power-law & $1-(2/p)$ & $6(1-n)$ & \\
$p=10$ & .80 & 1.2 & $3.4\times 10^{16}$ \\
\hline
`Natural' & $1-(2/r)$ & & \\
$r=10$ & .80 & $7\times10^{-6}$ & $1.2\times10^{15}$ \\
\hline
$R^2$ & $.96$ & $4\times 10\mthree$ & $8\times10^{15}$ \\
\hline
$\phi^\alpha$ & $1-.0083(\alpha+2)$ & $6(1-n)-0.1$  & \\
$\alpha=2$ & $.966$ & .10 & $1.9\times10^{16}$ \\
\hline
Two-scale & & &\\
$m=10^3\GeV$ & $1.0001$ & $10^{-20}$ & $4\times 10^{11}$ \\
$m=10^{13}\GeV$ & $1.14$ & $.007$ & $9 \times 10^{15}$ \\
\hline \hline
\end{tabular}
\end{table}

\newpage
\section*{Figure Captions}

\vspace{0.3cm}
\noindent
{\em Figure 1a}\\
The transfer function used to obtain results in this paper, following the
parametrisation of Efstathiou (1990) based on $4$ fitting parameters. This is
obtained by scaling the values obtained by Bond and Efstathiou (1984) for
h=0.75.

\vspace{0.3cm}
\noindent
{\em Figure 1b}\\
A selection of transfer functions plotted by fractional departure from that of
Efstathiou (1990), labelled $T_E$, as used in this paper. We show Bardeen {\it
et al} (1986), Davis {\it et al} (1985), Holtzman (1989), Starobinsky and
Sahni (1984) and an alternative from Bond and Efstathiou (1984) based on
simulations with $h = 0.5$. Each given as appropriate to CDM universes with
$\Omega = 1$, $h = 0.5$, low baryonic content and three neutrino species. They
feature $4$, $5$, $3$, $2$ and $4$ fitting parameters respectively. Not all
parametrisations have attempted to fit down to the smallest scales.
As we shall see later, spectra normalised to galaxies typically cross at $k =
0.1$ Mpc$^{-1}$. Consequently, the relative normalisation of the spectra is
well defined by the value of the transfer function at this point. The
different parametrisations show sizeable discrepencies.

\vspace{0.3cm}
\noindent
{\em Figure 2}\\
The normalisation $b_8\delta_H(k=1 \Mpc\mone)$ as a function of $n$. The solid
line is from the $\sigma_8$ normalisation (used in this paper), while the
dashed line shows the rival $J_3$ normalisation.

\vspace{0.3cm}
\noindent
{\em Figure 3a}\\
The present day spectra, as calculated in the linear approximation, for a
selection of values of $n$. One sees the additional large-scale power and the
deficit on short scales when one compares $n < 1$ to the standard CDM
spectrum. Note $k$ is in Mpc$^{-1}$, without a factor of $h$.

\vspace{0.3cm}
\noindent
{\em Figure 3b}\\
The dispersion $b_8\sigma(M)$ as a function of mass, for $n=1$ and $n=0.7$, and
with both choices of filter. The top hat filtered spectra are unity at $M \sim
10^{15}\msun$, as required by the normalisation. The gaussian filtering gives
significantly smaller answers than does the top hat, as its smearing gives a
higher contribution to the larger scales.

\vspace{0.3cm}
\noindent
{\em Figure 4}\\
This figure is reproduced from the paper of
Crittenden {\it et al} 1993, which cites
references for the various experiments. The top Figure shows the spectrum of
the cmb anisotropy for a slightly tilted standard CDM model with spectral
index $n=0.85$ and $\Omega_B=0.05$. As discussed in the text, the case $n=1$
is recovered for $l\gg1$ by multiplying the curves by a factor
$l^{(1-n)/2}=l^{.075}$. The contribution of an adiabatic density perturbation
is the middle line, labelled `S', and the contribution of gravitational waves
is the bottom line, labelled `T'. The light dashed line is the density
contribution with $\Omega_B=0.01$. For each curve, the quantity plotted is
$l(l+1) \Sigma_l^2$, normalised to 1 for the quadrupole, $l=2$. If the
gravitational wave contribution to the quadrupole is equal to that of the
density perturbation, as is roughly the case for power-law inflation with
$n=0.85$, the top curve indicates the total. On the other hand, it could well
be that the gravitational contribution is negligible. The bottom Figure shows
the filters $2F_l$
for various experiments, as defined in Section \ref{MWB}.1. SP89
denotes the South Pole experiment of Meinhold and Lubin (the UCB/UCSB MAX
experiment has a similar filter), Ten denotes the Tenerife experiment, MIT
denotes the balloon experiment of Page {\em et al} and SP91 denotes the Gaier
{\em et al} and Schuster {\em et al} experiment. OVRO22 is a planned
experiment to be undertaken using the 5 meter dish at Owens Valley. OVRO is
the original Readhead experiment.

\vspace{0.3cm}
\noindent
{\em Figure 5}\\
The triangles indicate the prediction for the mean quadrupole as a function of
$n$ for power-law inflation, and the stars for natural inflation. The dotted
lines plot the COBE observation for two choices of bias (note the bias
implicit in the $y$-axis). The CDM prediction for the mean at bias 1 is very
close to the COBE result. The vertical bars on the starred data indicate the
spread of the {\em pdf} for the quadrupole (for clarity the bars have been
omitted for the triangles --- they are the same size); 95\% of the {\em pdf}
is above the bottom of the bars, while 95\% is below the top. The $\chi_5^2$
distribution is not symmetric, so the bars are skewed to higher values
(somewhat concealed by the log plot). A value of $n$ is allowed at 95\%
exclusion if the observations cut through its vertical bar. Modelling the
observational errors (see text) gives an even looser criterion.

\vspace{0.3cm}
\noindent
{\em Figure 6}\\
The predictions for $b_8 \left. \frac{\Delta T}{T}\right|_{10^0}$, as defined
in
text, along with the COBE limits (at $1$-sigma) for bias $1$ and $2$. The
upper line represents the power-law inflation predictions, the lower those for
natural inflation. As for the quadrupole, the $n=1$ prediction at bias $1$ is
very close to the COBE result.

\vspace{0.3cm}
\noindent
{\em Figure 7}\\
The predicted angular correlation functions for a choice of $n$ are plotted
alongside the observational data from the APM survey (Maddox {\it et al}
1990). With the anticipated residual systematics, values of $n$ between about
0.3 and 0.6 provide reasonable fits, while the standard CDM curve falls well
below the data.

\vspace{0.3cm}
\noindent
{\em Figure 8}\\
The predicted {\em rms} velocity flows, when smoothed with a top hat of radius
$R_f$, for different choices of $n$.

\vspace{0.3cm}
\noindent
{\em Figure 9}\\
The predicted {\em rms} velocity flows in a configuration mimicking the POTENT
observational data. The velocity field is first smoothed with a $12h^{-1}$ Mpc
gaussian, reducing the short scale power, and then smoothed with top hat
filters of radius $R_f$, giving predictions smaller than in figure 9. The
solid lines indicate the predictions for $n=1$ and $n=0.7$. The stars indicate
the POTENT observational data at bias $1$ (read from figure 4 of Dekel
(1991)), and the triangles the same at bias $1.6$. The error bars on the data
(the last ones just overlap) are $1$-sigma. Finally, we emphasise that the
theoretical curves are averages over all observer points, whereas the
observations are a single realisation, with correlated errors due to long
wavelength domination of bulk flows.

\vspace{0.3cm}
\noindent
{\em Figure 10}\\
The asphericity parameter $x\mone$ plotted against $\nu$.

\vspace{0.3cm}
\noindent
{\em Figure 11}\\
The ratio $2n\sub{up}/n_\chi$ plotted against $\nu$.

\vspace{0.3cm}
\noindent
{\em Figure 12}\\
The ratio $V\sub{peak}/V_f$ plotted against $\nu$.

\vspace{0.3cm}
\noindent
{\em Figure 13}\\
The ratio $n\sub{up}/n\sub{P-S}$ plotted versus $\nu$.
The ratio is not very sensitive to $n$, and only the case
$n=1$ is plotted.

\vspace{0.3cm}
\noindent
{\em Figure 14}\\
The number density $n(>M)$ as a function of redshift. It is calculated
theoretically by equating it with the number density of peaks of the linearly
evolved density contrast, filtered on the mass scale $M$. Figures 16a--d
refer respectively to $M=10^{15}\msun$, $10^{12}\msun$ $10^{10}\msun$ and
$10^8\msun$. For each case there are three curves. They correspond to the
three choices $\{n,b\}=\{.7,1.6\},\{1,1.6\}$ and $\{1,1\}$ which mark the
corners of the more or less triangular region allowed by the QDOT and COBE
data (figure 15). The arrows on the vertical axes give the observed galaxy
number densities at the present epoch. The arrows on the horizontal axes
indicate very roughly the `observed' range of formation epochs, deduced from
the indicated range of virial velocities. Each curve ends at the epoch when
$\sigma(M)=1$, signalling the end of linear evolution. According to the theory
of biased galaxy formation, luminous galaxies form significantly before that
epoch. In the case of bright galaxies, the epoch can be calculated by
demanding that the resulting bias factor is equal to $b_8$, and it is indicated
by a star in figure 16 b. The theoretical and observational uncertainties in
are discussed in the text.

\vspace{0.3cm}
\noindent
{\em Figure 15}\\
The limits on the tilted CDM models in the $n$--$\sigma_8$ plane, where $b_8
\equiv 1/\sigma_8$. COBE limits are shown for two different inflation models,
and extrapolated to $n>1$ in the case with no gravitational waves. The COBE
and QDOT limits should be considered as $1$-sigma bands about an unplotted
mean, while the APM and pairwise velocities limits have greater significance.
See the text, Section \ref{CLUSTER}.3, for a detailed discussion.

\vspace{0.3cm}
\noindent
{\em Figure 16}\\
The dispersion of the density field is plotted for both tilted models ($n=0.7$
both with and without gravitational waves) and for mixed dark matter. The
dispersion is normalised to the standard CDM value. The ability of mixed dark
matter to fit the clustering data is due to the sharp fall-off of the relative
power on scales 20--50 Mpc.

\vspace{0.3cm}
\noindent
{\em Figure 17}\\
The dispersion of the density contrast, when normalised to COBE, as a function
of mass for both standard CDM and for mixed dark matter, utilising the
transfer function of Klypin {\it et al} (1992).

\end{document}